\documentclass[aps,preprintnumbers,nofootinbib,twocolumn,amsmath,amssymb,floatfix,superscriptaddress,pra]{revtex4-1}
\pdfoutput=1
\usepackage[utf8x]{inputenc}
\usepackage{amsmath}
\usepackage{amsfonts}
\usepackage{braket}
\usepackage{tensor}
\usepackage{amssymb}
\usepackage{graphicx}
\usepackage{titlesec}
\usepackage[titletoc,toc,title]{appendix}
\def\be{\begin{equation}}
\def\ee{\end{equation}}
\def\bea{\begin{eqnarray}}
\def\eea{\end{eqnarray}}

\def\XXint#1#2#3{{\setbox0=\hbox{$#1{#2#3}{\int}$}
         \vcenter{\hbox{$#2#3$}}\kern-.5\wd0}}

\renewcommand{\AA}{\mathbb{A}}

\usepackage[colorlinks=true,linkcolor=blue, citecolor=blue, urlcolor=blue, bookmarks]{hyperref}

\begin{document}

\title{Dynamics of large deviations in the hydrodynamic limit: Non-interacting systems}

\author{Gabriele Perfetto}	
\affiliation{SISSA --- International School for Advanced Studies, via Bonomea 265, 34136 Trieste, Italy.}
\affiliation{INFN, Sezione di Trieste, via Bonomea 265, 34136, Trieste, Italy.}
\author{Andrea Gambassi}
\affiliation{SISSA --- International School for Advanced Studies, via Bonomea 265, 34136 Trieste, Italy.}
\affiliation{INFN, Sezione di Trieste, via Bonomea 265, 34136, Trieste, Italy.}

\begin{abstract}
We study the dynamics of the statistics of the energy transferred across a point along a quantum chain which is prepared in the inhomogeneous initial state obtained by joining two identical semi-infinite parts thermalized at two different temperatures. In particular, we consider the transverse field Ising and harmonic chains as prototypical models of non-interacting fermionic and bosonic excitations, respectively. Within the so-called hydrodynamic limit of large space-time scales we first discuss the mean values of the energy density and current, and then, aiming at the statistics of fluctuations, we calculate exactly the scaled cumulant generating function of the transferred energy. From the latter, the evolution of the associated large deviation function is obtained. A natural interpretation of our results is provided in terms of a semi-classical picture of quasi-particles moving ballistically along classical trajectories. Similarities and differences between the transferred energy scaled cumulant and the large deviation functions in the cases of non-interacting fermions and bosons are discussed.
\end{abstract}

\maketitle

\section{Introduction}
\label{sec:intro}
Experimental progresses in the physics of cold atoms (see, e.g., Refs.~\cite{greiner2002collapse,kinoshita2006quantum,hofferberth2007non}) have turned the theoretical study of the non-equilibrium unitary dynamics of one-dimensional quantum statistical systems into a very active area of research (see, for instance Refs.~\cite{1742-5468-2016-6-064001,eisert2015quantum,polkovnikov2011colloquium,altman2015nonBIS,calabrese2015non}, for some reviews). 
A paradigmatic protocol for investigating non-equilibrium dynamics is the so-called homogeneous quantum quench \cite{calabrese2006time,calabrese2007quantum,calabrese2011quantum0,calabrese2012quantum1,calabrese2012quantum2,fagotti2013reduced} consisting in an instantaneous change of the value of a global parameter of the Hamiltonian governing the time evolution of the system. Despite the dynamics of the whole system remains unitary at all times, local observables are known to relax towards stationary values expressed, for integrable models, as an average over a generalized Gibbs ensemble (GGE), \cite{vidmar2016generalized} including local (and quasi-local) conserved quantities. Among the relevant phenomena occurring out of equilibrium, transport is usually studied theoretically and experimentally  in stationary states of open mesoscopic systems (see, e.g., Refs.~\cite{znidaric2010matrix,Prosen2011openXXZ,vznidarivc2011transport,ILIEVSKI2014,Carollo2017fluctuating,Carollo2018fluctuations}) and, in fact, it is now possible to measure the heat current flowing between two leads kept at different temperatures \cite{schwab2000measurement,jezouin2013quantum}.  

In order to study transport phenomena in isolated systems, it is more convenient to adopt the so-called partitioning protocol \cite{spohn1977stationary,vasseur2016nonequilibrium,bernard2016conformal}, whereby an homogeneous (i.e., translationally invariant in space) and stationary (translationally invariant in time) non-equilibrium steady state (NESS) is generated by gluing together at time $t=0$ two identical systems initially at thermal equilibrium at two different inverse temperatures $\beta_r$ and $\beta_l$. 
The resulting initial state is thus described by a density matrix $\rho_0$ which is inhomogeneous because of the two different temperatures, while the subsequent dynamics is determined by a translationally invariant Hamiltonian: accordingly, this protocol is referred to as inhomogeneous quench. Several aspects of the ensuing dynamics have been investigated in the literature. In particular, in one-dimensional quantum critical systems described by conformal field theory (CFT), the presence of independent right- and left- moving excitations renders the thermal transport ballistic and a universal expression for the energy current in the NESS has been derived \cite{bernard2012energy,bernard2015non,bernard2016conformal}.
Also in non-interacting models, quasi-particles excitations propagate independently without scattering and therefore the picture described by CFT carries over to these cases. This leads to a number of exact predictions concerning not only the NESS, \cite{bernard2013time,de2013nonequilibrium,de2015stationary,doyon2015non} but also, more generally, the so-called hydrodynamic, space-time or semi-classical limit within which the space coordinate $x$ along the chain and the time $t$ are both assumed to be large with fixed ratio $v=x/t$ \cite{antal1999transport,karevski2002scaling,platini2005scaling,collura2014non,collura2014quantum,allegra2016inhomogeneous,
viti2016inhomogeneous,bertini2016determination,eisler2016universal,perfetto2017ballistic,
kormos2017inhomogeneous,10.21468/SciPostPhys.6.1.004,mitra2018quantum}.
Remarkably, this analysis has been extended to interacting integrable models within the so-called generalized hydrodynamic theory (GHD), \cite{castro2016emergent,bertini2016transport} (see Ref.~\cite{doyon2019lecture} for a review). GHD is an integrability-Bethe ansatz based method to study the time evolution in the presence of spatial inhomogeneities. For the inhomogeneous initial state generated by the partitioning protocol introduced above, in particular, several exact results have been by now obtained for the dynamics of the mean values of charge densities and of the corresponding currents, for correlation functions, and entanglement entropy \cite{fagotti2016charges,doyon2017dynamics,piroli2017transport,bulchandani2017solvable,collura2018analytic,
AlviseGHD2018,DeNardis2018,kormossemiclassic2,DoyonLargeScaleCorrelations2018,bertini2018entanglement,EeEvolutionGHD2019,moller2020correlations}. 

Going beyond mean values, the full probability density function of the total energy $\Delta e(x,t)$ transferred across a point $x$ of the system up to time $t$ after the quench is of great physical relevance, as it encodes all the information about fluctuations of this quantity. Since in non-interacting models quasi-particles propagate ballistically, the transferred energy is expected to depend extensively on time $t$ and it is therefore convenient to focus on the associated intensive variable $J_E=\Delta e(x,t)/t$. The framework of large deviation theory \cite{touchette2009large} then provides the asymptotic behavior at large times $t$ of the PDF of the scaled transferred energy as $p(J_E) \sim \mbox{exp}[-t I(J_E)]$, where $I(J_E)$ is the so-called large deviation or rate function, which is non-negative and has a unique zero at the mean value $\langle J_E \rangle $, implying that the PDF at large times concentrates around the mean value with fluctuations exponentially suppressed as $t$ increases. The function $I(J_E)$ can be calculated from the knowledge of the associated scaled cumulant generating function (SCGF), which is equivalent, in the terminology used for quantum transport, to the knowledge of the full counting statistics (FCS).

Despite many investigations of the probability density function of various observables not related to transport, such as transverse and longitudinal magnetization in spin chains \cite{lamacraft2008order,eisler2013full,groha2018full,essler2020order,Collura2020,calabrese2020full,tortora2020relaxation}, particle number in the one-dimensional Bose gas \cite{armijo2010probing,pietraszewicz2017mesoscopic,lovas2017full,bastianello2018exact,bastianello2018sinh,arzamasovs2019full}, and work statistics \cite{silva2008statistics,gambassi2012large,sotiriadis2013statistics,smacchia2013work,rotondo2018singularities,
palmai2014quench,rylands2019loschmidt,perfetto2019quench,rylands2019quantum}, very few predictions exist for the scaled cumulant generating function of the transferred energy in inhomogeneous quenches. 
In particular, for one-dimensional critical systems a universal expression for the SCGF of the energy current in the NESS arising from the partitioning protocol has been obtained in Refs.~\cite{bernard2012energy,bernard2015non,bernard2016conformal} via CFT. For free fermionic theories in one dimension, an analytic expression for the FCS in the steady state has been originally determined by Levitov and Lesovik in Refs.~\cite{levitov1993charge,levitov1994quantum,levitov1996electron} and later rederived for both lattice and field theory models which can be mapped to free fermions \cite{klich2003elementary,schonhammer2007full,bernard2012full,de2013nonequilibrium,klich2014note,yoshimura2018full,Gamayun1,Gamayun2}.
As far as transport of bosons is concerned, instead, an expression for the NESS heat current and FCS has been obtained for the free Klein-Gordon field theory in arbitrary spatial dimension \cite{doyon2015non} and for a one-dimensional system of harmonic oscillators on the lattice in contact with two heat baths at different temperatures \cite{saito2007fluctuation}. The expression for the FCS can be considered as the equivalent of the Levitov-Lesovik formula for the free bosonic case.
Importantly, going beyond free models, in Refs.~\cite{doyonMyers2020,doyon2020fluctuations}, the expression of the FCS of the transferred energy in homogeneous and stationary GGEs, including the NESS which develops at long times in the partitioning protocol analyzed here, has been derived for one-dimensional interacting integrable systems by using the generalized hydrodynamics mentioned above.        

So far, however, the SCGF of the transferred energy $\Delta e(x,t)$ has been determined only for homogeneous and stationary states like the NESS, as mentioned above, where is independent of the space coordinate $x$ and the time $t$. For the inhomogeneous and dynamical state $\rho_0$, instead, the SCGF depends on $x$ and $t$ through the scaling variable $v=x/t$, with the particular value $v=0$ corresponding to the NESS. The dependence of the SCGF on $v$ is a consequence of the inhomogeneity of the state $\rho_0$ and therefore it is beyond reach of the analysis of Refs.~\cite{doyonMyers2020,doyon2020fluctuations}, which is restricted to homogeneous and stationary states. The complete dynamics in the hydrodynamic limit of the cumulant generating function as a function of $v$ for inhomogeneous initial states like $\rho_0$ has therefore never been addressed, neither for non-interacting, nor for interacting but possibly integrable models.
In this work we aim at filling this gap, starting from the simplest case, i.e., from the calculation in the hydrodynamic limit of the FCS for the inhomogeneous state $\rho_0$ in free fermionic and bosonic theories. Specifically, we will focus on the transverse field Ising chain (TFIC) realizing the former, and, on the harmonic oscillators chain, realizing the latter. In both cases we derive the analytical expression for the space-time scaling limit of the SCGF as a function of $v$, providing a simple semi-classical interpretation of the obtained results in terms of quasi-particles moving along classical trajectories. By taking the Legendre-Fenchel trasform \cite{touchette2009large} of the SCGF, the space-time scaling dynamics of the large deviation function $I(J_E,v)$ as a function of $v$ is derived and discussed. By comparing the fermionic and the bosonic large deviation functions it emerges that the quasi-particles statistics, which weakly affects the profile of the mean energy current, deeply influences the energy current fluctuations. In particular, the large deviation function $I(J_E,v)$, independently of the value of $v$, turns out to have support on a finite interval of energy currents $J_E$ for the TFIC model, while this support extends to all real values of the current for the harmonic chain. 

The rest of the presentation is organized as follows: In Sec.~\ref{firstsection} we briefly report the main formulas entering in the exact solution of the TFIC, presented in Subsec.~\ref{sec:Ising_solution}, and of the chain of harmonic oscillators, in Subsec.~\ref{sec:Oas_solution}. In Sec.~\ref{secondsection} we summarize known results concerning the mean energy current profile in the space-time scaling limit for the TFIC, Subsec.~\ref{sec:isingtransportsub}, while we first derive the prediction for the same quantities in the harmonic chain, in SubSec.~\ref{sec:oastransportsub}. In Sec.~\ref{thirdsection}, the hydrodynamic limit of the SCGF is first defined and then we recall known facts concerning the long-time limit of this function, i.e., in the NESS. Subsection \ref{sec:SCGF_general_result} presents the main result regarding the FCS in the space-time scaling limit and the corresponding derivation. In Subsec.~\ref{sec:semi_classics} we discuss a semi-classical picture for calculating the same function, in agreement with the previous section. In Subsecs.~\ref{sec:FCSTFIC} and \ref{sec:SCGF_bosons} we specialize the general expression of Sec.~\ref{sec:SCGF_general_result} to the fermionic and bosonic cases, respectively, and we determine the Legendre-Fenchel transform of the SCGF, discussing the resulting large deviation function $I(J_E,v)$ for both cases. Finally, we summarize our findings in Sec.~\ref{finalcomments}, while the technical aspects of the various calculations of the work are presented in the appendices.  

\section{Non-interacting models and their exact solutions} 
\label{firstsection}
The initial density matrix $\rho_0$ of the partitioning protocol \cite{spohn1977stationary,vasseur2016nonequilibrium,bernard2016conformal} is given by
\be
\rho_0 = e^{-\beta_r H_r} \otimes e^{-\beta_l H_l}/Z ,
\label{eq:intro-rho0}
\ee
where $H_r$ and $H_l$ are the Hamiltonians corresponding to the two parts of the system (e.g., two complementary but otherwise identical semi-infinite chains) initially at thermal equilibrium at the inverse temperatures $\beta_r$ and $\beta_l$, respectively, while $Z$ is the associated partition function. In Subsection \ref{sec:Ising_solution} we take for $H_{r,l}$ the transverse field Ising Hamiltonian, which corresponds to fermionic quasi-particle excitations, while in Subsection \ref{sec:Oas_solution} the harmonic chain Hamiltonian is considered, which is described by bosonic quasi-particle excitations.   

\subsection{The quantum Ising chain in a transverse field}
\label{sec:Ising_solution}
As anticipated above, in the partitioning protocol, two originally disconnected identical chains of length $N$ are joined at the initial time $t=0$. The right ($r$) and left ($l$) Hamiltonians before the quench are, respectively,
\begin{subequations}
\begin{align}
& H_r = -\frac{J}{2}\left[\sum_{n=1}^{N-1} \sigma_n^x \sigma_{n+1}^x  + h \sum_{n=1}^{N} \sigma_n^z \right], \\
& H_l = -\frac{J}{2}\left[\sum_{n=1}^{N-1} \sigma_{-n}^x \sigma_{-n+1}^x + h \sum_{n=0}^{N-1} \sigma_{-n}^z \right],
\end{align} 
\label{eq:chains}%
\end{subequations}
with $H_0 = H_r + H_l$ being the pre-quench Hamiltonian, $\sigma_n^{x,y,z}$ are the usual spin $1/2$ Pauli matrices at lattice site $n$, while $J$ and $h$ are the microscopic parameters of the model, i.e., the strength of the ferromagnetic interaction and the transverse field, respectively. The right chain is defined on the lattice sites labeled by $\{1, 2,\ldots, N\}$, while the left one on the sites $\{-N+1,-N+2, \ldots, 0\}$. Open boundary conditions are assumed for both chains. With a Jordan-Wigner transformation (see, e.g., Ref.~\cite{sachdev2007quantum}) one writes the Hamiltonians $H_{r,l}$ in terms of the Jordan-Wigner lattice fermionic operators $c_n$ at site $n$  
\begin{equation}
c_n = \left( e^{i \pi \sum_{m=1}^{n-1} {\sigma_m^- \sigma_m^{+}}}\right)  \sigma_n^+ = \left(\prod_{m=1}^{n-1} \sigma_m^z\right) \sigma_n^+, \label{eq:Jordan_Wigner}
\end{equation}
where $\sigma_m^{\pm} = (\sigma_m^x \pm i \sigma_m^y)/2$ are the spin raising and lowering operators. The model can then be mapped into one of free fermions by introducing, in the thermodynamic limit $N \rightarrow \infty$, the fermionic mode operators $\Phi_{r,l}(k)$ via a Bogoliubov rotation. We follow the notation of Ref.~\cite{perfetto2017ballistic}, where the operators $\Phi_{r,l}(k)$ in the thermodynamic limit are defined starting from their corresponding expressions at finite $N$ (see also Appendix \ref{app:appendix1} for additional details). In the thermodynamic limit, the Hamiltonian eventually reads:
\begin{equation}
H_{r,l} = \int_0^\pi dk \; \varepsilon (k) \Phi_{r,l}^\dagger (k) \Phi_{r,l} (k),
\label{eq:HRcont}
\end{equation}
with the single-particle energy spectrum
\begin{equation}
\varepsilon (k) = J \sqrt{h^2-2 h \,\mbox{cos} \, k +1}.
\label{eq:Isingspectrum}
\end{equation}

At time $t=0$ the two chains are instantaneously joined in order to form a unique, homogeneous chain with Hamiltonian:
\begin{eqnarray}
H &=& H_0 + \delta H = H_0 -\frac{J}{2} \sigma_0^x \sigma_1^x,  
\label{eq:postquenchH-0}
\end{eqnarray}
with $\delta H$ representing the local interaction determined by the junction of the left and right chains through their closest end points at $n=0$ and $n=1$, respectively. After the quench $H_0 \rightarrow H$, since there is no impurity and the two half-chains are equal, the Hamiltonian becomes translationally invariant, i.e., $[H,P_{tr}]=0$, where $P_{tr}$ is the translation operator 
\begin{equation}
\sigma_{n-1}^\alpha = P_{tr}^\dagger \sigma_n^\alpha P_{tr}, \quad \mbox{with} \quad \alpha =x,y,z. \label{eq:translation_operator_TFIC}
\end{equation} 
It is then possible to introduce two fermionic operators $\Psi_{R,L}(k)$, satisfying fermionic canonical anticommutation relations $[\Psi_{R,L}(k),\Psi_{R,L}^{\dagger}(k')]_{+} = \delta(k-k')$, for each value of the wavevector $k$ corresponding to right- and left-moving fermionic quasi-particles excitations, respectively, which acquire opposite phases under the action of the translation operator, i.e.,  
\begin{equation}
P_{tr}^\dagger \Psi_{R,L} (k) P_{tr} = e^{\mp ik} \Psi_{R,L} (k).
\label{eq:transinvariance}
\end{equation}
For the sake of completeness we report here the definition of $\Psi_{R,L}(k)$, in the thermodynamic limit $N \rightarrow \infty$, in terms of the Jordan-Wigner fermions of Eq.~\eqref{eq:Jordan_Wigner}, following the notation of Ref.~\cite{perfetto2017ballistic}:
\begin{equation}
\Psi_{R,L}(k) = \sum_{n=-\infty}^{+\infty} \left[c_n \omega^n_{R,L}(k) + c_n^{\dagger} \xi^n_{R,L}(k) \right],
\label{eq:post_quench_modes_TFIC}
\end{equation}
where
\begin{equation}
\omega^n_{R}(k) = \frac{1}{2}\frac{1}{\sqrt{2 \pi}} e^{-i n k +k}(1 + e^{-i f(k)}), \label{eq:post_quench_TFIC_functions_1} \\
\end{equation}
\begin{equation}
\xi^n_{R}(k) = \frac{1}{2}\frac{1}{\sqrt{2 \pi}} e^{-i n k +k}(1 - e^{-i f(k)}), \label{eq:post_quench_TFIC_functions_2}
\end{equation}
while $\omega^n_{L}(k)$ and $\xi^n_{L}(k)$ can be simply expressed in terms of the corresponding ``right" functions $\omega^n_{R}(k)$ and $\xi^n_{R}(k)$ as 
\begin{equation}
\omega^n_{L}(k) = \omega^n_{R}(-k) \, e^{i(k-f(k))} \, \, \, , \, \, \,  \xi^n_{L}(k) =  \xi^n_{R}(-k) \, e^{i(k-f(k))}, 
\label{eq:left_TFIC_post_quench_functions}
\end{equation}
with $f(k)$ given by 
\begin{equation}
f(k) = \mbox{arctan}\left( \frac{\mbox{sin} \, k}{\mbox{cos} \, k-h} \right).
\label{eq:Bogoliubov_angle}
\end{equation}
From Eq.~\eqref{eq:left_TFIC_post_quench_functions} it immediately follows that
\begin{equation}
\Psi_L(k) = e^{i(k-f(k))} \Psi_R(-k),
\label{eq:opposite_velocities}
\end{equation}
i.e., right-moving quasi-particles excitations with momentum $k$ have opposite momentum with respect to left moving ones $\Psi_L(k)$, but the same energy since $\varepsilon(k)$ in Eq.~\eqref{eq:Isingspectrum} is an even function of $k$. From Eq.~\eqref{eq:opposite_velocities} and the fermionic canonical anticommutation relations, the operators $\Psi_{R,L}(k)$, introduced after Eq.~\eqref{eq:translation_operator_TFIC}, satisfy the following anticommutation relation between $\Psi_R(k)$ and $\Psi_L(k)$: $[\Psi_{R}(k),\Psi_{L}^{\dagger}(k')]_{+}=\delta(k+k') \, \mbox{exp}(-i(k'-f(k')))$. On the basis of Eq.~\eqref{eq:opposite_velocities} one also obtains that $\Psi_R(k)$ and $\Psi_L(-k)$ commute, i.e., $[\Psi_R(k),\Psi_L(-k)]=0$. 
The post-quench Hamiltonian takes the diagonal form 
\begin{align}
H &= \int_0^\pi dk \; \, \varepsilon (k) \left[ \Psi_R^\dagger (k)  \Psi_R (k) + \Psi_L^\dagger (k) \Psi_L (k) \right] \nonumber  \\
  &\equiv H_R + H_L, \label{eq:postquenchhamiltonian}
\end{align}
which makes explicit the free-fermionic nature of the model.
\subsection{The harmonic chain}
\label{sec:Oas_solution}
The right ($r$) and left ($l$) Hamiltonians of the chains of harmonic oscillators are
\begin{subequations}
\label{oaschains}
\begin{align}
& H_r = \frac{1}{2} \sum_{x=1}^{N} \left(p_x^2 +m^2\phi_x^2\right) + \frac{1}{2} \sum_{x=0}^{N} \omega^2 (\phi_{x+1}-\phi_{x})^2,  \label{eq:prequenchoscillators_right} \\
& H_l = \frac{1}{2} \sum_{x=0}^{N-1} \left(p_{-x}^2 +m^2\phi_{-x}^2\right) + \frac{1}{2} \sum_{x=0}^{N} \omega^2 (\phi_{-x+1}-\phi_{-x})^2, \label{eq:prequenchoscillators}
\end{align} 
\end{subequations}
respectively, where the position operator $\phi_x$ and the momentum operator $p_x$ satisfy the equal-time canonical commutation relations $[\phi_x,p_y]= i\delta_{x,y}$, with all the other possible commutators vanishing, $m$ is the ``mass" of the oscillators and $\omega$ their angular frequency. As in the case of the Ising model discussed in the previous subsection, the right chain consists of $N$ lattice sites indexed by $\{1, 2,\ldots, N\}$, while the left chain is defined on the lattice sites $\{-N+1,-N+2, \ldots, 0\}$.  For both chains we assume Dirichlet boundary conditions, which read 
\be 
\phi_0 = \phi_{N+1} \equiv 0 \quad \mbox{and} \quad p_0 = p_{N+1} \equiv 0 \label{eq:fixedwall}
\ee  
for the right chain, while $\phi_1 = \phi_{-N} \equiv 0$ and $p_1 = p_{-N} \equiv 0$ for the left one. 

The first step for solving the model (here we provide some details for the right chain; the similar analysis for the left one is reported in Appendix \ref{app:appendix1}) is to introduce, in the thermodynamic limit $N \rightarrow \infty$, the operators $\hat{\phi}_r(k)$, $\hat{p}_r(k)$ for the right ($r$) chain (see, e.g., Ref.~\cite{lievens2008linear}), 
\begin{eqnarray}
\hat{\phi}_r(k) &=& \; \sqrt{\frac{2}{\pi}} \sum_{x=1}^{\infty} \mbox{sin}(k x) \phi_x,   \nonumber \\
\hat{p}_r(k) &=& \; \sqrt{\frac{2}{\pi}} \sum_{x=1}^{\infty} \mbox{sin}(k x) p_x, 
\label{eq:ftransf} 
\end{eqnarray}
in terms of which $\phi_x$ and $p_x$ are expressed as
\begin{eqnarray}
\phi_x &=& \; \sqrt{\frac{2}{\pi}} \int_{0}^{\pi} dk \, \mbox{sin}(k x) \hat{\phi}_r(k), \nonumber \\
p_x &=& \; \sqrt{\frac{2}{\pi}} \int_{0}^{\pi} dk \, \mbox{sin}(k x) \hat{p}_r(k).
\label{eq:inverseftransf}
\label{ftransfinv} 
\end{eqnarray}
In the thermodynamic limit the set of allowed values of $k$ is continuous within the interval $[0,\pi]$, due to the fact that for finite $N$ its values $k_n$ are discrete according to the integer $n=1,2,...N$, from the boundary condition Eq.~\eqref{eq:fixedwall}, with
\be
k_n = \frac{\pi n}{N+1}. \label{eq:discretekvalues}
\ee
Note that, as a consequence of the presence of the sine function in Eq.~\eqref{eq:inverseftransf}, the boundary condition for $\phi_x$ and $p_x$ in $0$ is automatically fulfilled. In terms of the operators $\hat{\phi}_r$ and $\hat{p}_r$, the usual creation and annihilation operators can be introduced,
\begin{eqnarray}
A_r^{\dagger}(k_n) = \frac{1}{\sqrt{2 \Omega(k)}}\left[\Omega(k) \hat{\phi}_r(k) - i \hat{p}_r(k)\right], \nonumber \\
A_r(k) = \frac{1}{\sqrt{2 \Omega(k)}}\left[\Omega(k) \hat{\phi}_r(k) + i \hat{p}_r(k)\right],  \label{eq:bosonicmodes}
\end{eqnarray}
which satisfy the canonical commutation relations $[A_r(k),A_r^{\dagger}(k')] = \delta(k-k')$, $H_r$ in Eq.~\eqref{eq:prequenchoscillators_right} then takes the diagonal form 
\be
H_r = \int_{0}^{\pi} dk \, \Omega(k) A_r^{\dagger}(k) A_r(k), \label{eq:diagonalrightbosons}
\ee
where $\Omega(k)$ denotes the single-particle dispersion relation given by 
\be
\Omega(k) = \sqrt{m^2 +2 \omega^2 (1-\mbox{cos} \, k)}, \label{eq:oasspectrum}
\ee
which has the same qualitative dependence on $k$ as Eq.~\eqref{eq:Isingspectrum} and becomes identical to it upon identifying $\omega \mapsto J \sqrt{h}$ and $m \mapsto J |h-1|$. 
Note that in Eq.~\eqref{eq:diagonalrightbosons} we have dropped the inconsequential zero-point energy term
\be
\sum_{k_n} \frac{\Omega(k_n)}{2} \rightarrow \frac{N}{2} \int_{-\pi}^{\pi} \frac{dk}{2 \pi} \Omega(k) \quad \mbox{for} \quad N \rightarrow \infty,
\ee
as it does not affect transport properties and their statistics (note that it can be anyhow removed by normal-ordering the initial Hamiltonians in Eq.~\eqref{oaschains}).

The quench occurring at time $t=0$ connects the chains via their end points at site $0$ resulting in the post-quench Hamiltonian $H= H_r+ H_l+  \delta H $, with $\delta H= -\omega^2 \phi_1 \phi_0$, and therefore
\be 
H =  \frac{1}{2} \sum_{x=-N+1}^{N} \left( p_x^2 +m^2\phi_x^2 \right) + \frac{1}{2} \sum_{x=-N}^{N} \omega^2 (\phi_{x+1}-\phi_{x})^2, \label{eq:post_quench_Oas_lattice}
\ee
with boundary conditions $\phi_{-N}=\phi_{N+1} \equiv 0$ and $p_{-N}=p_{N+1} \equiv 0$.
In the thermodynamic limit $N \rightarrow \infty$, the chain becomes translationally invariant, i.e., $[H,P_{tr}]=0$ with the translation operator $P_{tr}$ defined similarly to the Ising case (see Eq.~\eqref{eq:translation_operator_TFIC}) as 
\begin{equation}
P_{tr}^{\dagger} \phi_x P_{tr} = \phi_{x-1}, \, \, \, P_{tr}^{\dagger} p_x P_{tr} = p_{x-1}; \label{eq:translation_operator_OAS}
\end{equation}
the resulting model can be solved by means of Fourier transform as in the case with periodic boundary conditions, see, e.g., Ref.~\cite{calabrese2007quantum}, 
which yields
\begin{eqnarray} 
H &=& \int_{0}^{\pi} dk \; \; \Omega(k) \; \left[ \AA^{\dagger}(k)\AA(k) + \AA^{\dagger}(-k)\AA(-k) \right] \nonumber  \\
  &=& H_R+ H_L,  \label{eq:post_quench_Oas_diagonal}
\end{eqnarray}
where $k$ varies continuously within the interval $[-\pi,\pi]$ and
\be
\AA(k)= \frac{1}{\sqrt{2 \Omega(k)}}\left[\Omega(k)\hat{\phi}(k)+i\hat{p}(k)\right], \label{eq:postquenchdiagonal}
\ee
while
\begin{eqnarray}
\hat{\phi}(k) &=& \frac{1}{\sqrt{2 \pi}} \sum_{x=-\infty}^{+\infty} \mbox{e}^{-i k x} \phi_x, \nonumber \\ 
\hat{p}(k) &=& \frac{1}{\sqrt{2 \pi}} \sum_{x=-\infty}^{+\infty} \mbox{e}^{-i k x} p_x, \label{eq:Ftransform}
\end{eqnarray}
are the Fourier transformed operators. Note that by applying the definition in Eq.~\eqref{eq:translation_operator_OAS} to Eq.~\eqref{eq:postquenchdiagonal}, keeping into account Eqs.~\eqref{eq:Ftransform}, it follows that
\begin{equation}
P_{tr}^{\dagger} \AA (\pm k) P_{tr} = e^{\mp i k}, \label{eq:transinvariance2}
\end{equation}
i.e., analogously to Eq.~\eqref{eq:transinvariance} for the post-quench mode operators $\Psi_{R,L}(k)$ of the transverse field Ising chain, the operators $\AA(\pm k)$ having positive/negative wave vector $k$ can be interpreted as bosonic right/left moving quasi-particles excitations.
\section{Hydrodynamic limit of transport quantities} 
\label{secondsection}
The quantities related to transport which we focus on in this work are the energy density $u_x$ and current $j^E_x$ at a point $x$ of the chain. The former is defined from the Hamiltonian of the complete chain such that
\begin{equation}
H= \sum_{x=-N+1}^{N} u_x. \label{eq:energy_charge}
\end{equation}
The latter, instead, is defined such that
\begin{equation}
\frac{d u_x(t)}{d t} = i[H,u_x(t)] = j^E_x - j^E_{x+1}, \label{eq:energy_current}
\end{equation}
which is the continuity equation written at the operatorial level: the time derivative of the energy density $u_x$ equals the opposite of the discrete divergence of the energy current. This relationship ensures that the total energy $H$ in Eq.~\eqref{eq:energy_charge} is conserved in time. We emphasize here that both $u_x$ and $j^E_x$ are local operators, in the sense that they act non-trivially only on a finite number of sites around $x$. As explained in Sec.~\ref{firstsection}, within the partitioning protocol, the non-equilibrium dynamics is obtained by joining at time $t=0$ the two chains, which are initially independently thermalized so that the initial state $\rho_0$ is given by Eq.~\eqref{eq:intro-rho0}. Consequently, the mean values we are interested in are generically defined as 
\begin{equation}
O(x,t)  = \mbox{Tr}[\rho_0 \, o_x(t)], \label{eq:local_average}
\end{equation} 
where $o_x(t)$ is a local observable, e.g., $u_x$ or $j^E_x$, at site $x$ and evolved up to time $t$. 
Note that the initial state $\rho_0$ is neither stationary, i.e., invariant under time evolution with the post-quench Hamiltonian $H$, nor homogenous, i.e., invariant under space translations according to Eqs.~\eqref{eq:translation_operator_TFIC} or \eqref{eq:translation_operator_OAS}. As a consequence, $O(x,t)$ displays a non-trivial space and time dependence. In the present work we are interested in studying the dynamics of $O(x,t)$ in the limit where both $x$ and $t$ are much larger than the corresponding microscopic scales, with a fixed and finite ratio $v = x/t$. This regime is referred to as the hydrodynamic or space-time scaling or semi-classical limit in the literature of inhomogeneous quantum quenches, see, e.g., Refs.~\cite{antal1999transport,karevski2002scaling,platini2005scaling,collura2014non,collura2014quantum,allegra2016inhomogeneous,
viti2016inhomogeneous,bertini2016determination,eisler2016universal,perfetto2017ballistic,
kormos2017inhomogeneous,10.21468/SciPostPhys.6.1.004}. Accordingly, we will mostly use the term ``hydrodynamic limit'', although with reference to the theory of hydrodynamics \cite{spohn1977stationary}, the names ``Euler-scaling limit'' or ``ballistic scaling limit'' would be more accurate. In formulas the hydrodynamic limit $\mathcal{O}$ of the quantity $O$ is defined as      
\begin{equation}
\mathcal{O}(v) = \lim_{\substack{x,t\to\infty \\ v=x/t}} \lim_{N \rightarrow \infty} O(x,t)  = \mbox{Tr}[\rho(v) \, o_{x=0}(0)], \label{eq:hydro_limit}
\end{equation}
where, exploiting the definition of the translation operator in Eqs.~\eqref{eq:translation_operator_TFIC} or \eqref{eq:translation_operator_OAS}, we defined
\begin{equation}
\rho(v) =  \lim_{\substack{x,t\to\infty \\ v=x/t}} \lim_{N \rightarrow \infty} (P_{tr}^{\dagger})^x e^{-i H t} \rho_0 e^{i H t} (P_{tr})^x. \label{eq:hydro_state}
\end{equation}
Accordingly, the state $\rho(v)$ fully describes the hydrodynamic limit of any local observable $o_x(t)$. 

A complementary approach to Eqs.~\eqref{eq:hydro_limit} and \eqref{eq:hydro_state} for computing the hydrodynamic limit $\mathcal{O}$ consists in determining first the large space-time scaling of the operator $o(x,t)$, similarly to Eq.~\eqref{eq:hydro_state}, and then in taking the trace over the initial density matrix $\rho_0$ according to Eq.~\eqref{eq:local_average}. This scheme has been pursued in Refs.~\cite{perfetto2017ballistic,kormos2017inhomogeneous} to compute the space-time scaling limit $\mathcal{U}(v)$, $\mathcal{J}^E(v)$ of the energy density $u_x$ and current $j^E_x$, respectively. Compared to the latter, the advantage of the approach of Eqs.~\eqref{eq:hydro_limit} and \eqref{eq:hydro_state} is that, once the state $\rho(v)$ is known, the hydrodynamic limit of any local observable, not only of $u_x$ and $j^E_x$, can be readily obtained from Eq.~\eqref{eq:hydro_limit}. Furthermore, the knowledge of $\rho(v)$ is fundamental for the calculation of the transferred energy scaled cumulant generating function, as shown in Section \ref{thirdsection}. Accordingly, here we will proceed as in Eqs.~\eqref{eq:hydro_limit} and \eqref{eq:hydro_state}.

Moreover, as we see explicitly, see, c.f., Eq.~\eqref{eq:hydro_state_proved} for the TFIC and Eq.~\eqref{eq:hydro_state_proved_bosons} for the harmonic chain, the state $\rho(v)$ depends on an homogeneous and stationary combination of the post-quench mode operators $\Psi_R(k)$ and $\AA(k)$, being the dependence on $v$ brought in only by the coefficients. The existence of the limit in Eq.~\eqref{eq:hydro_limit} can be considered as a consequence, in the present context, of the so-called ``local entropy maximization principle" \cite{castro2016emergent,bertini2016determination,bertini2016transport}, which asserts that averages of local observables $o_x(t)$ over a state $\rho_0$, generically inhomogenous and non-stationary, can be replaced by averages of the same observable over the local equilibrium --- and therefore homogeneous and stationary --- state $\rho(x,t)$ at point $x$ and time $t$. Note that, while $\rho_0$ characterizes \textit{globally} the state of the system, $\rho(x,t)$ applies only \textit{locally} at the space-time point $(x,t)$ for the calculation of the average of local observables $o_x(t)$. 
Other quantities, such as the dynamical two-point functions, where the observables involve different space-time points, cannot be computed solely on the basis of the local equilibrium state $\rho(x,t)$, see, e.g., Refs.~\cite{DoyonLargeScaleCorrelations2018,moller2020correlations}. The ``local entropy maximization principle" is at the basis of the so-called generalized hydrodynamics description of integrable systems out of equilibrium \cite{castro2016emergent,bertini2016transport}, which allows the extension of the analysis underlying Eqs.~\eqref{eq:hydro_limit} and \eqref{eq:hydro_state} to the far more complex case of interacting integrable systems. In Eqs.~\eqref{eq:hydro_limit} and \eqref{eq:hydro_state} we are actually anticipating, c.f., Eq.~\eqref{eq:hydro_state_proved} for the TFIC and Eq.~\eqref{eq:hydro_state_proved_bosons} for the harmonic chain, that in the present case of a dynamics starting from the initial state $\rho_0$ in Eq.~\eqref{eq:intro-rho0}, the evolved state $\rho(x,t)$ is a scaling function of $v=x/t$.
In particular, in the long-time limit $t \rightarrow \infty$ with $x$ fixed and therefore $v \to 0$, the density matrix $\rho(v=0) \equiv \rho_{stat}$ describes the non-equilibrium steady state arising long after the quench. This stationary state has been extensively studied and for free models, with the notation of Eqs.~\eqref{eq:postquenchhamiltonian} and \eqref{eq:post_quench_Oas_diagonal}, takes the form \cite{de2013nonequilibrium,doyon2015non,bernard2016conformal}
\begin{equation}
\rho_{stat} = e^{-\beta_r H_L} \otimes e^{-\beta_l H_R} /Z. \label{eq:stationary_density_matrix}
\end{equation} 
Below we discuss separately the case of the TFIC in Sec.~\ref{sec:isingtransportsub} and the harmonic chain in Sec.~\ref{sec:oastransportsub}. For both the cases we write explicitly the density matrix $\rho(v)$, before computing the hydrodynamic limit of the energy density $u_x$ and current $j^E_x$, respectively, according to Eq.~\eqref{eq:hydro_limit}. The expressions for $\mathcal{U}(v)$ and $\mathcal{J}^E(v)$ are in agreement with the known results of Refs.~\cite{perfetto2017ballistic,kormos2017inhomogeneous}, for the TFIC, while the explicit expressions of $\rho(v)$ in, c.f., Eqs.~\eqref{eq:hydro_state_proved} and \eqref{eq:hydro_state_proved_bosons} are the primary results of this paper.
\subsection{The Ising chain in a transverse field}
\label{sec:isingtransportsub}
For the Ising chain we denote the energy density operator at site $x$ as $u_x$, which from Eqs.~\eqref{eq:chains} and \eqref{eq:energy_charge} takes the form 
\begin{equation}
u_x = -\frac{J}{4}(\sigma_x^x \sigma_{x+1}^x + \sigma_{x-1}^x \sigma_{x}^x) -\frac{J h}{2} \sigma_x^z, \label{eq:energy_density_TFIC}
\end{equation}
while the energy current $j_x^{E}$ consequently follows from Eq.~\eqref{eq:energy_current} 
\begin{equation}
j_x^{E} = \frac{J^2 h}{4} (\sigma_x^x \sigma_{x+1}^y - \sigma_x^y \sigma_{x+1}^x) = \frac{i h J^2}{2} (c_{x+1}^{\dagger} c_x -c_x^{\dagger} c_{x+1}), \label{eq:energy_current_TFIC}
\end{equation}
where in the last step we used the Jordan-Wigner transformation in Eq.~\eqref{eq:Jordan_Wigner} to write the energy current in terms of the lattice fermionic operators. 
In order to compute the hydrodynamic limit of the aforementioned quantities according to Eq.~\eqref{eq:hydro_limit} one first needs to construct the state $\rho(v)$ in Eq.~\eqref{eq:hydro_state}. To do this we write the initial state $\rho_0$ in Eq.~\eqref{eq:intro-rho0}, with $H_{r,l}$ expressed in terms of the pre-quench modes $\Phi_{r,l}(k)$ according to Eq.~\eqref{eq:HRcont}, as a function of the post-quench modes $\Psi_{R}(k)$ by means of the transformation 
\begin{equation}
\Phi_{\alpha}(k) = \int_{-\pi}^{\pi} dk' [\Psi_{R}(k') \, m_{+,\alpha}^{\ast}(k',k)+\Psi_{R}^{\dagger}(k') \, m_{-,\alpha}(k',k)], 
\label{eq:pre_quench_mode_post_quench_mode}
\end{equation}
with $\alpha \in \{r,,l \}$; the expressions for the coefficients $m_{\pm,\alpha}^{\ast}(k',k)$ are provided in Appendix \ref{app:appendix1} (see Eqs.~\eqref{eq:m_coefficients_Ising_left} and \eqref{eq:m_coefficients_Ising_right}). In terms of the post-quench operators, using Eq.~\eqref{eq:transinvariance} and, remembering that under the post-quench Hamiltonian $H$ the time evolution is trivial $e^{-iHt} \Psi_{R}(k) e^{iHt}=e^{i \varepsilon(k)t} \Psi_{R}(k)$, the space and time propagation $\rho(x,t)$ of the state $\rho$ according to Eq.~\eqref{eq:hydro_state} can be determined explicitly. As detailed in Appendix \ref{app:appendix2}, the leading space-time dependence of $\rho(x,t)$ in the semi-classical limit of Eq.~\eqref{eq:hydro_state} turns out to be
\be
\begin{split}
&\rho(x,t) = \frac{1}{Z}\mbox{exp} \left\{ - \int_{-\pi}^{\pi} dk' dk'' \Psi_R^{\dagger}(k') \Psi_R(k'') e^{i \varphi_{x,t}^{+}(k',k'')} \right. \\ 
&\ \qquad \qquad \qquad \qquad \times \left. [\beta_r I^r_{+,+}(k',k'')+ \beta_l I^l_{+,+}(k',k'')] \right \},
\label{eq:intermediate_result_rho_x,t}
\end{split}  
\ee
where the expressions of $\varphi_{x,t}^{+}(k',k'')$ and $I^{r,l}_{+,+}(k',k'')$ are reported in Appendix \ref{app:appendix2} (see Eqs.~\eqref{eq:phases_hydro_appendix} and \eqref{eq:I_plus_Integral_appendix}). The expression in Eq.~\eqref{eq:intermediate_result_rho_x,t} can be further simplified as $x,t \rightarrow \infty$ with fixed ratio $v=x/t$, by performing a stationary phase approximation \cite{viti2016inhomogeneous}: the procedure is completely analogous to the one followed in Refs.~\cite{perfetto2017ballistic,kormos2017inhomogeneous} (briefly reported in Appendix \ref{app:appendix2}) and it leads to the result 
\begin{align}
\rho(v)& = \frac{1}{Z} \mbox{exp} \left \{-\int_{-\pi}^{\pi} dk \, \beta(v,k) \varepsilon(k)\Psi_R^{\dagger}(k) \Psi_R(k)\right \}, \nonumber \\
\mbox{where}&  \, \, \, \, \beta(v,k) = \beta_r \Theta(v-v_g(k)) +\beta_l \Theta(v_g(k)-v), \label{eq:hydro_state_proved}
\end{align}
where $v_g(k) = d \varepsilon(k)/dk$ is the group velocity of the quasi-particles excitations with energy $\varepsilon(k)$ (see Eq.~\eqref{eq:Isingspectrum}) and $\Theta(x)=1$ if $x>0$ and 0 otherwise, being the Heaviside step function. Since $\rho(v)$ is diagonal in terms of the post-quench mode operators $\Psi_R(k)$, and the dependence on $v$ brought in only by the coefficients $\beta(v,k)$, it is effectively stationary and homogeneous, as anticipated in the discussion in Sec.~\ref{secondsection}. One can also notice that $\rho(v)$ is indeed a function of the scaling variable $v=x/t$ as the entire space-time dependence is encoded within the Heaviside function. Moreover, in the stationary limit $v=0$, it agrees with the known general expression of the non-equilibrium steady-state density matrix of Eq.~\eqref{eq:stationary_density_matrix}. The expression of $\rho(v)$ in Eq.~\eqref{eq:hydro_state_proved} generalizes the known result for the stationary state $\rho_{stat}$, thereby accounting for the whole dynamics of any local observable $o_x(t)$ along a ray in the space-time plane with fixed $v=x/t$. Outside the light cone, for $v>v_{max}$ ($v<-v_{max}$), $\rho(v)$ depends only on $\beta_r$ ($\beta_l$), as expected. It is, however, important to emphasize that $\rho(v)$ in Eq.~\eqref{eq:hydro_state_proved} does not reduce to $\rho_0$ outside the light cone ($|v|>v_{max}$). This is related to the fact that $\rho(v)$ is defined only locally at the space-time point $(x,t)$, as emphasized above in Sec.~\ref{secondsection}. This implies that $\rho(v)$ can be used for the calculation, in the hydrodynamic limit, of averages of local observables $\mathcal{O}(v)$ at the space-time point $(x,t)$ according to Eq.~\eqref{eq:hydro_limit}. For $|v|>v_{max}$ the average $\mathcal{O}(v)$ reduces to the corresponding average over the initial state $\rho_0$ in Eq.~\eqref{eq:intro-rho0} of the right or left chain. This can be explicitly checked in Eqs.~\eqref{eq:energy_current_hydro_TFIC} and \eqref{eq:energy_density_hydro_TFIC} (and Eqs.~\eqref{eq:energycurrentoas_hydro} and \eqref{eq:energydensityoas_hydro} for the harmonic chain) for the energy current and density, respectively.  
It is immediate to calculate the average over $\rho(v)$ of any fermionic bilinear function of the post-quench operators $\Psi_R^{\dagger}(k) \Psi_R(k')$, taking into account that 
\begin{equation}
\mbox{Tr}[\rho(v) \Psi_R^{\dagger}(k) \Psi_R(k')] = \delta(k-k') n^{+}(v,k), \label{eq:mode_1}
\end{equation}
where we introduced
\begin{equation}
n^{+}(v,k)= f_{\beta_r}^{+}(k) \Theta (v-v_g(k))+f_{\beta_l}^{+}(k)\Theta(v_g(k)-v) \label{eq:mode_GHD_TFIC}
\end{equation}
and $f_{\beta}^{+}(k)= 1/(e^{\beta \varepsilon(k)}+1)$ denotes the Fermi-Dirac distribution at inverse temperature $\beta$. The physical meaning of $n^{+}(v,k)$ is simple: the state $\rho(v)$ is determined by ballistically propagating quasi-particles capable of crossing the ray in the space-time diagram with fixed $v=x/t$: for the right half chain ($x>0$) this requires $v>v_g(k)$ while for the left one (with $x<0$) $v_g(k)>v$. Since these quasi-particles do not experience scattering, they maintain their initial thermal distribution $f_{\beta_r}^{+}(k)$ for the right chain and $f_{\beta_l}^{+}(k)$ for the left, from which Eq.~\eqref{eq:mode_GHD_TFIC} follows. To make contact with the GHD formalism of Refs.~\cite{castro2016emergent,bertini2016transport} we note that Eq.~\eqref{eq:mode_GHD_TFIC} represents the solution for a free theory of the GHD equation for the mode occupation $n^{+}(v,k)$ with the initial state of Eq.~\eqref{eq:intro-rho0}.      
 
Accordingly, concerning the calculation of mean values, the knowledge of $\rho(v)$ allows one to determine not only the space-time scaling limit of the transport quantities introduced in Section \ref{secondsection}, but, more generally, the hydrodynamic limit $\mathcal{O}(x,t)$ of any local observable $o_x(t)$, as dictated by Eq.~\eqref{eq:hydro_limit}. In practice, one should simply write the latter in terms of the post-quench mode operators $\Psi_R(k)$ and then use Eqs.~\eqref{eq:mode_1} and \eqref{eq:mode_GHD_TFIC}. Specializing to the energy current $j^E_{x=0}$ in Eq.~\eqref{eq:energy_current_TFIC} and the energy density $u_{x=0}$ in Eq.~\eqref{eq:energy_density_TFIC}, the results for the corresponding mean values $\mathcal{J}^E$ and $\mathcal{U}$ are in agreement with those of Refs.~\cite{perfetto2017ballistic,kormos2017inhomogeneous}: 
\begin{eqnarray}
\mathcal{J}^E (v) &=& \int_{-\pi}^{\pi} \frac{dk}{2 \pi} \varepsilon(k) v_g(k) n^+(v,k), \label{eq:energy_current_hydro_TFIC} \\
\mathcal{U} (v) &=& \int_{-\pi}^{\pi} \frac{dk}{2 \pi} \varepsilon(k) n^+(v,k). \label{eq:energy_density_hydro_TFIC} 
\end{eqnarray}
The physical interpretation of Eqs.~\eqref{eq:energy_current_hydro_TFIC} and \eqref{eq:energy_density_hydro_TFIC} is clear in terms of quasi-particles produced in the initial thermal state with statistics $f_{\beta_l}^{+}$ and $f_{\beta_r}^{+}$ for the left and right chain, respectively; these excitations propagate ballistically with velocity $v_g(k)$ without undergoing scattering since the model is non-interacting and translationally invariant, and they contribute with $\varepsilon(k) v_g(k) d k$ to the flux of energy. The edge of the profile $\mathcal{J}^E(v)$, beyond which the mean current vanishes, is determined by the maximal velocity $v_{max}$ of the quasi-particles, which for the TFIC, reads
\be 
v_{max} = J \, \mbox{min}(h,1) = \frac{\varepsilon_{max}-\varepsilon_{min}}{2}, \label{eq:maxvelocityTFIC}
\ee 
where we identified the maximum $\varepsilon_{max}=J(h+1)$ and the minimum $\varepsilon_{min}=J|h-1|$ of the dispersion relation $\varepsilon(k)$ in Eq.~\eqref{eq:Isingspectrum}.

\begin{figure}[h!]
\includegraphics[width=1\columnwidth]{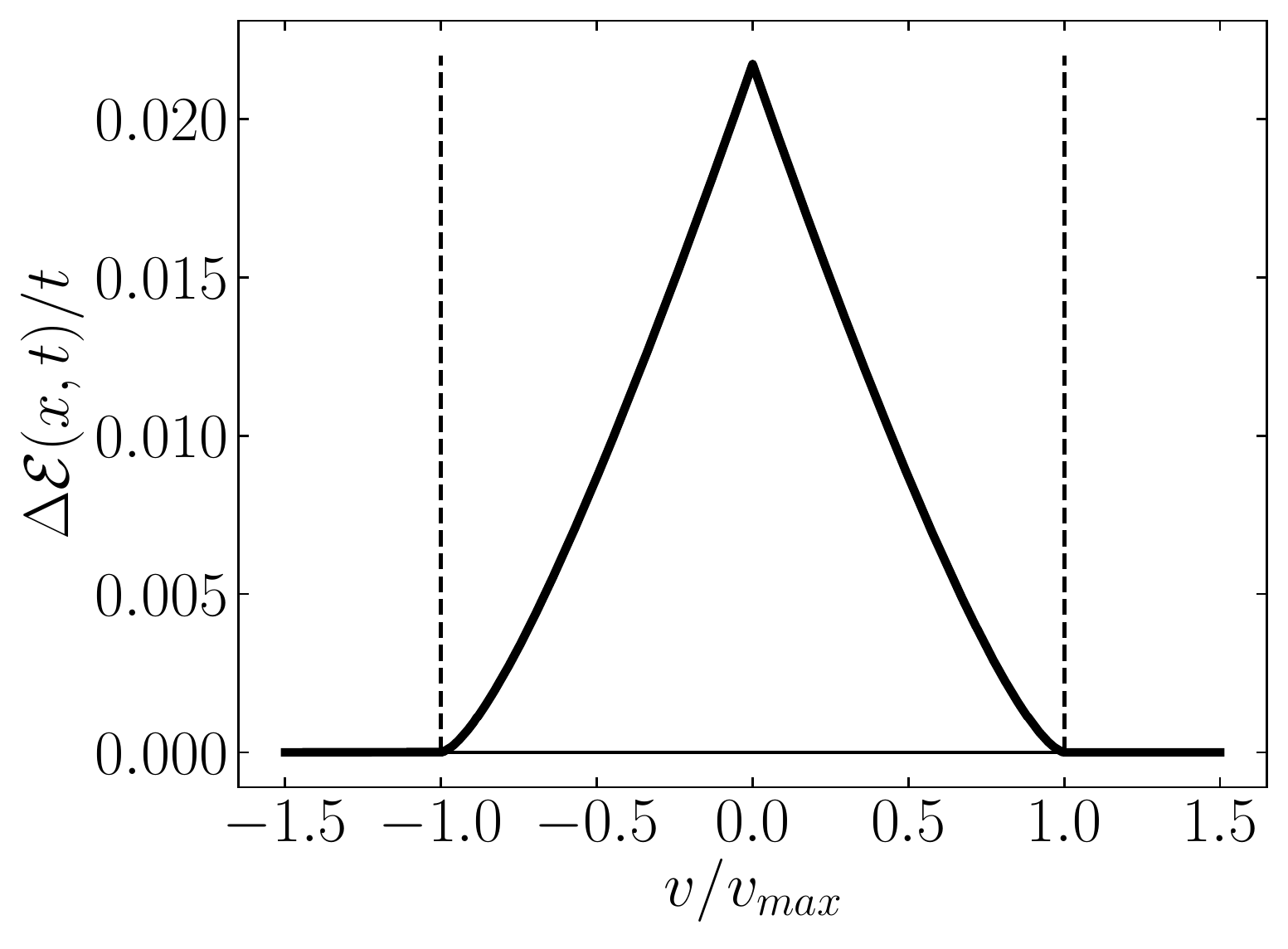}
\caption{Transferred energy $\Delta \mathcal{E}(x,t)$ rescaled by $t$ in the semi-classical limit as a function of $v/v_{max}$ for $J=1$ and $h=1.2$. The inverse temperatures of the initial inhomogeneous state are $\beta_l=2$ and $\beta_r=4$.}
\label{fig:deltaE}	
\end{figure}

Based on the knowledge of the mean energy current $\mathcal{J}^E(v)$ it is immediate to determine the total energy transferred across point $x$ in the time interval $[0,t]$, whose definition as an operator is
\be
\Delta e(x,t)= \int_{0}^{t} ds \, j^E_x(s); \label{eq:transferredenergyoperator}
\ee
its mean $\Delta \mathcal{E}(x,t)$, in the hydrodynamic limit, is given by (see Eq.~\eqref{eq:energy_current_hydro_TFIC})
\be
\begin{split}
 \Delta \mathcal{E}(x,t) &= t \int_{0}^{\pi} \frac{dk}{2 \pi} \varepsilon(k)(v_g(k) -|v|) \\
&\ \ \times \left[f_{\beta_l}^{+}(k)-f_{\beta_r}^{+}(k)\right] \Theta(v_g(k)-|v|). \label{transferredenergy}
\end{split}
\ee
Note that, as expected, the transferred energy grows extensively upon increasing time $t$. This property is fundamental for studying fluctuations of this observable within the large deviation theory, as shown in Sec.~\ref{thirdsection}. By rescaling the transferred energy by the time $t$, one obtains the scaling function of $v$ reported in Fig.~\ref{fig:deltaE} for a representative choice of the parameters.
In concluding this section, we emphasize that the expression of $\rho(v)$ in Eq.~\eqref{eq:hydro_state_proved} derived here, on the one hand, reproduces the known results of Refs.~\cite{perfetto2017ballistic,kormos2017inhomogeneous} for the mean values of transport quantities, see Eqs.~\eqref{eq:energy_current_hydro_TFIC} and \eqref{eq:energy_density_hydro_TFIC}, and, on the other hand, it allows the determination of the fluctuations of the transferred energy in Eq.~\eqref{eq:transferredenergyoperator} beyond the mean value in Eq.~\eqref{transferredenergy}, as discussed in Sec. \ref{thirdsection}.

\subsection{The harmonic chain}
\label{sec:oastransportsub}
For the harmonic chain, the energy density $u_x$ at lattice site $x$, from on Eq.~\eqref{eq:prequenchoscillators}, is given by
\begin{equation}
u_x = \frac{1}{2} p_x^2 + \frac{1}{2} m^2 \phi_x^2  + \frac{1}{4}\omega^2(\phi_{x+1}-\phi_x)^2 + \frac{1}{4}\omega^2 (\phi_{x-1}-\phi_x)^2,  \label{eq:oscillatorenergydensity}
\end{equation}
while the energy current $j^E_x$ at site $x$ is consequently defined according to the continuity equation in Eq.~\eqref{eq:energy_current}, i.e.,
\begin{equation}
j^E_x = \frac{\omega^2}{2}(\phi_{x-1} - \phi_{x})(p_{x-1}+p_x).
\label{eq:oscillatorcurrent}
\end{equation}
In order to compute the hydrodynamic limit of these observables the procedure to construct the state $\rho(v)$ of Eq.~\eqref{eq:hydro_state} is completely analogous to the one presented above for the quantum Ising chain and therefore we report here the final result, leaving all the details of the derivation in Appendix \ref{app:appendix2}. 
In the hydrodynamic limit $x,t \rightarrow \infty$ with fixed ratio $v=x/t$, one finds 
\begin{equation}
\rho(v) = \frac{1}{Z} \mbox{exp} \left \{-\int_{-\pi}^{\pi} dk \, \beta(v,k) \Omega(k) \AA^{\dagger}(k) \AA(k)\right \},  
\label{eq:hydro_state_proved_bosons}
\end{equation}
where $\beta(v,k)$ has the same formal expression as in the case of the TFIC in Eq.~\eqref{eq:hydro_state_proved}, but with the group velocity $v_g(k)= d \Omega(k)/dk$ determined by the dispersion relation $\Omega(k)$ in Eq.~\eqref{eq:oasspectrum}. As far as the mean of a bilinear function of the mode operators $\AA^{\dagger}(k)$ and $\AA (k')$ is concerned, one finds
\begin{equation}
\mbox{Tr}[\rho(v) \AA^{\dagger}(k) \AA(k')] = \delta(k-k') n^{-}(v,k), \label{eq:mode_1_bosons}
\end{equation}
where 
\begin{equation}
n^{-}(v,k)= f_{\beta_r}^{-}(k) \Theta (v-v_g(k))+f_{\beta_l}^{-}(k)\Theta(v_g(k)-v), \label{eq:mode_GHD_bosons}
\end{equation}
and $f_{\beta}^{-}(k)=1/(e^{\beta \Omega(k)}-1)$ is the Bose-Einstein occupation, with the dispersion relation $\Omega(k)$ of the harmonic chain defined in Eq.~\eqref{eq:oasspectrum}; the important difference between Eqs.~\eqref{eq:hydro_state_proved_bosons}, \eqref{eq:mode_1_bosons}, and \eqref{eq:mode_GHD_bosons} and the corresponding formulas in the fermionic case (see  Eqs.~\eqref{eq:hydro_state_proved}, \eqref{eq:mode_1}, and \eqref{eq:mode_GHD_TFIC})) is the fact that post-quench modes $\AA(k)$ have bosonic statistics and therefore they obey canonical commutation relations. This is also signaled by the appearance of $f_{\beta}^{-}(k)$ within the mode occupation function $n^{-}(v,k)$.
 
\begin{figure}[h!] 
\centering
\includegraphics[width=1\columnwidth]{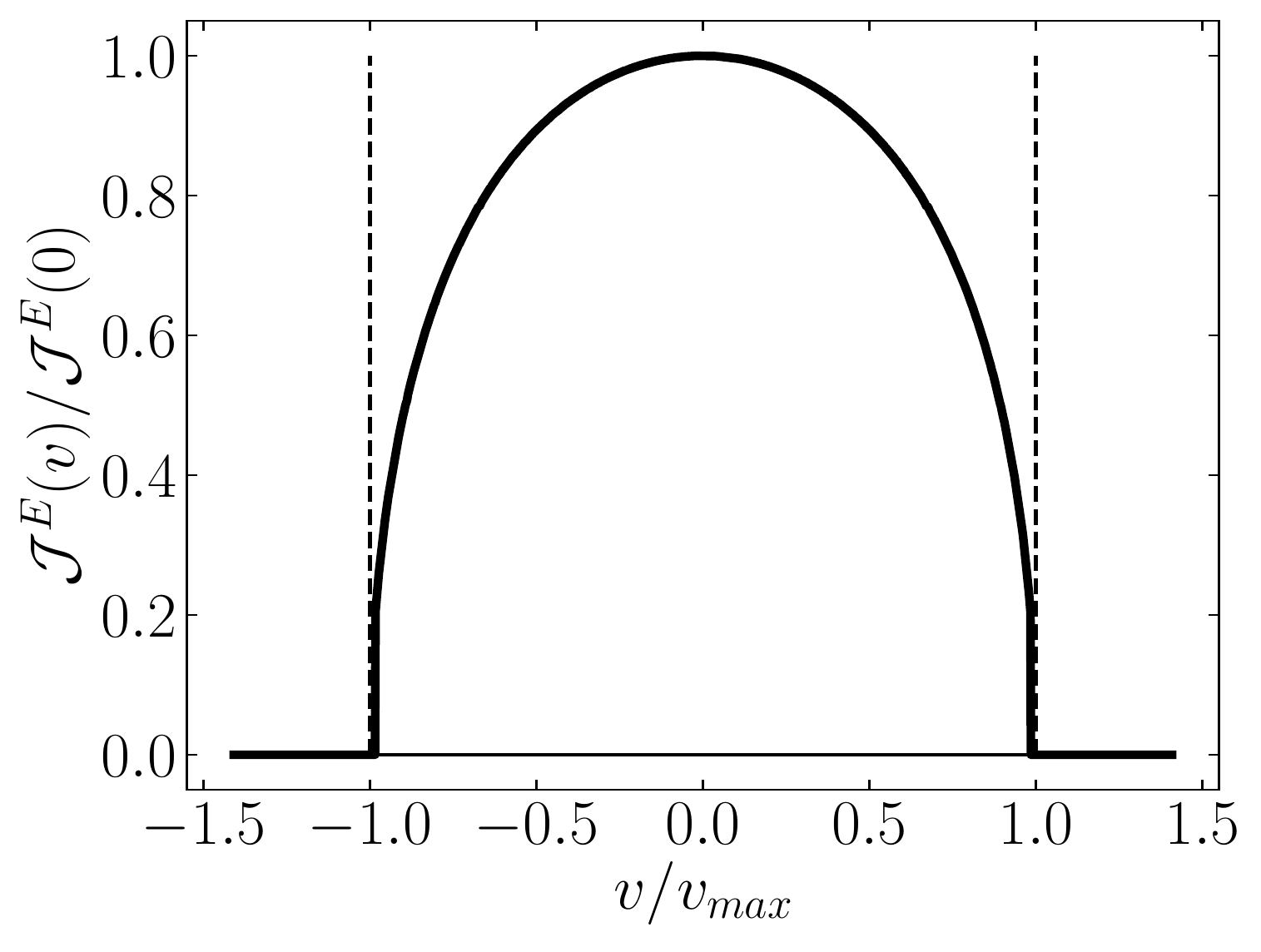}\\[-1mm]	
	\, \, \, (a)\\[2mm]
\includegraphics[width=1\columnwidth]{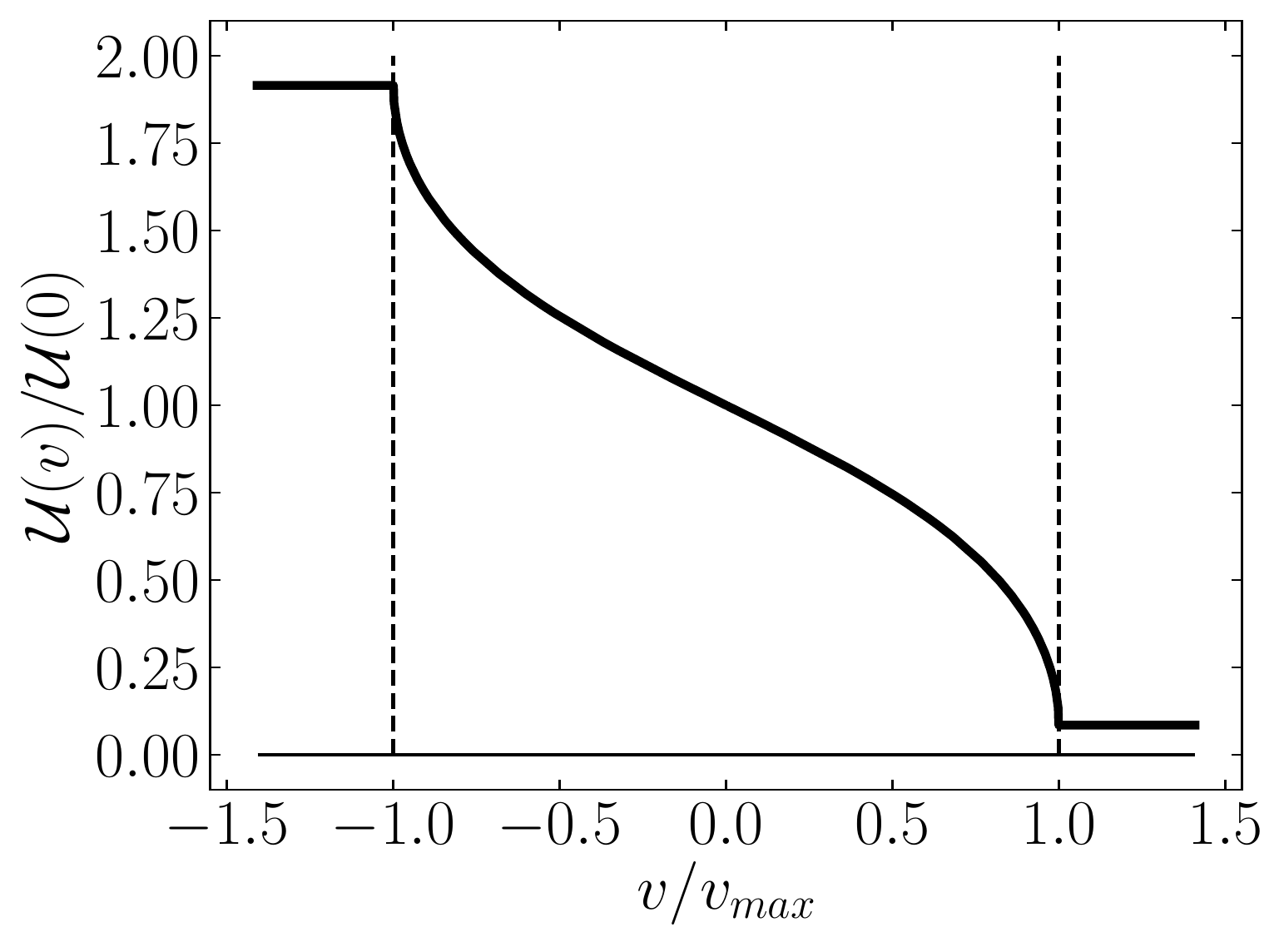}\\[-1mm]
\, \, \, (b)
\caption{Scaling form of (a) the energy current $\mathcal{J}^E(v)/\mathcal{J}^E(0)$ and (b) the energy density $\mathcal{U}(v)/\mathcal{U}(0)$, both have been normalized by the corresponding stationary values in $v=0$, as functions of $v/v_{max}$ for $\omega=1$ and $m=0.7$, resulting in a $v_{max} \simeq 0.71$ according to Eq.~\eqref{eq:maxvelocityOAS}. The inverse temperatures are chosen to be $\beta_l=2$ and $\beta_r=5$.} 
\label{fig:scaling}
\end{figure}

The hydrodynamic limit of the mean energy density $\mathcal{U}(v)$ and mean current $\mathcal{J}^E(v)$ then follows as 
\begin{eqnarray}
\mathcal{J}^E (v) &=&  \int_{-\pi}^{\pi} \frac{dk}{2 \pi} \Omega(k) v_g(k) n^-(v,k), \label{eq:energycurrentoas_hydro} \\
\mathcal{U}(v) &=& \int_{-\pi}^{\pi} \frac{dk}{2 \pi} \Omega(k) n^-(v,k),  \label{eq:energydensityoas_hydro}
\end{eqnarray}
which have precisely the same form as Eqs.~\eqref{eq:energy_current_hydro_TFIC} and \eqref{eq:energy_density_hydro_TFIC}, respectively. This shows that the form of the profile of $\mathcal{J}^E$ and $\mathcal{U}$ in the hydrodynamic limit is universal to a large extent since the only remaining microscopic ingredients characteristic of the model are the spectrum ($\varepsilon(k)$ in Eq.~\eqref{eq:Isingspectrum} for the quantum Ising chain and $\Omega(k)$ in Eq.~\eqref{eq:oasspectrum} for the harmonic chain) and the statistics of the involved quasi-particles ($f_{\beta}^{+}(k)$ for the fermionic case and $f_{\beta}^{-}(k)$ in the bosonic one). Moreover, it is easy to check explicitly that, as expected, $\mathcal{J}^E (v)$ and $\mathcal{U}(v)$ satisfy the continuity equation (for the quantum Ising chain this has been already observed in Refs.~\cite{perfetto2017ballistic,kormos2017inhomogeneous}), 
\begin{equation}
\frac{\partial \mathcal{U}(x,t)}{\partial t} = -\frac{\partial \mathcal{J}^E(x,t)}{\partial x}. \label{eq:continuity_equation}
\end{equation}
The plot of Eqs.~\eqref{eq:energycurrentoas_hydro} and \eqref{eq:energydensityoas_hydro} for a representative choice of the parameters is reported in Fig.~\ref{fig:scaling}.
The curves in Fig.~\ref{fig:scaling} are, as expected, qualitatively similar to those corresponding to the same physical quantities in the quantum Ising chain (see the plots of Eqs.~\eqref{eq:energy_current_hydro_TFIC} and \eqref{eq:energy_density_hydro_TFIC} in Fig.~(1) of Ref.~\cite{perfetto2017ballistic} and in Figs.~(1) and (3) of Ref.~\cite{kormos2017inhomogeneous}) and clearly reveals the ballistic nature of the transport occurring in the harmonic chain. The singular edge at $v=v_{max}$ in the profile of $\mathcal{J}^E (v)$ --- beyond which the mean energy current vanishes identically --- is, however, a model-specific quantity depending on the microscopic parameters of the Hamiltonian in Eq.~\eqref{oaschains} and it is therefore different from Eq.~\eqref{eq:maxvelocityTFIC}. In the harmonic chain one has
\begin{equation}
v_{max} = \frac{1}{2} \left(\sqrt{m^2 + 4 \omega^2}-m \right) = \frac{1}{2}(\Omega_{max}-\Omega_{min}), \label{eq:maxvelocityOAS}
\end{equation}
where $\Omega_{max}=\sqrt{m^2 + 4 \omega^2}$ and $\Omega_{min}=m$ are the maximum and the minimum, respectively, of the dispersion relation $\Omega(k)$ in Eq.~\eqref{eq:oasspectrum}. 
The integral over $k$ in Eq.~\eqref{eq:energycurrentoas_hydro} can be calculated analytically, as detailed in Appendix \ref{app:appendix2}, leading to
\begin{equation}
\mathcal{J}^E(v) = \Theta(v_{max}-|v|) [\mathcal{Y}(\beta_l,v)-\mathcal{Y}(\beta_r,v)], \label{eq:integrated_oas_current_1}
\end{equation} 
where
\begin{equation}
\mathcal{Y}(\beta,v) = \frac{Y(\beta \Omega_{-}(v)) - Y(\beta \Omega_{+}(v))}{2 \pi \beta^2}, \label{eq:integrated_oas_current_2}
\end{equation}
with 
\begin{equation}
Y(x) = \mbox{Li}_2(e^{-x})-x \, \mbox{ln}(1-e^{-x}), \label{eq:integrated_oas_current_3}
\end{equation}
$\mbox{Li}_2$ is the polylogarithm of order $2$ while $\Omega_{\pm}(v)$ are given in Eqs.~\eqref{eq:Omega_1_2} in Appendix \ref{app:appendix2} and they depend only on $v^2$, implying that $\mathcal{J}^E(v)$ is an even function of $v$ as one realizes from Fig.~\ref{fig:scaling}. For $v=0$, Eq.~\eqref{eq:integrated_oas_current_1} reduces to the steady state current supported by the stationary state $\rho_{stat}$ in Eq.~\eqref{eq:stationary_density_matrix}. In this case, the expression in Eq.~\eqref{eq:energycurrentoas_hydro} agrees with the result of Ref.~\cite{saito2007fluctuation} for the steady-state energy current flowing in a translationally invariant harmonic chain as in Eq.~\eqref{oaschains}, where the mass $m$ and the angular frequency $\omega$ are the same at every lattice site. Note that the dispersion relation $\Omega(k)$ in Ref.~\cite{saito2007fluctuation} takes arbitrary real values, while here $\Omega(k) \in (\Omega_{min},\Omega_{max})$ from Eqs.~\eqref{eq:oasspectrum} and \eqref{eq:maxvelocityOAS}. In addition, the system considered in Ref.~\cite{saito2007fluctuation} is open, as the harmonic chain is connected to two external baths at temperature $T_{l}$-$T_{r}$, which are modeled as an infinite collection of harmonic oscillators. In the present work, instead, with the partitioning protocol, the heat baths are provided by portions of the system itself, so that, as a whole, it evolves unitarily. Our result in Eq.~\eqref{eq:energycurrentoas_hydro} therefore shows the independence, in the hydrodynamic limit, of the energy current profile from the actual setting adopted to obtain the non-equilibrium steady state.

Figure \ref{fig:scaling} also shows that, as it happens for the TFIC (see Refs.~\cite{perfetto2017ballistic,kormos2017inhomogeneous}), the energy current $\mathcal{J}^E(v)$ approaches the edge at $v=v_{max}$ of the propagating front with a non-analytic behavior, which can be determined from Eqs.~\eqref{eq:integrated_oas_current_1}, \eqref{eq:integrated_oas_current_2} and \eqref{eq:integrated_oas_current_3}: for $v\rightarrow \pm v_{max}^{\mp}$, it turns out to be
\begin{equation}
\mathcal{J}^E(v) = C_1 \sqrt{v_{max}^2-v^2}+ \mathcal{O}\left((v_{max}-|v|)^{3/2} \right),  \label{eq:edge_ballistic_bosons}
\end{equation}
with the constant $C_1$ given in Eq.~\eqref{eq:C_1_constant} of Appendix \ref{app:appendix2}. Interestingly enough, when the mass ``$m$ '' is set to zero and thus the spectrum $\Omega(k)$ in Eq.~\eqref{eq:oasspectrum} becomes gapless, the qualitative form of the edge singularity in Eq.~\eqref{eq:edge_ballistic_bosons} is unchanged, with $C_1=(\beta_l^{-1} -\beta_r^{-1})/\pi$. This is in stark contrast with the case of the quantum Ising chain for which, as shown in Ref.~\cite{perfetto2017ballistic}, the qualitative behavior of the edge changes from the one analogous to Eq.~\eqref{eq:edge_ballistic_bosons} to a functional form $(v_{max}^2-v^2)^{3/2}$ when the transverse field is set to its critical value $h=1$.

\begin{figure}[h!] 
\centering
\includegraphics[width=1\columnwidth]{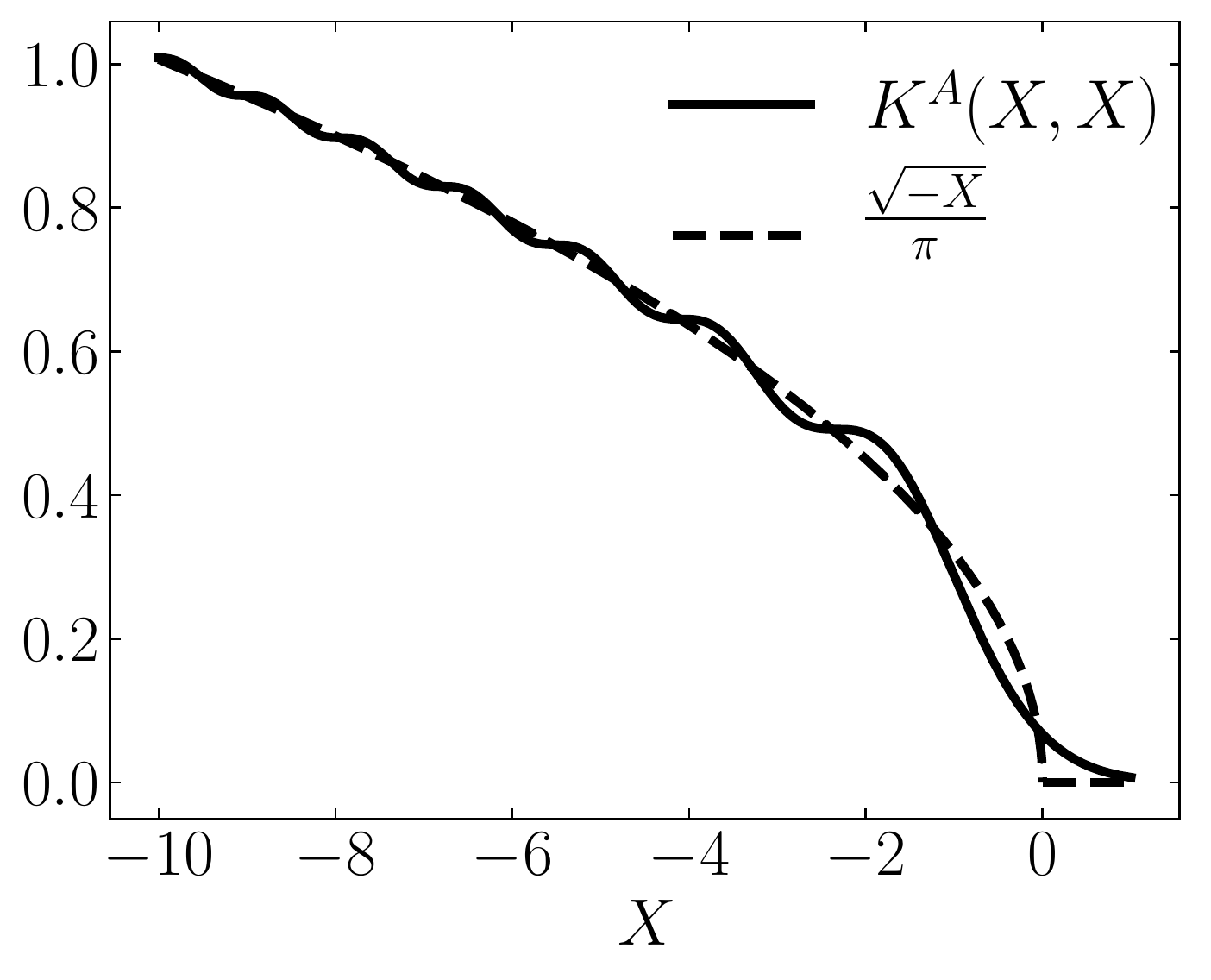}\\[-1mm]	
\caption{On the vertical axis the rescaled energy current $\mathcal{J}^E(X,t) \, (v_{max}t/2)^{1/3}/(\Omega(k_s) v_{max} (f_{\beta_l}^{-} -f_{\beta_r}^{-}))$ is plotted as a function of $X$. We compare the edge behavior of the energy current in Eq.~\eqref{eq:edge_ballistic_bosons} (dashed line) at the ballistic scale and the edge asymptotic in Eqs.~\eqref{eq:Airy_kernel_bosons_1}, \eqref{eq_Airy_kernel_bosons_2}, and \eqref{eq:Airy_kernel_bosons_3} of the same quantity including sub-diffusive corrections (solid line). The dashed line is obtained by expressing Eq.~\eqref{eq:edge_ballistic_bosons} as a function of $X$ (see Eq.~\eqref{eq:scaling_variable_Airy}). The solid line is the Airy kernel $K^A(X,X)$ in Eq.~\eqref{eq:Airy_kernel_bosons_3}. Sub-diffusive corrections introduce oscillations on top of the ballistic edge profile. These oscillations vanish in the limit $X \rightarrow -\infty$ corresponding to the hydrodynamic scaling.}
\label{fig:Airy_edge_comparison}
\end{figure}

Close to the edges $|v| \simeq v_{max}$, it has been shown in free fermionic systems \cite{eisler2013full,viti2016inhomogeneous,allegra2016inhomogeneous} that the propagating front exhibits a finer structure within a distance $\Delta x$ from the edge $x \simeq \pm v_{max}t$ which scales as $\Delta x \sim t^{1/3}$. This behavior is classified as sub-diffusive, as it grows slower than the typical diffusive scaling $\Delta x \sim t^{1/2}$. Note that in non-interacting systems, such as those considered here, diffusion does not occur, as shown in Refs.~\cite{Fagotti2017Higher,DeNardis2018}. The leading correction to the hydrodynamic scaling is therefore sub-diffusive with a relative width $\Delta x/x \sim t^{-2/3}$, which vanishes in the limit $t \to \infty$. In particular, this behavior has been shown to be described by a universal function, the Airy kernel \cite{tracy1994level}. We show here that the latter characterizes also the sub-diffusive corrections to the hydrodynamic scaling of the front edge for the bosonic chain. In fact, introducing the scaling variable $X$,
\begin{equation}
X = (x-v_{max}t)\left(\frac{2}{v_{max}t} \right)^{1/3}, \label{eq:scaling_variable_Airy}
\end{equation}
for the energy current $\mathcal{J}^E(X,t)$ at the right edge $x \simeq v_{max}t$
we have (the derivation is presented in Appendix \ref{app:appendix2.5})     
\begin{equation}
\mathcal{J}^E(X,t) = \Omega(k_s) v_{max} [n_l(X,t)-n_r(X,t)],
 \label{eq:Airy_kernel_bosons_1}
\end{equation}
with
\begin{equation}
n_{l,r}(X,t) = \left(\frac{2}{v_{max}t} \right)^{1/3} f_{\beta_{l,r}}^{-}(k_s) \, K^{A}(X,X), \label{eq_Airy_kernel_bosons_2}
\end{equation}
where $k_s$ is the solution of the stationary phase equation (see Eq.~\eqref{eq:systemsolution} in Appendix \ref{app:appendix2})
\begin{equation}
v_g(k_s) = v_{max}, \label{eq:saddle_point_Airy}
\end{equation}    
and the Airy kernel $K^{A}(X,X)$ is defined as 
\begin{equation}
K^A(X,X) = [\mbox{Ai}'(X)]^2 -X[\mbox{Ai}(X)]^2, \label{eq:Airy_kernel_bosons_3}
\end{equation} 
where $\mbox{Ai}$ is the Airy function. The same formula applies to the left edge $x \simeq -v_{max} t$ with $x$ replaced by $-x$ in Eq.~\eqref{eq:scaling_variable_Airy} as $\mathcal{J}^E(-x,t)= \mathcal{J}^E(x,t)$. Note that in the hydrodynamic limit with fixed $v=x/t$ and large $x$ and $t$, the scaling variable $X$ in Eq.~\eqref{eq:scaling_variable_Airy} behaves as $X \sim t^{2/3}(v-v_{max}) \rightarrow -\infty$ and, by using the corresponding asymptotic behavior of the Airy kernel \cite{NIST:DLMF} $K^A(X,X) \rightarrow \sqrt{-X}/\pi$, one realizes that Eq.~\eqref{eq:Airy_kernel_bosons_1} reduces to Eq.~\eqref{eq:edge_ballistic_bosons}. This is shown in Fig.~\ref{fig:Airy_edge_comparison}, where we plot on the vertical axis the rescaled energy current $\mathcal{J}^E(X,t) \, (v_{max}t/2)^{1/3}/(\Omega(k_s) v_{max} (f_{\beta_l}^{-} -f_{\beta_r}^{-}))$ as a function of $X$ in Eq.~\eqref{eq:scaling_variable_Airy} for the ballistic limit in Eq.~\eqref{eq:edge_ballistic_bosons}, dashed line, and in the sub-diffusive case of Eq.~\eqref{eq:Airy_kernel_bosons_1}, solid line. The latter displays a typical staircase structure: for a free fermionic chain starting from a domain-wall initial state this staircase has been interpreted in Ref.~\cite{eisler2013full} by establishing a correspondence between the counting statistics of free fermions and the eigenvalues statistics in random matrix theory.
For the transverse field Ising chain it has been shown in Refs.~\cite{perfetto2017ballistic,kormos2017inhomogeneous} that when the magnetic field is set to its critical value $h=1$ the kernel describing the edge behavior as in Eq.~\eqref{eq:Airy_kernel_bosons_1} is no longer the Airy kernel $K^A(X,X)$ but a different one (see Eqs.~(64), (66), and (68) of Ref.~\cite{perfetto2017ballistic}; note that a factor $1/2$ is missing in front of Eq.~(64)\footnote{We are grateful to M.~Kormos for pointing out this issue with Eq.~(64).}) lacking the staircase structure of the Airy kernel. In the bosonic case, however, this is not the case, since when the mass $m$ is set to zero, from Eq.~\eqref{eq:Airy_kernel_bosons_1} one obtains (details are provided in Appendix \ref{app:appendix2.5})
\begin{equation}
\mathcal{J}^E(X,t) = v_{max} \left(\frac{8}{v_{max}t}\right)^{1/3} \left(\frac{1}{\beta_l}-\frac{1}{\beta_r} \right) K^A(X,X) \label{eq:Airy_critical}
\end{equation}
with the scaling variable $X$ defined, in this case, as  
\begin{equation}
X= (x-v_{max}t)\left(\frac{8}{v_{max}t} \right)^{1/3}, \label{eq:scaling_variable_critical}
\end{equation}
which is therefore proportional to that corresponding to $m \neq 0$ (see Eq.~\eqref{eq:scaling_variable_Airy}).
Accordingly, for free bosonic systems, the edge behavior does not qualitatively change at criticality $m=0$, in contrast with what happens in free fermionic systems. The different edge behavior of $\mathcal{J}^E$ in critical fermionic and bosonic systems is heuristically related to the fact that in the latter case the zero-momentum mode $k_s=0$, which solves Eq.~\eqref{eq:saddle_point_Airy} for $v \rightarrow \pm v_{max}^{\mp}$, can be populated by an arbitrarily large number of quasi-particles, since $f_{\beta}^{-}(k_s)$ diverges at $k_s=0$ for $m=0$. In the fermionic case, instead, the occupation of the mode $k_s=0$ remains finite at criticality $h=1$, as $f_{\beta}^{+}(k_s)=1/2$. For bosonic systems this discussion is similar to the one done for the Bose-Einstein condensation of the ideal Bose gas, see, e.g., Ref.~\cite{huang2009introduction}, where the zero-momentum mode in the condensed phase becomes macroscopically populated. The precise connection between the edge behavior of the energy current $\mathcal{J}^E$ in critical non-interacting bosonic systems, and the occurrence of the Bose-Einstein condensation goes, however, beyond the scope of the present manuscript and we leave it for future investigation.

The hydrodynamic limit of the total energy $\Delta e(x,t)$ flowing through point $x$, defined in Eq.~\eqref{eq:transferredenergyoperator}, takes a form analogous to Eq.~\eqref{transferredenergy}, i.e.,
\be 
\begin{split}
\Delta \mathcal{E}(x,t)  &= t \int_{0}^{\pi} \frac{dk}{2 \pi} \varepsilon(k)(v_g(k) -|v|) \\
&\ \ \times \left[f^{-}_{\beta_l}(k)-f^{-}_{\beta_r}(k)\right] \Theta(v_g(k)-|v|). \label{eq:transferredenergybosons}
\end{split}
\ee 
Accordingly, as far as the mean value of the transferred energy $\Delta \mathcal{E}(x,t)$ is concerned, a free bosonic theory is actually very similar to a free fermionic theory. In the next section, however, we will show that the full counting statistics of the operator $\Delta e(x,t)$ --- which takes into account also higher order cumulants --- strongly differs in the two cases.  

\section{Scaled cumulant generating function and large deviations in the hydrodynamic limit}
\label{thirdsection}
The analysis of the previous section focused on the mean value of the transferred energy operator $\Delta e(x,t)$ in Eq.~\eqref{eq:transferredenergyoperator} within the hydrodynamic scaling limit. However, to get information about fluctuations beyond mean values, one needs to study higher-order cumulants of this quantity. This is conveniently done by defining the scaled cumulant generating function $G(\lambda,v)$ (SCGF; see, e.g., Ref.~\cite{touchette2009large}) at the hydrodynamic scale with $x,t \rightarrow \infty$ and fixed $v=x/t$, which for the transferred energy $\Delta e(x,t)$ reads:
\be 
G(\lambda,v) \equiv \lim_{\substack{x,t\to\infty \\ v=x/t}} \frac{1}{t} \mbox{ln} \, \mbox{Tr} \{\rho_0 \, \mbox{exp}[{-\lambda \Delta e(x,t)}]\}, \label{eq:SCGF_hydro_euler}
\ee
where we are anticipating the fact that $G(\lambda,v)$ depends on $x$ and $t$ only via the scaling variable $v$. Note that the operator $\Delta e(x,t)$, differently from the energy current $j^E_x$ and the density $u_x$ and according to its very definition in Eq.~\eqref{eq:transferredenergyoperator}, is not local and therefore the average over $\rho$ in Eq.~\eqref{eq:SCGF_hydro_euler} cannot be taken directly as in Eq.~\eqref{eq:hydro_limit}. Moreover, the trace in Eq.~\eqref{eq:SCGF_hydro_euler} is taken with respect to the initial density matrix $\rho_0$ in Eq.~\eqref{eq:intro-rho0} which is, as already stated in Sec.~\ref{secondsection}, non-stationary and inhomogeneous. This causes $G(\lambda,v)$ to have a non-trivial dependence on $v$.  

As noticed after Eq.~\eqref{transferredenergy}, $\Delta \mathcal{E}(x,t)$ grows extensively upon increasing the time $t$ and it is therefore convenient to focus on the intensive quantity $J_E=\Delta e(x,t)/t$. According to the large deviation principle (see, e.g., Ref.~\cite{touchette2009large}), this kind of intensive quantities have a probability density function $p(J_E,v)$ that for $t \rightarrow \infty$ peaks exponentially around the mean value $\langle J_E \rangle = \Delta \mathcal{E}(x,t)/t$ as 
\begin{equation}
p(J_E,v) \sim \mbox{exp}[{-t I(J_E,v)}], \label{eq:large_deviation_principle}
\end{equation}   
where $I(J_E,v)$ is referred to as the large deviation or rate function. This function is convex, non-negative, with a unique zero at the mean and most probable value $\langle J_E \rangle$, i.e., $I(\langle J_E \rangle,v)=0$. The rate function $I$ can be determined from $G(\lambda,v)$ via the Legendre-Fenchel transform \cite{touchette2009large}
\begin{equation}
I(J_E,v) = \sup_{\lambda} \left[-\lambda J_E -G(\lambda,v)  \right]. \label{eq:legendrefenchel}
\end{equation} 
Moreover, when $G(\lambda,v)$ is strictly convex (i.e., convex with no linear parts), as in the cases that will be analyzed in Secs.~\ref{sec:FCSTFIC} and \ref{sec:SCGF_bosons}, the Legendre-Fenchel transform reduces to the well-known Legendre transform with the rate function related to the SCGF by the Legendre duality, i.e., 
\begin{equation}
\frac{\partial G(\lambda, v)}{\partial \lambda} = -J_E; \quad \mbox{and} \quad \frac{\partial I(J_E, v)}{\partial J_E} = -\lambda , \label{eq:legendreduality}
\end{equation}
from which it follows that the slope of $G(\lambda,v)$ as a function of $\lambda$ equals $-J_E$ and, vice versa, the slope of $I(J_E,v)$ as a function of $J_E$ equals $-\lambda$.

To our knowledge, all the available predictions for the scaled cumulant generating function of the transferred energy $\Delta e(x,t)$ are obtained by computing the trace in Eq.~\eqref{eq:SCGF_hydro_euler} over the stationary density matrix $\rho_{stat}$ in Eq.~\eqref{eq:stationary_density_matrix}, i.e.,
\begin{equation}
G(\lambda) = \lim_{t \rightarrow \infty} \frac{1}{t} \mbox{ln} \, \mbox{Tr} \{\rho_{stat} \, \mbox{exp}[{-\lambda \Delta e(x,t)}]\}; \label{eq:SCGF_stationary_state}
\end{equation}
since $\rho_{stat}$ is homogeneous and stationary $G(\lambda)$ does not depend, in this case, on space or time.
In particular, for free-fermions models, $G(\lambda)$ can be determined via the celebrated Levitov and Lesovik formula \cite{levitov1993charge,levitov1994quantum,levitov1996electron}, which, with the notation of this work, reads
\begin{align}
G(\lambda) = \int_{}^{} \frac{d \varepsilon}{2 \pi} \; &\mbox{ln}\left\{1 +T(\varepsilon)[(e^{-\lambda \varepsilon}-1)f_{\beta_l}^{+}(\varepsilon)(1-f_{\beta_r}^{+}(\varepsilon)) \right.  \nonumber \\ 
& \left. + (e^{\lambda \varepsilon}-1)f_{\beta_r}^{+}(\varepsilon)(1-f_{\beta_l}^{+}(\varepsilon))]\right\}, \label{eq:LevitovLesovik}
\end{align}
where the integral runs over the energy spectrum $\varepsilon \in (\varepsilon_{min},\varepsilon_{max})$ of the model ($\varepsilon_{min}$ and $\varepsilon_{max}$ have been defined after Eq.~\eqref{eq:maxvelocityTFIC} for the Ising chain) and $T(\varepsilon)$ denotes the transmission probability of a particle from the left to the right chain and vice versa. 
In a similar way, $G(\lambda)$ can be computed for a free bosonic theory \cite{saito2007fluctuation} and its analytic expression, obtained via the Keldysh formalism, turns out to have a structure similar to that of Eq.~\eqref{eq:LevitovLesovik}:
\begin{align}
G(\lambda) = -\int_{\Omega_{min}}^{\Omega_{max}} \frac{d \Omega}{4 \pi} \; &\mbox{ln}\left\{1 +T(\Omega)[(e^{-\lambda \Omega}-1)\right. \nonumber \\ 
\times f_{\beta_l}^{-}(\Omega)f_{\beta_r}^{-}(-\Omega) & \left. + (e^{\lambda \Omega}-1)f_{\beta_r}^{-}(\Omega)f_{\beta_l}^{-}(-\Omega)]\right\}, \label{eq:daarLevitov}
\end{align}
with $\Omega_{min}$ and $\Omega_{max}$ given in Eq.~\eqref{eq:maxvelocityOAS} for the harmonic chain. Remarkably, a recent formula for $G(\lambda)$ for interacting integrable models in homogenous stationary states has been found in Refs.~\cite{doyonMyers2020,doyon2020fluctuations} on the basis of generalized hydrodynamics techniques. This formula is valid for homogeneous and stationary GGEs \cite{vidmar2016generalized}, which include the non-equilibrium steady states of the form in Eq.~\eqref{eq:stationary_density_matrix} obtained from the partitioning protocol and, in fact, it renders Eqs.~\eqref{eq:LevitovLesovik} and \eqref{eq:daarLevitov} when specialized to free fermions and bosons, respectively.

In spite of this important progress, a formula for the SCGF $G(\lambda,v)$ of the transferred energy $\Delta e(x,t)$ over an inhomogeneous state $\rho_0$ is still missing, even in the case of free models. In particular, the analysis of Refs.~\cite{doyonMyers2020,doyon2020fluctuations} cannot describe the dependence of $G(\lambda,v)$ on $v$ since it applies only to homogeneous and stationary states, where $G(\lambda,v)=G(\lambda)$ is independent of $v$, as emphasized above after Eq.~\eqref{eq:SCGF_stationary_state}. Aiming at filling this gap, we therefore begin in Sec.~\ref{sec:SCGF_general_result} with the exact calculation of the SCGF $G(\lambda,v)$ as a function of $v$ at the hydrodynamic scale according to Eq.~\eqref{eq:SCGF_hydro_euler}, while in Sec.~\ref{sec:semi_classics} a simple semi-classical interpretation of these results is provided. In Secs.~\ref{sec:FCSTFIC} and \ref{sec:SCGF_bosons} the general result of Sec.~\ref{sec:SCGF_general_result} is eventually specialized for the transverse field Ising chain, and the harmonic chain, respectively. 

\subsection{The scaled cumulant generating function in the hydrodynamic limit} 
\label{sec:SCGF_general_result}
The derivation of the scaled cumulant generating function presented here is similar to that of $G(\lambda)$ in Eq.~\eqref{eq:SCGF_stationary_state} done in CFT \cite{bernard2012energy,bernard2015non,bernard2016conformal}, on the lattice for the TFIC \cite{de2013nonequilibrium}, and, more recently, for interacting integrable models \cite{doyonMyers2020,doyon2020fluctuations}. \\
In order to determine $G(\lambda,v)$ we start by taking the derivative with respect to $\lambda$ of Eq.~\eqref{eq:SCGF_hydro_euler}, i.e.,
\begin{equation}
-\frac{\partial G(\lambda,v)}{\partial \lambda} = \! \! \lim_{\substack{x,t\to\infty \\ v=x/t}} \frac{1}{t} \int_{0}^{t} ds \frac{\mbox{Tr}\{\rho_0 \, j^E_x(s) \, \mbox{exp}[{-\lambda \Delta e(x,t)}]\}}{\mbox{Tr}\{\rho_0 \, \mbox{exp}[{-\lambda \Delta e (x,t)}]\}}. \label{eq:derivative_SCGF_Hydro}
\end{equation} 
By using the definition of the translation operator $P_{tr}$ in Eq.~\eqref{eq:translation_operator_TFIC} (or in Eq.~\eqref{eq:translation_operator_OAS} for the harmonic chain) and the time evolution under $H$ we can write
\begin{equation}
j_x^{E}(s) = e^{i H s}(P_{tr})^x j_0^{E} (0) (P_{tr}^{\dagger})^x e^{-i H s},
\end{equation}
and therefore, by cyclicity of the trace, Eq.~\eqref{eq:derivative_SCGF_Hydro} becomes
\begin{equation}
-\frac{\partial G(\lambda,v)}{\partial \lambda} = \lim_{\substack{x,t\to\infty \\ v=x/t}} \frac{1}{t} \int_{0}^{t} ds \frac{\mbox{Tr}[ \rho(x,s,\lambda) j^E_0(0)]}{\mbox{Tr}[\rho(x,s,\lambda)]}, \label{eq:tilted_average} 
\end{equation}   
where we defined
\begin{equation}
\rho(x,s,\lambda) \! \equiv \! \mbox{exp}[{-\lambda \Delta e (0;s,t)}] (P_{tr}^{\dagger})^x e^{-i H s} \rho_0 e^{i H s} P_{tr}^x, \label{eq:non_local_tilting_1}
\end{equation}
and
\begin{align}
\mbox{exp}[{-\lambda \Delta e (0;s,t)}] & \equiv   e^{-i H s} \mbox{exp}[{-\lambda \Delta e(0,t)}] e^{i H s} \nonumber \\
 &= \mbox{exp}\left({-\lambda \int_{-s}^{t-s} ds' j_0^E(s')}\right). \label{eq:non_local_tilting_2}
\end{align}
Note that, for $\lambda=0$, Eq.~\eqref{eq:non_local_tilting_1} reduces in the hydrodynamic limit to Eq.~\eqref{eq:hydro_state_proved} for the Ising chain, and to Eq.~\eqref{eq:hydro_state_proved_bosons} for the harmonic one, while Eq.~\eqref{eq:tilted_average} is just the hydrodynamic limit of the mean of the energy current $j_0^E(0)$ given in Eq.~\eqref{eq:energy_current_hydro_TFIC} for the fermionic case, and in Eq.~\eqref{eq:energycurrentoas_hydro} for the bosonic one.      
The physical interpretation of Eqs.~\eqref{eq:tilted_average}, \eqref{eq:non_local_tilting_1}, and \eqref{eq:non_local_tilting_2} is therefore that the insertion of the exponential of the time-integrated current biases the statistical measure, from $\rho(v)$ to $\rho(x,t,\lambda)$, with respect to which the energy current is averaged. 
The key point to proceed in the calculation is that this $\lambda$-tilted ensemble $\rho(x,t,\lambda)$ has still the same form as Eq.~\eqref{eq:hydro_state_proved} for fermions (and Eq.~\eqref{eq:hydro_state_proved_bosons} for bosons) with $\beta(v,k)$ acquiring an additional dependence on $\lambda$ as $\beta(v,k,\lambda)$.

To see this, we consider the hydrodynamic limit of Eq.~\eqref{eq:non_local_tilting_2}, which can be readily determined by writing $j_0^{E}(0)$ in terms of post-quench mode operators $\Psi_R(k)$ for the Ising case, and $\AA(k)$ for the harmonic oscillators (see Eq.~\eqref{eq:current_operator_modes} in Appendix \ref{app:appendix2}); then we consider the time evolution up to time $s'$, integrating according to Eq.~\eqref{eq:non_local_tilting_2} and then doing a stationary phase approximation analogous to the one done after Eq.~\eqref{eq:intermediate_result_rho_x,t}, with $x$ set to zero. An analogous analysis has been done for $\Delta e(0;t/2,t)$ in Ref.~\cite{de2013nonequilibrium}. For the Ising model, this results in (the corresponding equation for the harmonic chain can be obtained by replacing $\Psi_R(k) \rightleftarrows \AA(k)$ and $\varepsilon(k) \rightleftarrows \Omega(k)$)
\begin{equation}
\Delta e (0;s,t) = \int_{-\pi}^{\pi} dk \, \mbox{sgn}(v_g(k)) \varepsilon(k) \Psi_R^{\dagger}(k) \Psi_{R}(k), \label{eq:tilting_sign}
\end{equation}         
with ${\rm sgn}(x>0) = +1$ and ${\rm sgn}(x<0)= -1$. By plugging Eq.~\eqref{eq:tilting_sign} into Eq.~\eqref{eq:non_local_tilting_2} and then into Eq.~\eqref{eq:non_local_tilting_1}, we get a state $\rho(x,t,\lambda)=\rho(v,\lambda)$ equal to the one in Eq.~\eqref{eq:hydro_state_proved} (or Eq.~\eqref{eq:hydro_state_proved_bosons} for the harmonic chain) with the replacement 
\begin{equation}
\beta(v,k) \longrightarrow \beta(v,k,\lambda) = \beta(v,k) + \lambda \, \mbox{sgn}(v_g(k)).  \label{eq:flow_equation}
\end{equation} 
In Eq.~\eqref{eq:tilted_average}, with the $\rho(v,\lambda)$ determined by Eq.~\eqref{eq:flow_equation}, one can directly calculate the average of $j_0^E(0)$  in the hydrodynamic limit, as in Eq.~\eqref{eq:hydro_limit}, since now only the local operator $j_0^{E}(0)$ appears inside the trace. Using the expression in Eqs.~\eqref{eq:energy_current_hydro_TFIC} and \eqref{eq:energycurrentoas_hydro} into Eq.~\eqref{eq:derivative_SCGF_Hydro}, we get 
\begin{equation}
\frac{\partial G(\lambda,v)}{\partial \lambda} = -\frac{1}{t}  \int_{0}^{t} ds \int_{-\pi}^{\pi} \frac{dk}{2 \pi} \varepsilon(k) v_g(k) n^{+}\left(\frac{x}{s},k,\lambda \right), \label{eq:SCGF_hydro_almost_done}    
\end{equation}
where
\begin{equation}
n^{+}(v,k,\lambda) = f_{\beta_r(\lambda)}^{+}(k) \Theta (v-v_g(k))+f_{\beta_l(\lambda)}^{+}(k)\Theta(v_g(k)-v),
\end{equation}     
with $\beta_{r,l}(\lambda)=\beta_{r,l} + \lambda \, \mbox{sgn}(v_g(k))$. Integrating Eq.~\eqref{eq:SCGF_hydro_almost_done} over $\lambda$ with the initial condition $G(\lambda=0,v)=0$, after simple algebraic manipulations, one obtains a final compact expression for $G(\lambda,v)$ with $v>0$ valid for both fermions and bosons 
\begin{eqnarray}
G(\lambda,v) = G_{\beta_r}(\lambda) &-& \! \int_{\epsilon_{min}}^{\epsilon_{max}} \frac{d \varepsilon}{2 \pi} \Theta(v_g(\varepsilon)-v) \left(1-\frac{v}{v_g(\varepsilon)}\right) \nonumber \\ 
             && \left\{[F((\beta_l+\lambda)\, \varepsilon)-F(\beta_l \, \varepsilon)]   \right. \nonumber \\ 
             &&-\left.[F((\beta_r+\lambda) \, \varepsilon)-F(\beta_r \, \varepsilon)]\right\}, \nonumber \\ \label{eq:finalresultFCS_general}
\end{eqnarray}
where we introduced 
\begin{align}
G_{\beta}(\lambda) = -\int_{\epsilon_{min}}^{\epsilon_{max}} \frac{d \varepsilon}{2 \pi} \left\{[F((\beta+\lambda) \, \varepsilon)-F(\beta \, \varepsilon)]   \right. \nonumber \\ 
             +\left.[F((\beta-\lambda) \, \varepsilon)-F(\beta \, \varepsilon)]\right\},  \label{eq:reservoir_FCS_general}
\end{align}
with the function $F(\varepsilon)$ depending on the statistics of the quasi-particles as
\be
F(\varepsilon) = \left\{
  \begin{array}{lr}
    -\mbox{ln}(1+e^{-\varepsilon}) \; \; \; \;  \mbox{for fermions}; \\
    \mbox{ln}(1-e^{-\varepsilon}) \; \; \; \; \; \, \, \, \mbox{for bosons.}
  \end{array}
\right. \label{free_energy_TBA}
\ee
In the previous expressions $\epsilon_{min}$ and $\epsilon_{max}$ are the minimum and the maximum of the single-particle energy spectrum (see Eq.~\eqref{eq:maxvelocityTFIC} for the Ising chain and Eq.~\eqref{eq:maxvelocityOAS} for the harmonic chain). For $v<0$, one gets from Eq.~\eqref{eq:SCGF_hydro_almost_done} a formula similar to Eq.~\eqref{eq:finalresultFCS_general} with the replacements $v \rightarrow -v$, $l \rightleftarrows r$ and $\lambda \rightarrow -\lambda$, i.e., 
\begin{eqnarray}
G(\lambda,v) = G_{\beta_l}(\lambda) &-& \int_{\epsilon_{min}}^{\epsilon_{max}} \frac{d \varepsilon}{2 \pi} \Theta(v_g(\varepsilon)+v) \left(1+\frac{v}{v_g(\varepsilon)}\right) \nonumber \\ 
             && \left\{[F((\beta_r-\lambda)\, \varepsilon)-F(\beta_r \, \varepsilon)]   \right. \nonumber \\ 
             &&-\left.[F((\beta_l-\lambda) \, \varepsilon)-F(\beta_l \, \varepsilon)]\right\}. \nonumber \\ \label{eq:finalresultFCS_general_v_negative}
\end{eqnarray}
Equations \eqref{eq:finalresultFCS_general} and \eqref{eq:finalresultFCS_general_v_negative} are the main results of this paper.

One can see that for $v>v_{max}$ ($v<-v_{max}$) (where $v_{max}$ is given in Eq.~\eqref{eq:maxvelocityTFIC} or \eqref{eq:maxvelocityOAS} depending on the model considered), the second term in Eqs.~\eqref{eq:finalresultFCS_general} and \eqref{eq:finalresultFCS_general_v_negative} vanishes and one is left with $G(\lambda,v)= G_{\beta_r}(\lambda)$ ($G_{\beta_l}(\lambda)$). The physical interpretation of this result is straightforward since outside the light-cone $v>v_{max}$ ($v<-v_{max}$) the system is described by a reservoir at inverse temperature $\beta_r$ ($\beta_l$), which is not affected by the dynamics. Correspondingly, the temperature in this region is homogeneous and the mean current $\mathcal{J}^E(v)$ vanishes, while due to thermal fluctuations, its higher-order cumulants do not and they are described by the SCGF of the reservoir at the initial temperature of that part of the chain. On the other hand, the NESS can be retrieved as a particular case of Eqs.~\eqref{eq:finalresultFCS_general} and \eqref{eq:finalresultFCS_general_v_negative}. Setting $v=0$ in Eq.~\eqref{eq:finalresultFCS_general} (Eq.~\eqref{eq:finalresultFCS_general_v_negative}) and using Eq.~\eqref{eq:reservoir_FCS_general} for $G_{\beta_r}$ ($G_{\beta_l}$), one finds
\begin{align}
G(\lambda, v=0) = -\int_{\epsilon_{min}}^{\epsilon_{max}} \frac{d \varepsilon}{2 \pi} \left\{[F((\beta_{l}+\lambda) \, \varepsilon)-F(\beta_{l} \, \varepsilon)]   \right. \nonumber \\ 
             +\left.[F((\beta_{r}-\lambda) \, \varepsilon)-F(\beta_{r} \, \varepsilon)]\right\}.  \label{eq:NESS_FCS_general}
\end{align} 
In Secs.~\ref{sec:FCSTFIC} and \ref{sec:SCGF_bosons}, we show that for non-interacting fermions and bosons Eq.~\eqref{eq:NESS_FCS_general} coincides with Eqs.~\eqref{eq:LevitovLesovik}, and \eqref{eq:daarLevitov} respectively, with unitary transmission coefficient, as expected from the fact that, after the quench, the Hamiltonian is translational invariant and therefore no reflection occurs at the junction. For generic values of $v$, Eqs.~\eqref{eq:finalresultFCS_general} and \eqref{eq:finalresultFCS_general_v_negative} provide the complete dynamics of the SCGF, and of all the cumulants of the transferred energy $\Delta e(x,t)$, in the hydrodynamic limit, extending the known results in the literature about the NESS.   

Note that the expressions in Eqs.~\eqref{eq:finalresultFCS_general} and \eqref{eq:finalresultFCS_general_v_negative} for $G(\lambda,v)$ satisfy the important relation
\begin{align}
\frac{\partial G(\lambda,v)}{\partial \lambda} = & - \frac{1}{t}\Delta \mathcal{E}(x,t)\left|_{\beta_l+\lambda,\beta_r-\lambda}\right.  \nonumber \\ 
&-\frac{1}{t}\Delta\mathcal{E}(x,t)\left|_{\beta_r-\lambda,\beta_r+\lambda}\right. + \mathcal{J}^{E}_{NESS}\left|_{\beta_r -\lambda,\beta_r+\lambda}\right., \label{eq:generalized_EFR_hydro}
\end{align}
where we denoted by $\Delta\mathcal{E}(x,t)\left|_{\beta_{l},\beta_{r}}\right.$ the mean in Eqs.~\eqref{transferredenergy} and \eqref{eq:transferredenergybosons} of the transferred energy operator $\Delta e(x,t)$ at the hydrodynamic scale. The first subscript $\beta_{l}$ of $\Delta\mathcal{E}(x,t)\left|_{\beta_{l},\beta_{r}}\right.$ refers to the inverse temperature of the first Fermi-Bose function $f_{\beta_{l,r}}^{\pm}$ appearing on the right hand side of Eqs.~\eqref{transferredenergy} and \eqref{eq:transferredenergybosons} with positive sign, while the second subscript $\beta_{r}$ denotes the inverse temperature of the second Fermi-Bose factor appearing in the same equations with negative sign. 
$\mathcal{J}^{E}_{NESS}\left|_{\beta_r -\lambda,\beta_r+\lambda}\right.$ is the stationary-state energy current obtained upon setting $v=0$ in Eqs.~\eqref{eq:energy_current_hydro_TFIC} and \eqref{eq:energycurrentoas_hydro} and by replacing $\beta_l \rightarrow \beta_r-\lambda$ and $\beta_r \rightarrow \beta_r+\lambda$. In particular, for $v=0$ the two terms on the second line of Eq.~\eqref{eq:generalized_EFR_hydro} cancel each other and one obtains 
\begin{equation} 
\frac{\partial G(\lambda,v=0)}{\partial \lambda} = -\mathcal{J}^E_{NESS} \left |_{\beta_l + \lambda,\beta_r - \lambda} \right., \label{eq:EFR}
\end{equation}  
which is known in the literature as the extended fluctuation relation; it was proved in Ref.~\cite{bernard2013time} for the NESS limit of the SCGF $G(\lambda)$ in Eq.~\eqref{eq:SCGF_stationary_state}, and in particular it is known to apply to free particles models \cite{de2013nonequilibrium,bernard2015non,yoshimura2018full} and conformal field theory \cite{bernard2012energy,bernard2015non,bernard2016conformal}. Recently, a generalization of Eq.~\eqref{eq:EFR} for homogeneous stationary states of interacting integrable models has been proved in Refs.~\cite{doyonMyers2020,doyon2020fluctuations} where the SCGF has been expressed as an integral over $\lambda$ of the mean energy current with Lagrange parameters $\beta (\lambda)$ depending on $\lambda$. In the absence of interactions the dependence of this $\beta(\lambda)$ on $\lambda$ reduces to a shift by $\lambda $ as in Eq.~\eqref{eq:EFR}. 
Our result in Eq.~\eqref{eq:generalized_EFR_hydro} therefore represents an extension of the extended fluctuation relation to the space-time scaling limit $v=x/t$ of the SCGF $G(\lambda,v)$ in Eq.~\eqref{eq:SCGF_hydro_euler}. The relation in Eq.~\eqref{eq:generalized_EFR_hydro} and generalizations thereof are important for generalizing the calculation of the cumulant generating function in Eq.~\eqref{eq:SCGF_hydro_euler} to the more complex case of interacting integrable models. In the latter, in fact, one only knows from the generalized hydrodynamics formalism of Refs.~\cite{bertini2016transport,castro2016emergent} the expression of the mean energy current. Then, by exploiting the extended fluctuation relation, one can derive the SCGF just by integrating the current with appropriately modified Lagrange parameters $\beta$ as a function of $\lambda$, therefore providing access to an expression otherwise extremely difficult to obtain. 

\subsection{Semi-classical interpretation of the expression of the scaled cumulant generating function}
\label{sec:semi_classics}
We have seen in Sec.~\ref{thirdsection}, that Eqs.~\eqref{eq:energy_current_hydro_TFIC} and \eqref{eq:energy_density_hydro_TFIC} for the Ising chain, and Eqs.~\eqref{eq:energycurrentoas_hydro} and \eqref{eq:energydensityoas_hydro} for the harmonic chain, can be simply interpreted in terms of quasi-particles excitations generated by the post-quench mode operators $\Psi_R^{\dagger}(k)$ or $\AA^{\dagger}(k)$ in Eqs.~\eqref{eq:post_quench_modes_TFIC} or \eqref{eq:post_quench_Oas_diagonal} with wave-vector $k \in [0,\pi)$, which travel ballistically with velocity $\pm v_g(k)$, defined after Eqs.~\eqref{eq:hydro_state_proved} and \eqref{eq:hydro_state_proved_bosons}. The quasi-particle picture is, indeed, expected to give exact results in the hydrodynamic limit $x,t \rightarrow \infty$ at fixed $v=x/t$ for the average of local observables, as shown for instance in Refs.~\cite{antal1999transport,karevski2002scaling,platini2005scaling,collura2014non,collura2014quantum,allegra2016inhomogeneous,
viti2016inhomogeneous,bertini2016determination,eisler2016universal,perfetto2017ballistic,
kormos2017inhomogeneous,10.21468/SciPostPhys.6.1.004}. This picture has been further corroborated in interacting integrable systems within the generalized hydrodynamics \cite{castro2016emergent,bertini2016transport}. Within the latter theory, the semiclassical quasi-particle picture has been, however, so far used only for the calculation of mean values of currents and densities, see, e.g., Refs.~\cite{kormossemiclassic1,kormossemiclassic2}. Similarly, the results of Refs.~\cite{doyonMyers2020,doyon2020fluctuations}, in which GHD is applied to the large-deviation theory of ballistically transported quantities, do not seem to have a simple interpretation in terms of the quasi-particle picture.
In this Subsection we provide, to our knowledge, the first application of the quasi-particle picture to the calculation of the SCGF $G(\lambda,v)$ in Eq.~\eqref{eq:SCGF_hydro_euler} of the time-integrated energy current. We therefore show that this picture can be non-trivially extended in order to exactly capture the \textit{fluctuations} of the transferred energy $\Delta e(x,t)$ beyond the mean value. The possibility of accounting for fluctuations via the quasi-particle picture resides in the fact that in finite-temperature states, as in the partitioning protocol considered in this manuscript, fluctuations are dominated by classical effects, i.e., they are essentially due to the statistical distribution of excitations in the initial state. Note, however, that these fluctuations, albeit being of classical nature, carry memory of the quantumness of the system through the function $F(\varepsilon)$ in Eq.~\eqref{free_energy_TBA}, which encodes the quantum statistics of the underlying quasi-particle excitations.

Quasi-particles with velocity $+v_g(k)$ 
propagate rightwards (right mover), while those with velocity $-v_g(k)$ propagate leftwards (left mover). 
The occupation of each mode $k$ is determined by the statistics of the initial state. For the one resulting from the partitioning protocol in Eq.~\eqref{eq:intro-rho0}, this occupation is thermal at inverse temperatures $\beta_l$ and $\beta_r$ for $y<0$ and $y>0$, respectively, where $y$ is the spatial coordinate along the chain. 
As a consequence, in order to represent in a semi-classical way the quasi-particles corresponding to the modes $\Psi_R^{\dagger}(k)$ and $\AA^{\dagger}(k)$, 
one defines the number $n_{\beta(y)}(k)$ of quasi-particles with wave vector $k$ initially ``located'' at site $y$ as a classical random variable with a probability distribution $P(n_{\beta(y)}(k))$ determined by the thermal distribution at inverse temperature $\beta(y)$. According to elementary statistical mechanics \cite{huang2009introduction}, for fermionic quasi-particles this distribution is given by
\begin{equation}
P(n_{\beta(y)}(k)= n) = \frac{e^{-\beta(y)\varepsilon(k)n}}{1+e^{-\beta(y)\varepsilon(k)}} \quad \mbox{with} \quad n=0,1, 
\label{eq:semiclassicprob_fermions}
\end{equation} 
while in the bosonic case
\begin{equation}
P(n_{\beta(y)}(k)=n) = e^{-\beta(y) \Omega(k)n}(1-e^{-\beta(y) \Omega(k)}),
\label{eq:semiclassicprob_bosons}
\end{equation}
with $n={0,1,...\infty}$ and
\begin{equation}
\beta(y) = \beta_r \Theta(y) + \beta_l \Theta(-y). \label{eq:semiclassicprob_0}
\end{equation}
The random variables $n_{\beta(y')}(k')$ and $n_{\beta(y)}(k)$ at lattice sites $y' \neq y$ and with wave vector $k' \neq k$ are taken to be independent since in free-particle models the various modes evolve independently and therefore Eqs.~\eqref{eq:semiclassicprob_fermions}, \eqref{eq:semiclassicprob_bosons}, and \eqref{eq:semiclassicprob_0} specify completely the probability of a given configuration of the quasi-particles along the chain after the quench.        

As a consequence of the independence of the variables $n_{\beta(y)}(k)$ for different values of $k$, one can write the scaled cumulant generating function $G(\lambda,x,t)$ of the total transferred energy $\Delta e(x,t)$ as
\be
G(\lambda,x,t) = \lim_{\substack{x,t\to\infty \\ v=x/t}} \frac{1}{t} \, \int_{0}^{\pi} \frac{dk}{2 \pi} \, \mbox{ln} \, \, g(\lambda,x,t;k) \label{eq:semiclassicalSCGF},
\ee
where $g(\lambda,x,t;k)$ is the moment generating function of the contribution $\Delta e(x,t;k)$ to the total transferred energy $\Delta e(x,t)$ due to the quasi-particles with wave vector $k$, defined as
\be
g(\lambda,x,t;k) = \langle \mbox{e}^{-\lambda \Delta e (x,t;k)} \rangle_{sc}, \label{eq:semiclassicalFCS}
\ee 
where the subscript \textit{``sc''} denotes the semi-classical average according to the mode distributions in Eqs.~\eqref{eq:semiclassicprob_fermions}, \eqref{eq:semiclassicprob_bosons}, and \eqref{eq:semiclassicprob_0}. We emphasize that, within the semi-classical description presented here, $\Delta e(x,t;k)$ is considered  as a classical random variable depending on $n_{\beta(y)}(k)$ and it is simply related to the total transferred energy $\Delta e(x,t)$ as
\be    
\Delta e(x,t) = \int_{0}^{\pi} \frac{dk}{2 \pi} \Delta e(x,t;k). \label{eq:decompostion_delta_e_semiclassics}
\ee
This formula expresses again the fact that modes with different $k$ contribute independently to $\Delta e(x,t)$ and therefore to $G(\lambda,x,t)$, as one can also see from Eq.~\eqref{eq:semiclassicalSCGF}.

\begin{figure}[h!] 
\centering
\includegraphics[width=0.99\columnwidth]{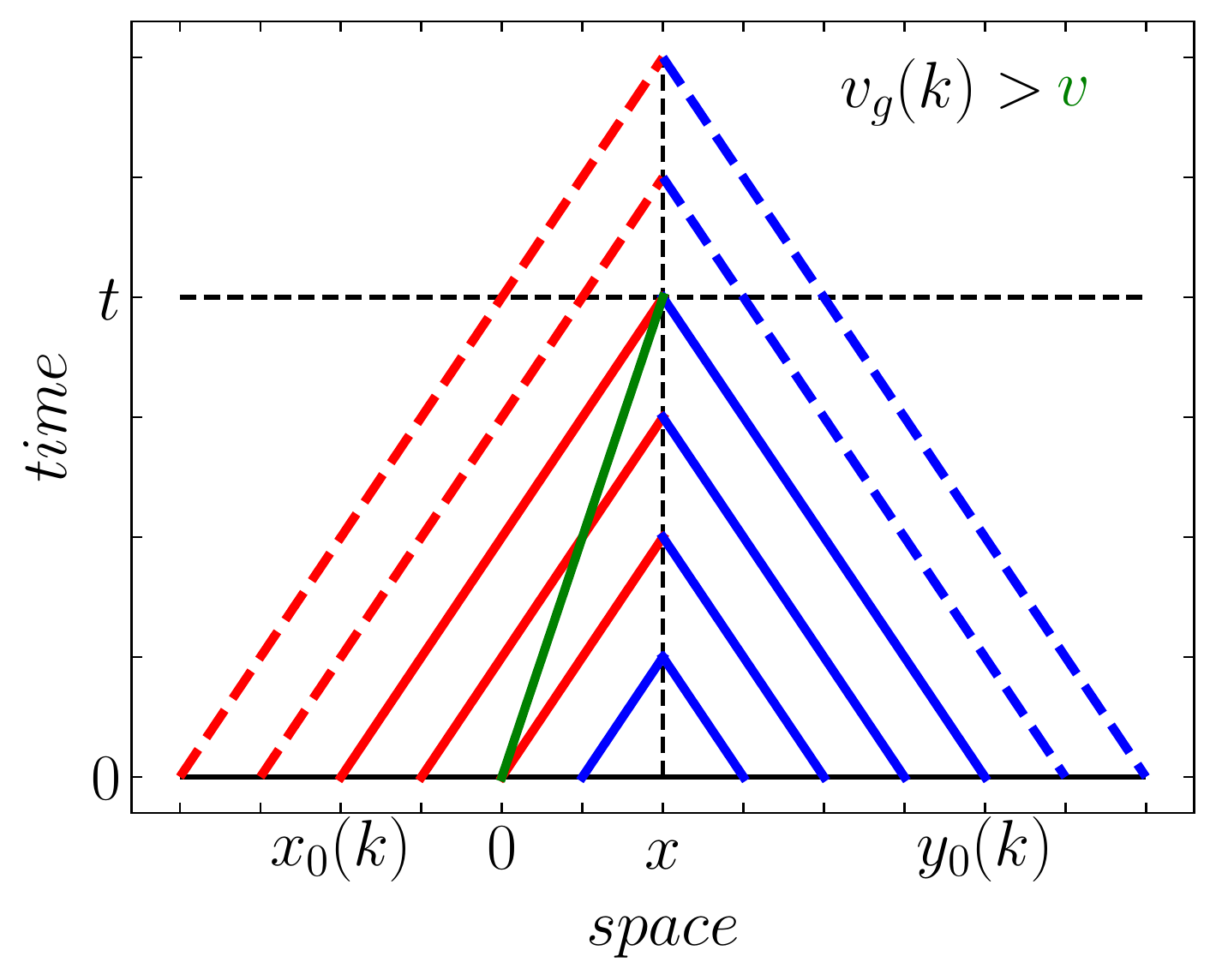}\\[-1mm]	
	(a)\\[2mm]
\includegraphics[width=0.99\columnwidth]{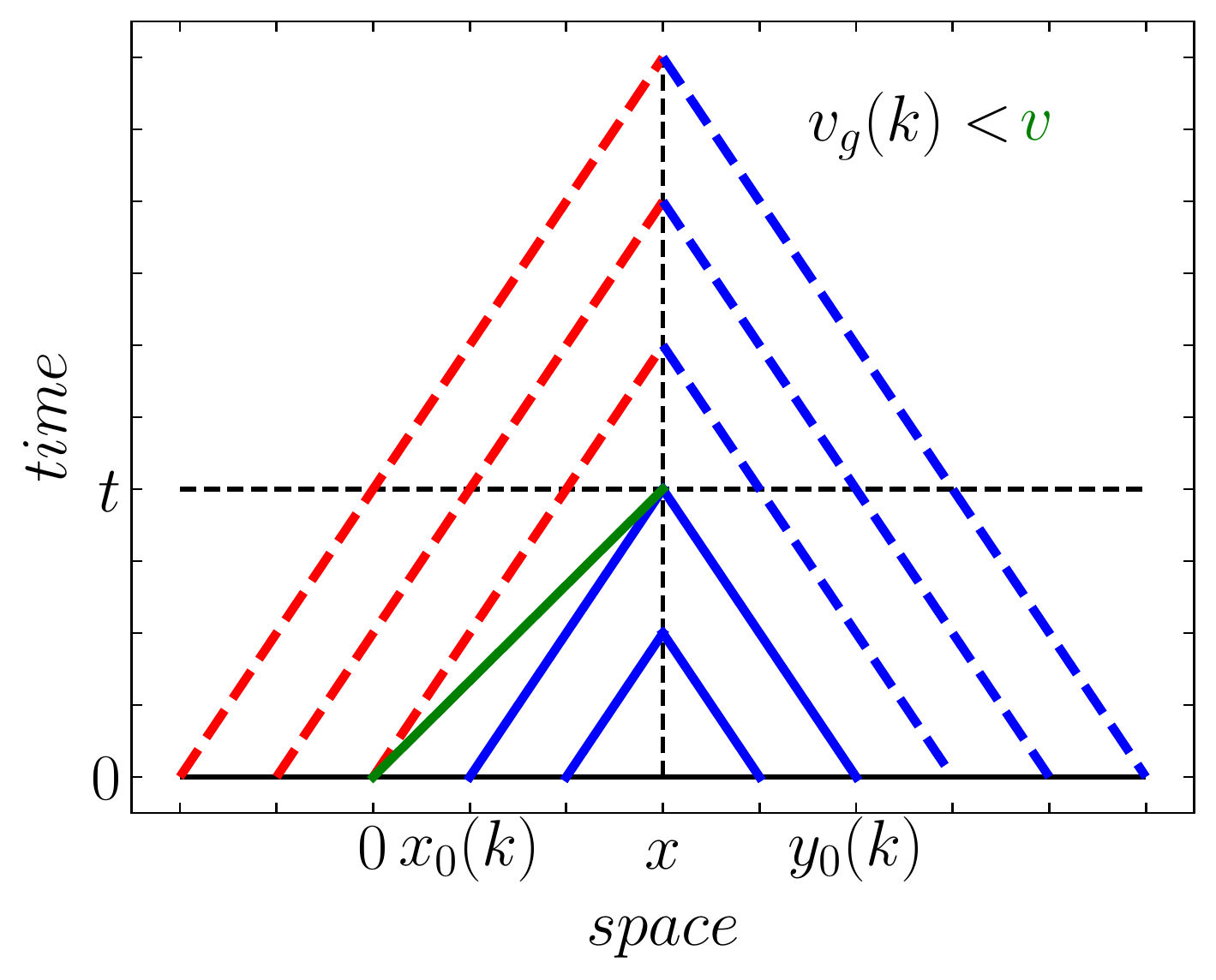}\\[-1mm]
(b)
\caption{Quasi-particles interpretation of the transferred energy $\Delta e(x,t;k)$ determined by the mode $k$ for a point at position $x=2$ and time $t=4$, corresponding to $v=x/t=0.5$. 
%
%
In panel (a) we consider a value of $k$ such that $v_g(k)>v$, choosing, as an example, $v_g(k)=1$. The light-ray with constant $v=x/t$ is reported in green. Right-moving quasi particles with $v_g(k)>0$ initially generated at points $y$ with $x_0(k) \leq y \leq  x$ cross the point $x$ within the interval $[0,t]$ and therefore contribute to the statistics of $\Delta e(x,t;k)$; in the sketch, their light-rays are indicated as red or blue solid lines depending on their inverse temperature being $\beta_l$ (for $y<0$) or $\beta_r$ (for $y>0$), respectively. 
Quasi-particles arriving in $x$ after time $t$, instead, do not contribute to $\Delta e(x,t;k)$ and the corresponding light-rays are indicated by dashed lines. Similarly, left-moving quasi particles contribute if they come from the interval $x \leq y \leq  y_0(k)$. In panel (b) we consider the case $x=3$, $t=4$, corresponding to $v=0.75$ and a value of $k$ such that $v_g(k)<v$, choosing $v_g(k)=0.5$. The same interpretation as panel (a) applies with the difference that only quasi particles coming from the right chain ($y>0$) determine now the statistics of the transferred energy.} 
\label{fig:quasiparticlesFCS}
\end{figure}

Since the transferred energy $\Delta e(x,t;k)$ is a time-integrated observable, 
it is determined not only by the flux of quasi-particles with wave vector $k$ arriving in $x$ at time $t$, but also by all the excitations crossing $x$ at times earlier than $t$, i.e., within the temporal interval $(0,t)$. 
Given that the quasi-particles propagate ballistically with velocity $\pm v_g(k)$, it is straightforward to express $\Delta e(x,t;k)$ in terms of the random variables $n_{\beta(y)}(k)$ in Eq.~\eqref{eq:semiclassicprob_0}. In particular, assuming $v\equiv x/t>0$, the quasi-particles with $v_g(k)>v$ and coming from the left chain are always able to reach the point $x$ within the time interval of interest, 
as sketched in Fig.~\ref{fig:quasiparticlesFCS}(a), while those with $v_g(k)<v$ contribute to the total energy change $\Delta e(x,t;k)$ only if coming from the right chain, see Fig.~\ref{fig:quasiparticlesFCS}(b).
%
%
%
The semi-classical expression of the energy transferred by the mode $k \in[0,\pi)$ for a generic value of $v$ is therefore given by the difference between the flux of quasi-particles initially generated in the interval $[x_0(k),x]$ and that of the quasi-particles generated in $[x,y_0(k)]$, with $x_0(k)=x-v_g(k)t$ and $y_0(k)=x+v_g(k)t$, as shown in Fig.~\ref{fig:quasiparticlesFCS}. In formulas,   
\be 
\Delta e(x,t;k) = \sum_{y=x_0(k)}^{x} \varepsilon(k) n_{\beta(y)}(k) - \sum_{y=x}^{y_0(k)} \varepsilon(k) n_{\beta(y)}(k), \label{eq:fastrandomvariable}
\ee
where $n_{\beta(y)}(k)$ is defined before Eq.~\eqref{eq:semiclassicprob_fermions}, and the energy $\varepsilon(k)$ carried by a mode $k$ in Eq.~\eqref{eq:Isingspectrum} for the Ising chain, while for the harmonic chain one has to replace $\varepsilon(k) \leftrightarrows \Omega(k)$, with $\Omega(k)$ given in Eq.~\eqref{eq:oasspectrum}. 
The moment generating function $g(\lambda,x,t;k)$ in Eq.~\eqref{eq:semiclassicalFCS}
can be then computed (see Appendix \ref{app:appendix3}), starting from Eqs.~\eqref{eq:fastrandomvariable}, \eqref{eq:semiclassicprob_fermions}, \eqref{eq:semiclassicprob_bosons}, and \eqref{eq:semiclassicprob_0} given that 
\begin{align}
\langle \mbox{e}^{-\lambda \varepsilon(k) n_{\beta(y)}(k)} \rangle_{sc} &= 1 + f_{\beta(y)}^{+}(k) (\mbox{e}^{-\lambda \varepsilon(k)}-1) \nonumber \\ 
&= \mbox{exp}[F(\beta(y)\varepsilon)-F((\beta(y)+\lambda)\varepsilon)], \label{eq:Bernoulliaverage}
\end{align}
for fermionic excitations, while 
\begin{align}
\langle \mbox{e}^{-\lambda \Omega(k) n_{\beta(y)}(k)} \rangle_{sc} &= [1 + f_{\beta(y)}^{-}(k) (1-\mbox{e}^{-\lambda \Omega(k)})]^{-1}    \nonumber \\ 
&=\mbox{exp}[F(\beta(y)\Omega)-F((\beta(y)+\lambda)\Omega)] \nonumber \\
& \qquad \qquad \qquad \qquad \qquad  \, \mbox{for} \, \, \,  \lambda>-\beta(y),  \label{eq:extremesdomain}
\end{align}
and otherwise infinite, in the bosonic case. 
The function $F(\varepsilon)$ has been defined in Eq.~\eqref{free_energy_TBA} and it depends on the statistics of the quasi-particles. Inserting the expression of $g(\lambda,x,t;k)$ into Eq.~\eqref{eq:semiclassicalSCGF} and after taking the space-time scaling limit, the expression for the SCGF turns out to be, as expected, a scaling function of $v=x/t$, i.e., $G(\lambda,x,t) \equiv G(\lambda,v)$ the expression of which coincides with the Eqs.~\eqref{eq:finalresultFCS_general}, for $v>0$, and with \eqref{eq:finalresultFCS_general_v_negative} for $v<0$. 
The semi-classical picture of ballistically propagating quasi particles 
is therefore not only capable of exactly capturing the mean value of the energy current and density at the hydrodynamic scale, i.e., of predicting $\mathcal{J}^E(v)$ and $\mathcal{U}(v)$ in Eqs.~\eqref{eq:energy_current_hydro_TFIC}, \eqref{eq:energy_density_hydro_TFIC} (or Eqs.~\eqref{eq:energycurrentoas_hydro}, \eqref{eq:energydensityoas_hydro}), respectively, but it also provides an exact prediction for the SCGF $G(\lambda,v)$ in Eq.~\eqref{eq:SCGF_hydro_euler}, thereby accounting for all higher-order cumulants of the transferred energy $\Delta e(x,t)$.
In particular, the semi-classical picture provides a natural explanation of the structure of Eq.~\eqref{eq:generalized_EFR_hydro}, which we have already recognized as a generalization of the extended fluctuation relation of Eq.~\eqref{eq:EFR} in the hydrodynamic limit.
Indeed in Ref.~\cite{bernard2013time} such a relation has been proved under the assumption of pure transmission, i.e., assuming that the energy of left(right)-moving quasi-particles coming from the far right (left) of the system flows towards its far right (left) part without experiencing reflection. 
Also in the case analysed here quasi-particles do not experience scattering; 
%
%
however, as shown in Fig.~\ref{fig:quasiparticlesFCS}(a), for times $t$ comparable to the space coordinate $x$ not only the right (left) moving particles coming from the left (right) chain contribute to the statistics of the transferred energy $\Delta e(x,t)$, but also the right-moving particles from the space interval $[0,x]$, which result in the additional terms in the second line of Eq.~\eqref{eq:generalized_EFR_hydro}. In the NESS, the contribution from particles generated within the interval $[0,x]$ vanishes and only right (left) moving particles from the left (right) chain matter, recovering the extended fluctuation relation in Eq.~\eqref{eq:EFR}.
\subsection{The quantum Ising chain in a transverse field: SCGF and large deviations}
\label{sec:FCSTFIC}
For the quantum Ising chain in a transverse field, the SCGF  can be calculated explicitly by inserting in the general expression \eqref{eq:finalresultFCS_general} the function $F(\varepsilon)$ specified in the first line of Eq.~\eqref{free_energy_TBA}, $\epsilon_{min}=\varepsilon_{min}$ and $\epsilon_{max}=\varepsilon_{max}$ defined after Eq.~\eqref{eq:maxvelocityTFIC}, with the result
\begin{eqnarray}
G(\lambda,v) = G_{\beta_r}(\lambda) &+& \int_{\varepsilon_{min}}^{\varepsilon_{max}} \frac{d \varepsilon}{2 \pi} \Theta(v_g(\varepsilon)-v) \left(1-\frac{v}{v_g(\varepsilon)}\right) \nonumber \\ 
             && \left\{\mbox{ln}[1+f_{\beta_l}^{+}(\varepsilon)(e^{-\lambda \varepsilon}-1)]\right. \nonumber \\ 
             &&-\left.\mbox{ln}[1+f_{\beta_r}^{+}(\varepsilon)(e^{-\lambda \varepsilon}-1)]\right\}, \nonumber \\ \label{eq:finalresultFCS_Levitov}
\end{eqnarray}
where
\begin{align}
G_{\beta}(\lambda) = \int_{\varepsilon_{min}}^{\varepsilon_{max}} \frac{d \varepsilon}{2 \pi} \left\{\mbox{ln}[1+f_{\beta}^{+}(\varepsilon)(e^{-\lambda \varepsilon}-1)]\right. \nonumber \\ 
             +\left.\mbox{ln}[1+f_{\beta}^{+}(\varepsilon)(e^{\lambda \varepsilon}-1)]\right\}, \label{eq:Levitov_reservoir} 
\end{align}
for $v>0$, while for $v<0$ one gets from Eq.~\eqref{eq:finalresultFCS_general_v_negative} the same result with the replacements $v \rightarrow -v, l \leftrightarrows r$ and $\lambda \rightarrow -\lambda $. The general considerations done in Sec.~\ref{sec:SCGF_general_result} about the dependence of $G(\lambda,v)$ on $v$ 
applies. In particular, for $v>v_{max}$, with $v_{max}$ given by Eq.~\eqref{eq:maxvelocityTFIC}, $G(\lambda,v) = G_{\beta_r}(\lambda)$: after simple algebraic manipulations, it is easy to show that this expression coincides with the Levitov-Lesovik formula for non-interacting fermions in Eq.~\eqref{eq:LevitovLesovik}, with the two parts of the system having equal inverse temperatures set to $\beta_r$. 
Accordingly, $G(\lambda,v>v_{max})$ describes the energy current fluctuations in the right thermal reservoir. 
Upon setting $v=0$ in Eq.~\eqref{eq:Levitov_reservoir}, instead, we get the NESS limit of the SCGF, which for non-interacting fermions models is again provided by the Levitov-Lesovik formula in Eq.~\eqref{eq:LevitovLesovik} with unitary transmission coefficient, as already noted after Eq.~\eqref{eq:NESS_FCS_general}. 
From the latter equation, performing explicitly the integral over the energy spectrum one finds 
\be
G(\lambda,v=0) = g_{\beta_l}^+(\lambda)-g_{\beta_l}^+(0) + g_{\beta_r}^+(-\lambda)-g_{\beta_r}^+(0), \label{eq:vitiresult}
\ee
where
\be
g_{\beta}^+(\lambda)= \frac{\mbox{Li}_2(-e^{-(\beta+\lambda)\varepsilon_{max}})-\mbox{Li}_2(-e^{-(\beta+\lambda)\varepsilon_{min}})}{2 \pi (\beta +\lambda)},
\ee
which agrees with the expression found (under the assumption $h>1$) in Ref.~\cite{de2013nonequilibrium}, see Eqs.~(33) and (34) therein, 
for the stationary limit of the SCGF of the transferred energy following an inhomogeneous quench of two Ising chains according to the very same protocol considered in this work and calculated by evaluating  Eq.~\eqref{eq:SCGF_stationary_state}.

\begin{figure}[h!] 
\centering
\includegraphics[width=1\columnwidth]{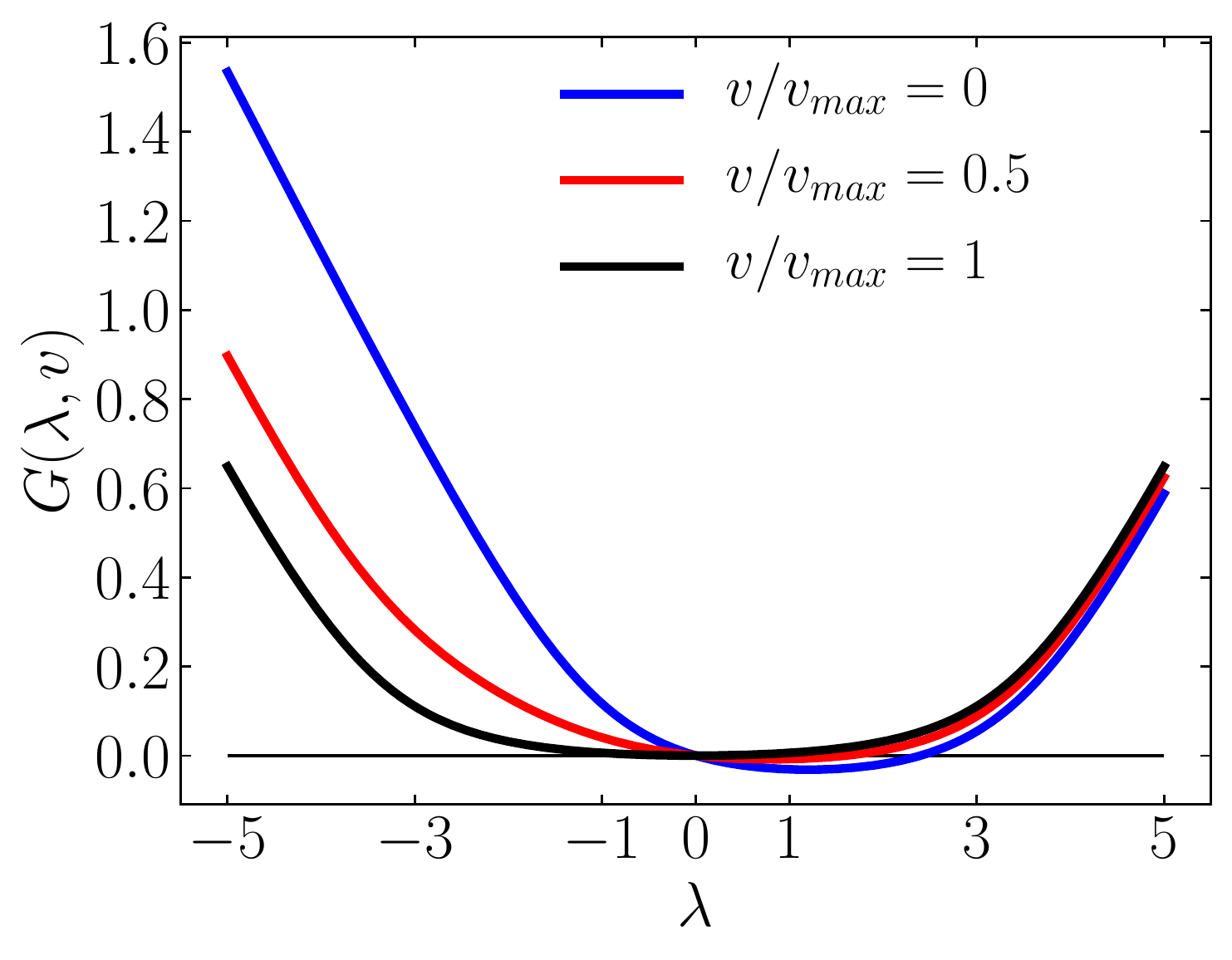}\\[-1mm]	
	\, \, \, (a)\\[2mm]
\includegraphics[width=1\columnwidth]{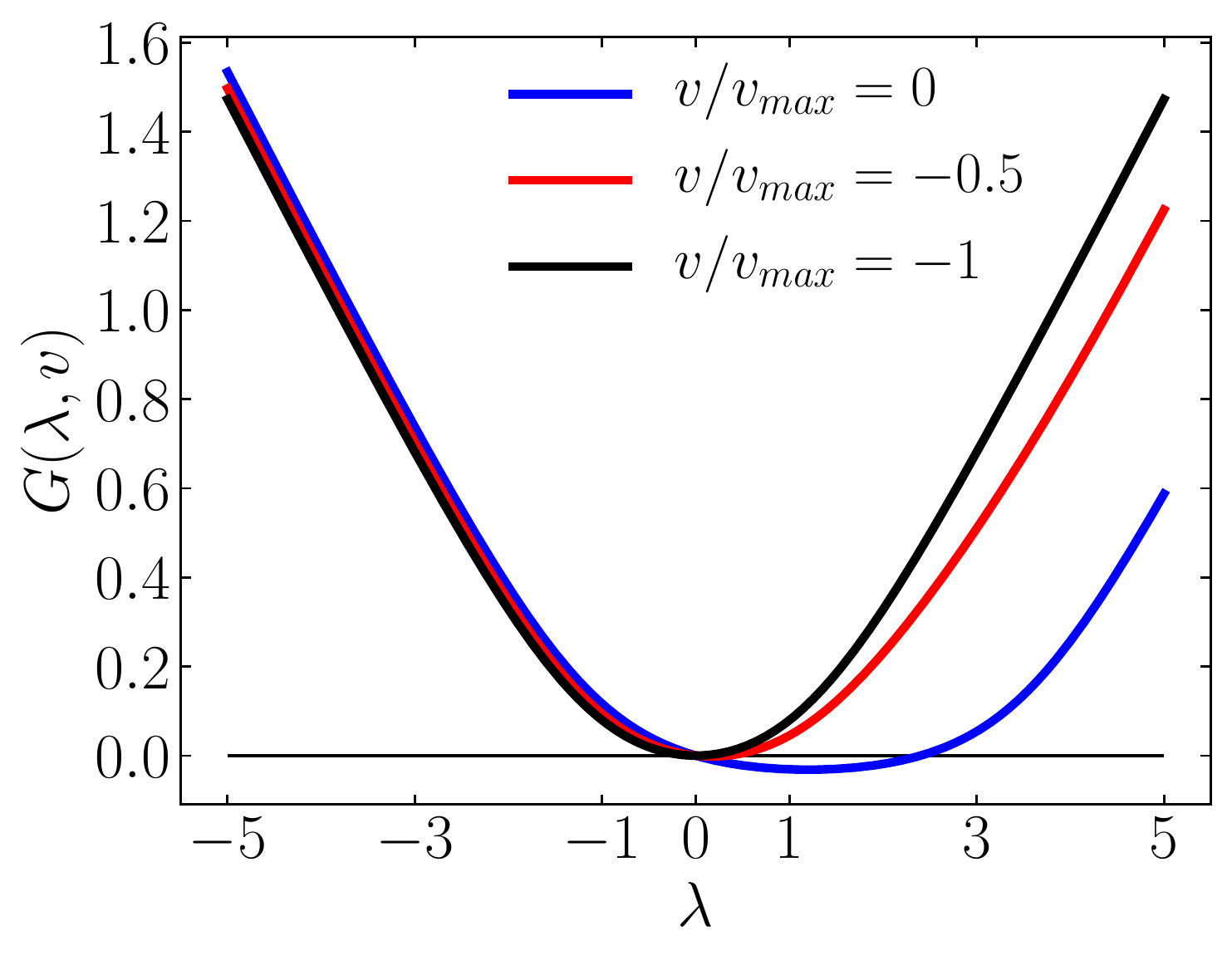}\\[-1mm]
\, \, \, (b)
\caption{Scaled cumulant generating function $G(\lambda,v)$ as a function of $\lambda$ and fixed $v$ for the quantum Ising chain with $\beta_l=1.1$, $\beta_r=3.5$, $h=1.3$, $J=1$. In particular, panel (a) corresponds to positive values of $v/v_{max}=0, 0.5, 1$ (from top to bottom) while panel (b) to negative values $v/v_{max}=0, -0.5, -1$ (from bottom to top). $v_{max}$ is given in Eq.~\eqref{eq:maxvelocityTFIC}.} 
\label{fig:FCSVdependent}
\end{figure}

\begin{figure}[b] 
\centering
\includegraphics[width=1\columnwidth]{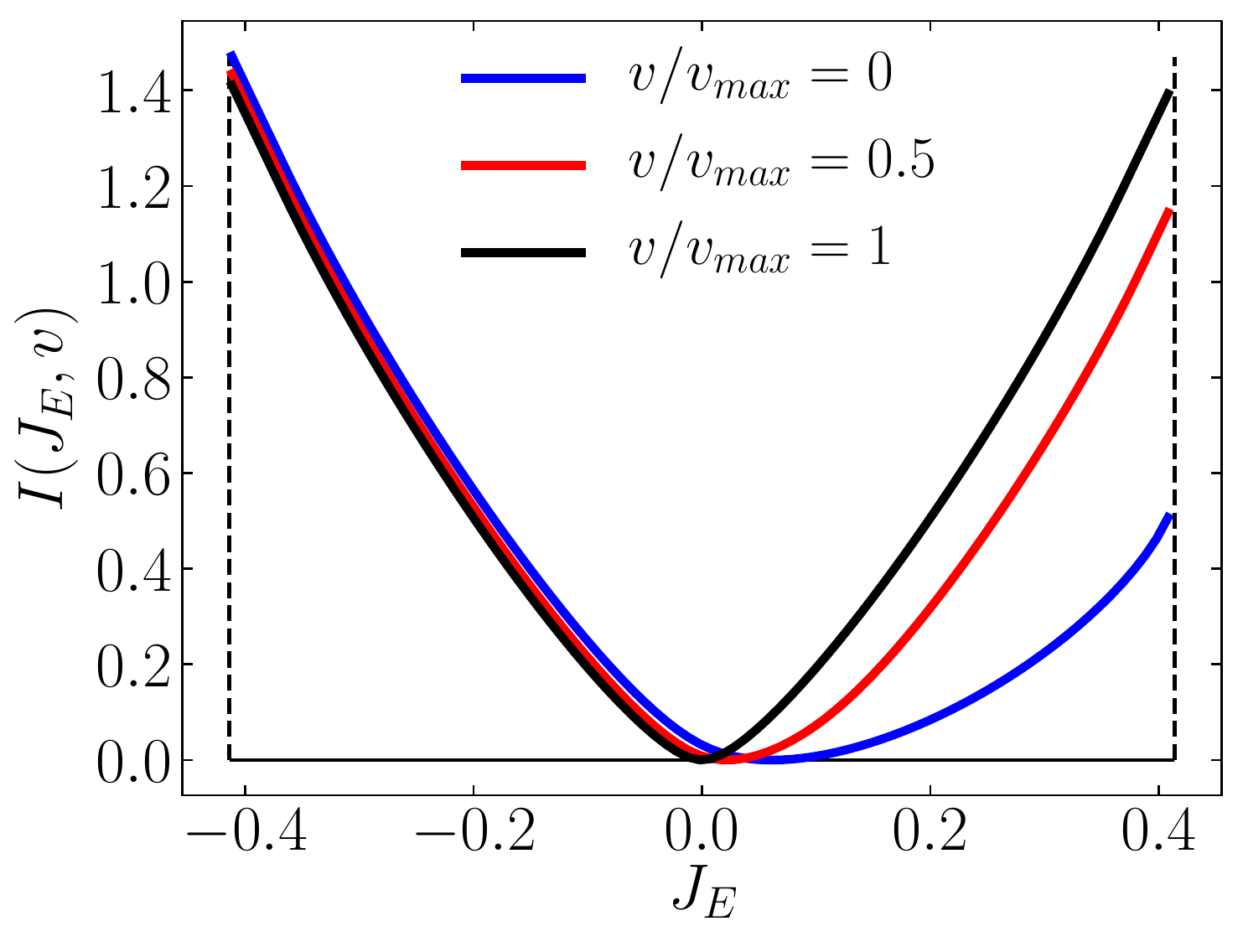}\\[-1mm]	
	\, \, \, (a)\\[2mm]
\includegraphics[width=1\columnwidth]{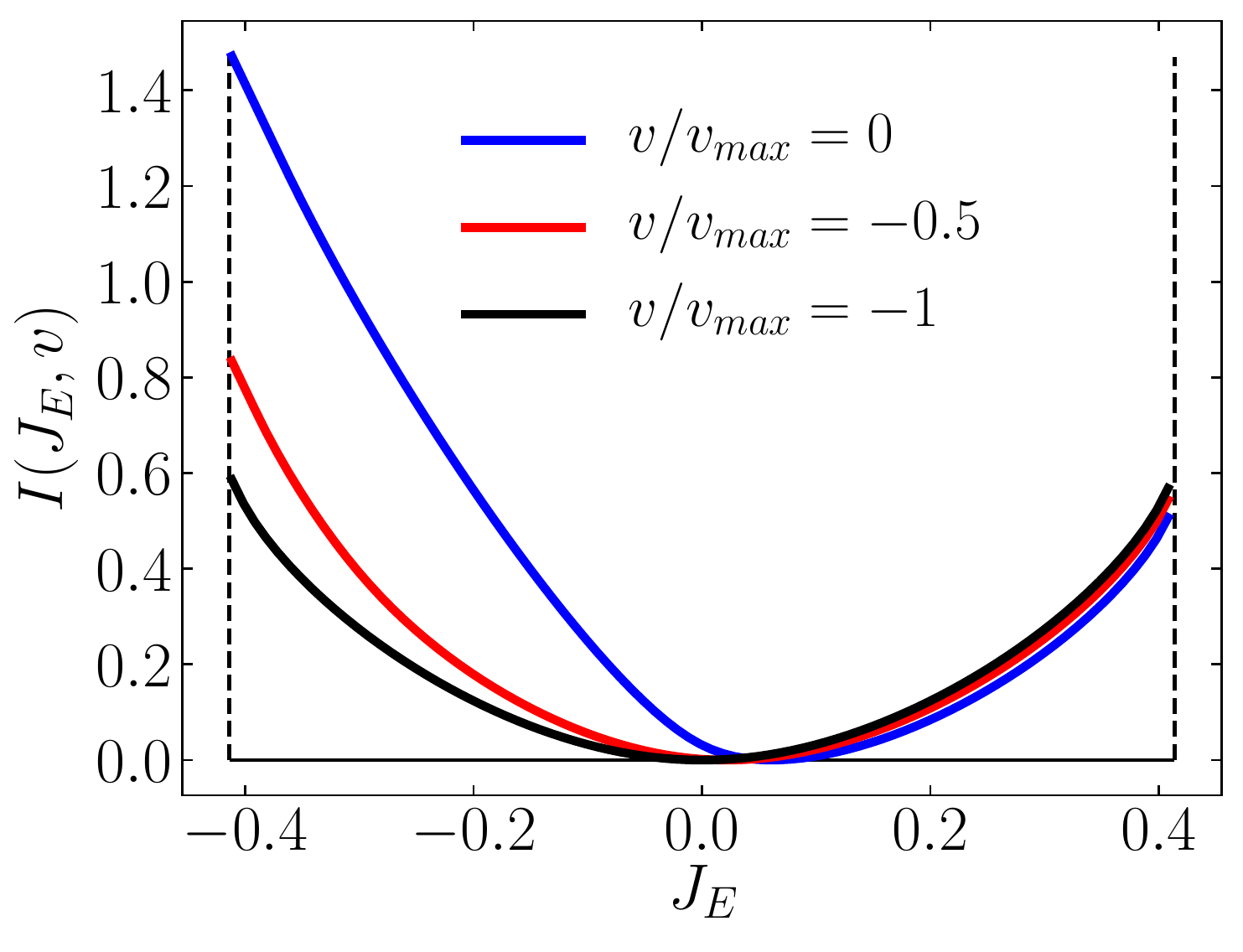}\\[-1mm]
\, \, \, (b)
\caption{Large deviation function $I(J_E,v)$ as a function of $J_E$ and fixed $v$ for the quantum Ising chain with the same values of parameters as in Fig.~\ref{fig:FCSVdependent}. In particular, panel (a) corresponds to positive values of  $v/v_{max}=0, 0.5, 1$ (from bottom to top) while panel (b) to negative values $v/v_{max}=0, -0.5, -1$ (from top to bottom). $v_{max}$ is given in Eq.~\eqref{eq:maxvelocityTFIC}. 
The vertical dashed lines correspond to the maximal and minimal values of the current $\pm \mathcal{J}^E_{max}$ in Eq.~\eqref{eq:extremes}: the rate function is finite only within the interval $(-\mathcal{J}^E_{max},\mathcal{J}^E_{max})$ while it is infinite outside it.} 
\label{fig:ratefunctionVdependent}
\end{figure}

The plot of $G(\lambda,v)$ in Eq.~\eqref{eq:finalresultFCS_Levitov} as a function of $\lambda$ for various fixed values of $v$ is reported in Fig.~\ref{fig:FCSVdependent}(a) for $v>0$ and in Fig.~\ref{fig:FCSVdependent}(b) for $v<0$. The corresponding large-deviation function $I(J_E,v)$, obtained by taking the Legendre-Fenchel transform of $G(\lambda,v)$, is reported in Fig.~\ref{fig:ratefunctionVdependent} for the same values of parameters as in Fig.~\ref{fig:FCSVdependent}. A different choice of the parameters $\beta_{l,r}$ does not alter the qualitative features of the plot, but it changes the zero of $I$, i.e., $I(\langle J_E \rangle,v)=0$, where $\langle J_E \rangle = \Delta \mathcal{E}(x,t)/t$ is the mean and typical value. In particular, for $\beta_r>\beta_l$, $\langle J_E \rangle$ is positive as the typical flow of energy is from the left (hotter) to the right (colder) chain, according to the initial temperature gradient. In the opposite case, $\beta_r<\beta_l$, one has $\langle J_E \rangle<0$ and the zero of $I$ is consequently negative. 
%
%
As far as the dependence on $v$ of $G(\lambda,v)$ and $I(J_E,v)$ is concerned, we note that the statistics of the rare fluctuations with $J_E$ smaller (larger) than $\langle J_E \rangle$ does not depend significantly on the value of $v>0$ ($v<0$). 
This fact can be understood in terms of the quasi-particles picture sketched in Fig.~\ref{fig:quasiparticlesFCS}: in the case $v>0$, for instance, in order to have a current smaller than the mean one, e.g., a negative value with the current flowing against the temperature gradient, one needs a fluctuation in the number of left-moving particles coming from the right chain, in particular those initially generated within the space interval $[x,y_0(k)]$, with $y_0(k)$ defined after Eq.~\eqref{eq:fastrandomvariable}; given that these excitations are entirely produced in the right part of the chain, at inverse temperature $\beta_r$, the corresponding fluctuations are practically time-independent because the properties of the reservoir have not been affected by the dynamics.
Concerning the dependence on $\lambda$, instead, the SCGF is defined over the whole real axis for all values of $v$ and it is asymptotically linear as $\lambda \rightarrow \pm \infty$, with slopes $\mathcal{J}^E_{max}$ and $\mathcal{J}^E_{min} = -\mathcal{J}^E_{max}$, respectively, which are independent of $v$:
\begin{equation}
\mathcal{J}^E_{max} = -\frac{\partial G(\lambda,v)}{\partial \lambda}\Bigr|_{\lambda \rightarrow -\infty} = \frac{J^2 h}{\pi}. 
\label{eq:extremes}
\end{equation}  
Accordingly, by using the Legendre duality relations in Eqs.~\eqref{eq:legendreduality}, an asymptotic linear behavior of $G(\lambda \rightarrow \pm \infty,v)$ such as that displayed by $G(\lambda,v)$ in Fig.~\ref{fig:FCSVdependent} implies that $I(J_E,v)$ diverges for values of $J_E$ outside the interval delimited by the slopes of $G(\lambda \rightarrow -\infty,v)$ and $G(\lambda \rightarrow \infty,v)$  and, correspondingly, the probability vanishes. 
This means that the values $\mathcal{J}^{E}_{min}$ and $\mathcal{J}^{E}_{max}$ identified above actually coincide with the minimal and maximal possible values, respectively, of $J_E=\Delta e(x,t)/t$. 
%
%
%
%
Accordingly, the rate function $I(J_E,v)$ is finite only within the interval $J_E \in [\mathcal{J}^E_{min},\mathcal{J}^E_{max}]$, with $\mathcal{J}^E_{min}$ and $\mathcal{J}^E_{max}$ given in Eq.~\eqref{eq:extremes}, while it diverges outside this interval, meaning that the corresponding values of the transferred energy cannot be observed in the system. In fact, because the transport is determined by fermionic quasi-particles, the exclusion principle requires that each mode $k$ has at most an occupation 1 and therefore the modulus of the energy current $|\mathcal{J}^E(v)|$ in Eq.~\eqref{eq:energy_current_hydro_TFIC} can never exceed the value $\mathcal{J}^E_{max}$ obtained by setting all these occupation numbers to 1. 
This can be seen quantitatively by starting from the expression of $\mathcal{J}^E(v)$ in Eq.~\eqref{eq:energy_current_hydro_TFIC}. Remembering that $\mathcal{J}^E(v=0)$ is the value of the energy current in the NESS, and that $\mathcal{J}^E(v) < \mathcal{J}^E(0)$ since as time increases the current along the chain approaches the steady state value from below since transport is ballistic, one has 
\be
|\mathcal{J}^E(v)| < |\mathcal{J}^E(0)| < \int_{0}^{\pi}\frac{dk}{2 \pi} \varepsilon(k) v_g(k)|f_{\beta_l}^{+}(k)-f_{\beta_r}^+(k)|.
\ee
By observing that, due to the fermionic statistics,
\be
|f_{\beta_l}^+(k)-f_{\beta_r}^+(k)|<1, \label{eq:fermionbound}
\ee
it follows that 
\begin{eqnarray}
|\mathcal{J}^E(v)| &<& \int_{0}^{\pi}\frac{dk}{2 \pi} \varepsilon(k) v_g(k) = \int_{\varepsilon_{min}}^{\varepsilon_{max}}\frac{d\varepsilon}{2 \pi} \varepsilon \nonumber \\
                           &=& \frac{J^2 h}{\pi},
\end{eqnarray}
which is indeed the value in Eq.~\eqref{eq:extremes} of the asymptotic slope of $G(\lambda,v)$ for $\lambda \rightarrow \infty$. 

\subsection{The harmonic chain: SCGF and large deviations}
\label{sec:SCGF_bosons}

In the harmonic chain, $F(\varepsilon)$ in Eq.~\eqref{eq:finalresultFCS_general} is given by the second line of Eq.~\eqref{free_energy_TBA}, while $\epsilon_{min}=\Omega_{min}$ and $\epsilon_{max}=\Omega_{max}$ are defined after Eq.~\eqref{eq:maxvelocityOAS}.
Accordingly, Eq.~\eqref{eq:finalresultFCS_general} becomes
\begin{align}
G(\lambda,v) = - &\int_{\Omega_{min}}^{\Omega_{max}} \frac{d \varepsilon}{2 \pi} \Theta(v_g(\varepsilon)-v) \left(1-\frac{v}{v_g(\varepsilon)}\right) \nonumber \\ 
             & \left\{\mbox{ln}[1+f_{\beta_l}^-(\varepsilon)(1-e^{-\lambda \varepsilon})]\right. \nonumber \\ 
             &-\left.\mbox{ln}[1+f_{\beta_r}^-(\varepsilon)(1-e^{-\lambda \varepsilon})]\right\} + G_{\beta_r}(\lambda),  \label{eq:finalresultFCSbosons}
\end{align}
where
\begin{align}
G_{\beta}(\lambda) = -\int_{\Omega_{min}}^{\Omega_{max}} \frac{d \varepsilon}{2 \pi} \left\{\mbox{ln}[1+f_{\beta}^-(\varepsilon)(1-e^{-\lambda \varepsilon})]\right. \nonumber \\ 
             +\left.\mbox{ln}[1+f_{\beta}^-(\varepsilon)(1-e^{\lambda \varepsilon})]\right\}. \label{eq:SCGFreservoirbosons} 
\end{align}
For $v<0$ a similar expression can be written starting from Eq.~\eqref{eq:finalresultFCS_general_v_negative} with the replacements $v \rightarrow -v$, $\lambda \rightarrow -\lambda$ and $l \leftrightarrows r$. Similarly to the case of the Ising model, the term $G_{\beta}(\lambda)$ encodes thermal fluctuations of the right reservoir. 
Upon setting $v=0$ in Eq.~\eqref{eq:finalresultFCSbosons} the NESS scaled cumulant generating function can obtained by directly performing the integration over the energy spectrum $\Omega$, which renders  
\be
G(\lambda, v=0) = g_{\beta_l}^-(\lambda) - g_{\beta_l}^-(0) + g_{\beta_r}^-(-\lambda) - g_{\beta_l}^-(0) \label{eq:NESSbosons}
\ee
where
\be
g_{\beta}^-(\lambda) = \frac{\mbox{Li}_2(e^{-\Omega_{min}(\beta + \lambda)}) - \mbox{Li}_2(e^{-\Omega_{max}(\beta + \lambda)})}{2 \pi (\beta + \lambda)}. 
\ee
The result in Eq.~\eqref{eq:NESSbosons} is consistent with the one obtained in Ref.~\cite{saito2007fluctuation} for the SCGF of a chain of harmonic oscillators coupled to two external heat baths at temperatures $T_l$ and $T_r$. In addition, it shows that the SCGF and the cumulants of the transferred energy $\Delta e(x,t)$ are independent of the protocol chosen to get the non-equilibrium steady state, as it happens for the mean value of the energy current reported after Eqs.~\eqref{eq:integrated_oas_current_1}, \eqref{eq:integrated_oas_current_2}, and \eqref{eq:integrated_oas_current_3}.

\begin{figure}[h!] 
\centering
\includegraphics[width=1\columnwidth]{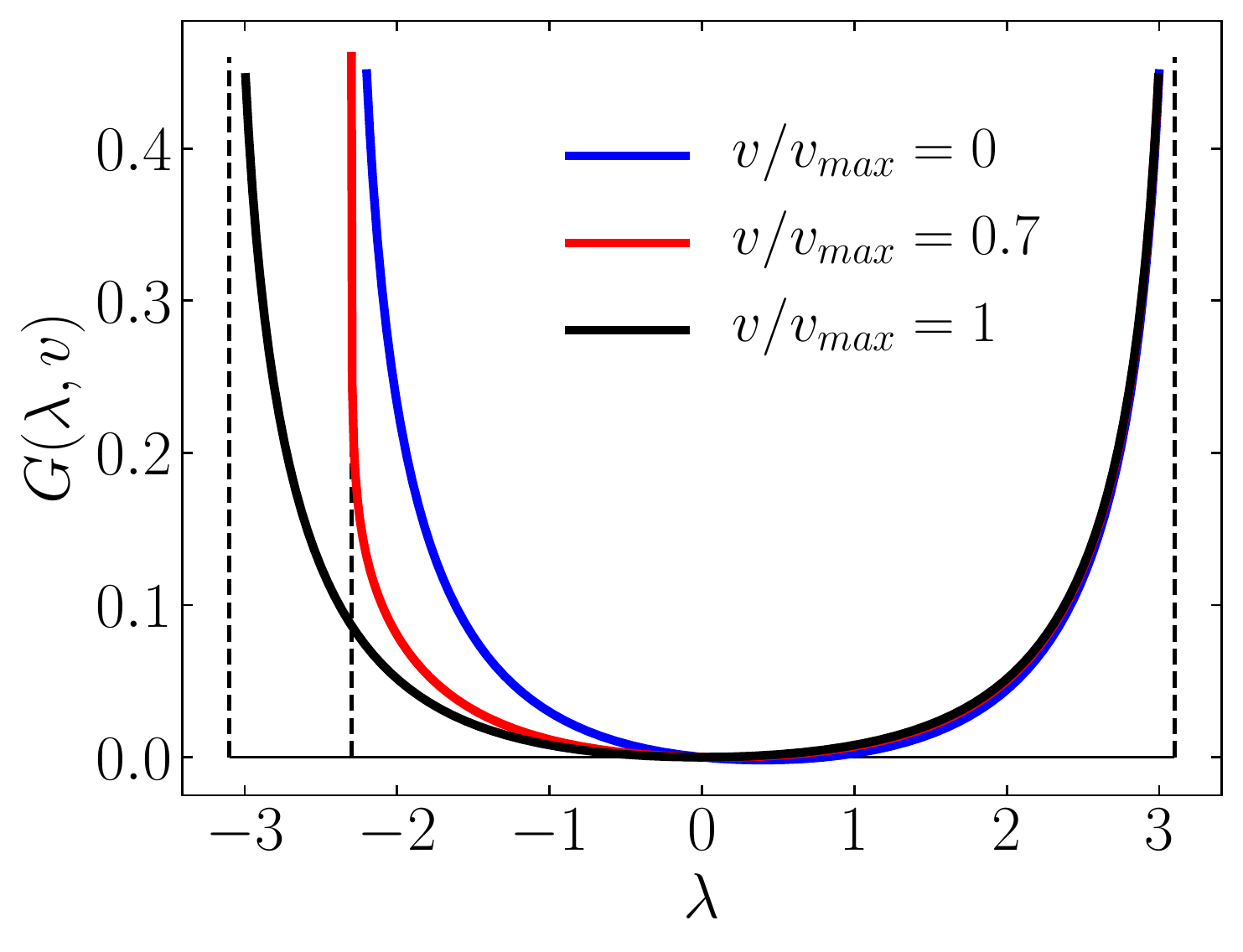}\\[-1mm]	
	\, \, \, (a)\\[2mm]
\includegraphics[width=1\columnwidth]{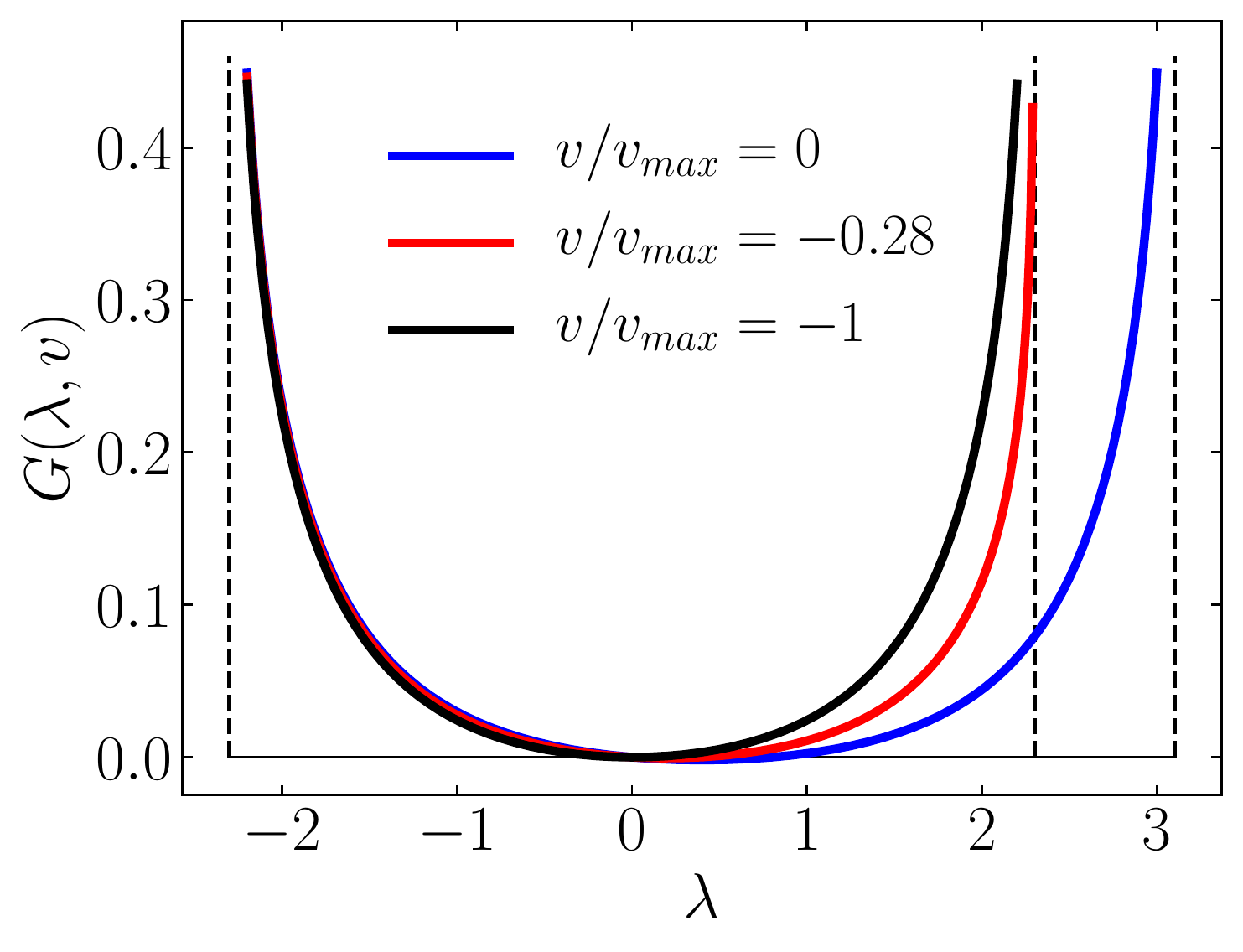}\\[-1mm]
\,\, \, (b)
\caption{Scaled cumulant generating function $G(\lambda,v)$ as a function of $\lambda$ and fixed $v$ for the  harmonic chain with $\beta_l=2.3$, $\beta_r=3.1$, $m=0.7$, $\omega=1$, $v_{max} \simeq 0.71$ according to Eq.~\eqref{eq:maxvelocityOAS}. In particular, panel (a) corresponds to positive values of $v/v_{max}=0, 0.7, 1$ (from top to bottom) while panel (b) to negative values $v/v_{max}=0, -0.28, -1$ (from bottom to top). The vertical dashed lines correspond to the boundaries of the domain of $G(\lambda,v)$ according to Eqs.~\eqref{eq:domainbosons} and \eqref{eq:domainbosonsextremes}.
} 
\label{fig:FCSVdependentbosons}
\end{figure}

\begin{figure}[h!] 
\centering
\includegraphics[width=1\columnwidth]{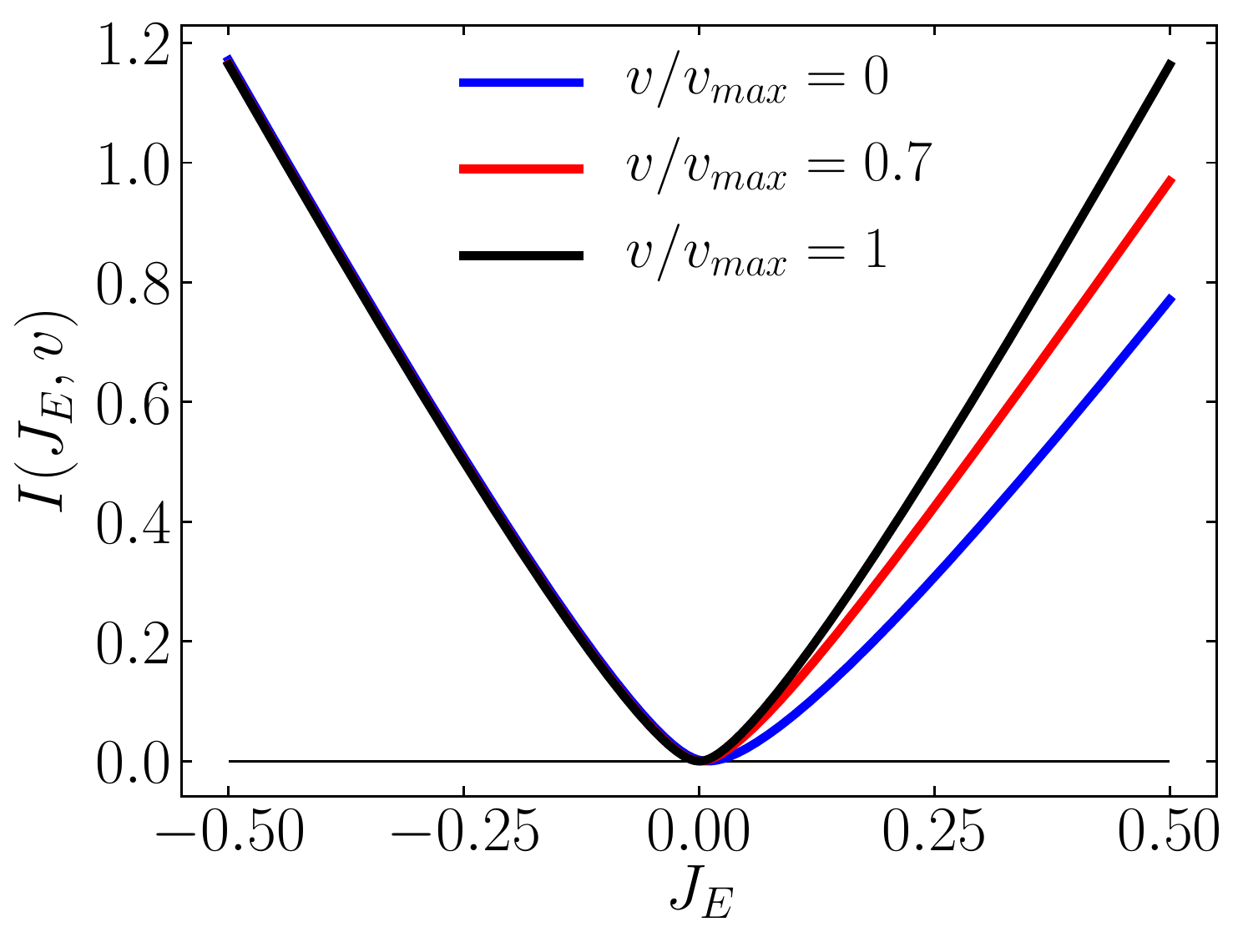}\\[-1mm]	
	\, \, \, (a)\\[2mm]
\includegraphics[width=1\columnwidth]{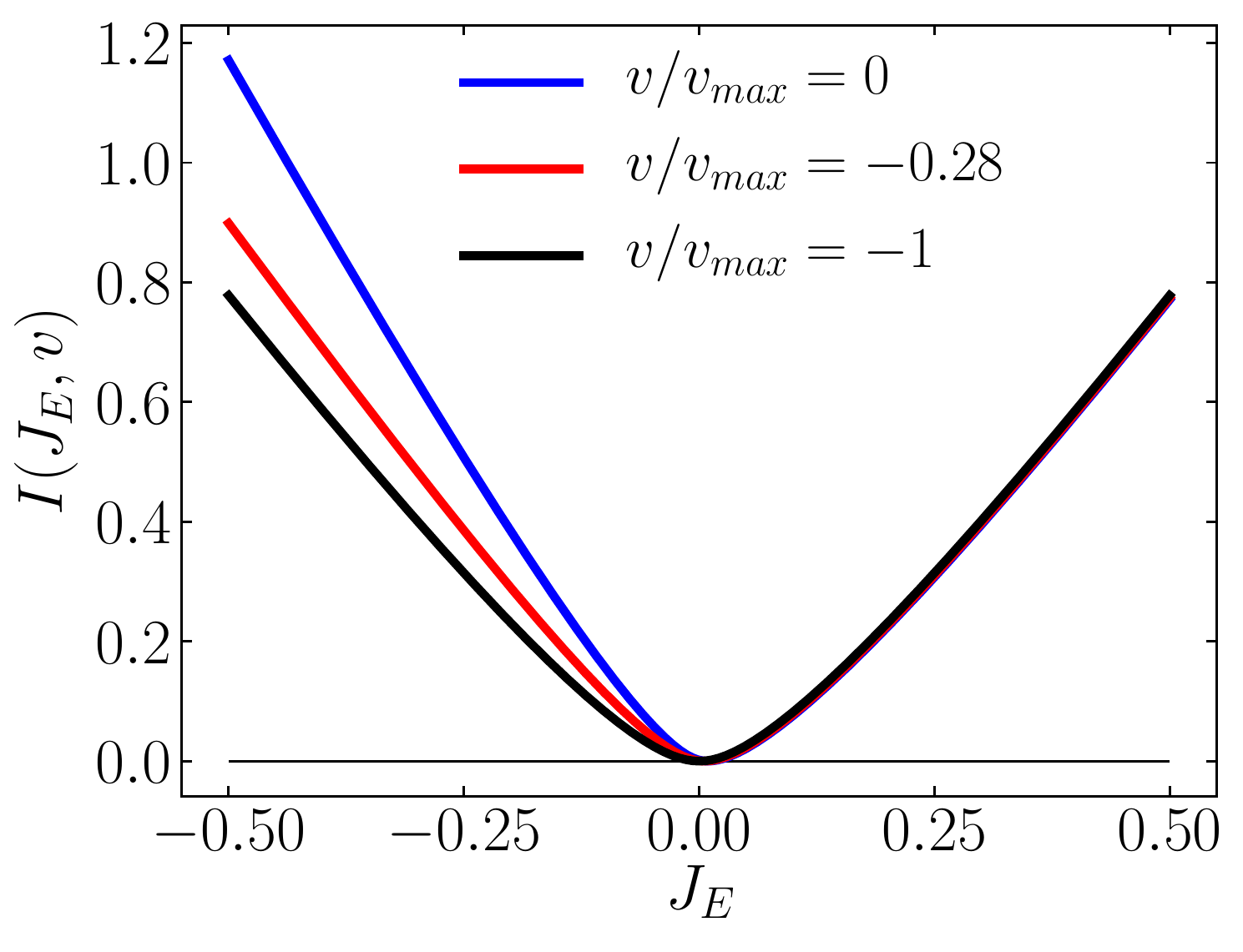}\\[-1mm]
\, \, \, (b)
\caption{Large deviation function $I(J_E,v)$ as a function of $J_E$ and fixed $v$ for the quantum harmonic chain with the same values of parameters as in Fig.~\ref{fig:FCSVdependentbosons}. In particular, panel (a) corresponds to positive values of $v/v_{max}=0, 0.7, 1$ (from bottom to top) while panel (b) to negative values $v/v_{max}=0, -0.28, -1$ (from top to bottom).
}
\label{fig:ldevdependentbosons}
\end{figure}

The plot of $G(\lambda,v)$ in Eq.~\eqref{eq:finalresultFCSbosons} as a function of $\lambda$ for various fixed values of $v$ is reported in Fig.~\ref{fig:FCSVdependentbosons}(a) for $v>0$ and in Fig.~\ref{fig:FCSVdependentbosons}(b) for $v<0$. The corresponding large-deviation function $I(J_E,v)$, obtained by taking the Legendre-Fenchel transform of $G(\lambda,v)$, is reported in the two panels of Fig.~\ref{fig:ldevdependentbosons} for the same values of parameters as in Fig.~\ref{fig:FCSVdependentbosons}. As in the case of the Ising chain, the qualitative features of the plot are unaltered upon changing the parameters of model, the only difference being in the sign of the mean transferred energy $\langle J_E \rangle$ which is positive for $\beta_r>\beta_l$, as it is the case for Fig.~\ref{fig:ldevdependentbosons}, and negative otherwise.

%
%
%
%
%
%
The most important difference with respect to the SCGF of the fermionic case reported in Fig.~\ref{fig:FCSVdependent} is that $G(\lambda,v)$ as a function of $\lambda$ is defined on a finite interval, the extremes of which depend on the value of the variable $v$, i.e., it is finite for
\begin{align}
&\lambda \in [-\mbox{min}(\beta_l,\beta_r),\beta_r] \quad \mbox{for} \quad  0<v<v_{max}, \nonumber \\
&\lambda \in [-\beta_l,\mbox{min}(\beta_l,\beta_r)] \quad \mbox{for} \quad  -v_{max}<v<0, \label{eq:domainbosons}
\end{align}
while it is otherwise infinite. 
In the non-equilibrium stationary state, corresponding to setting $v=0$, and in the cases $v>v_{max}$ and $v<-v_{max}$, with $v_{max}$ given in Eq.~\eqref{eq:maxvelocityOAS}, the domain of the SCGF is, instead,
\begin{align}
&\lambda \in [-\beta_r,\beta_r] \quad \mbox{for} \quad  v\geq v_{max}, \nonumber \\
&\lambda \in [-\beta_l,\beta_l] \quad \mbox{for} \quad  v\leq -v_{max}, \nonumber \\ 
&\lambda \in [-\beta_l,\beta_r] \quad \mbox{for} \quad  v=0.     \label{eq:domainbosonsextremes}
\end{align} 
In particular, the dependence of the domain of $G(\lambda,v)$ on $v$, as we can see from Fig.~\ref{fig:FCSVdependentbosons}, turns out to be discontinuous; namely in the case $v>0$ of Fig.~\ref{fig:FCSVdependentbosons}(a) the domain is $\lambda\in [-\beta_l,\beta_r]$ for $0<v<v_{max}$ since $\beta_r > \beta_l$, while, in the case $v>v_{max}$ it becomes the one of the the SCGF of the right reservoir in Eq.~\eqref{eq:SCGFreservoirbosons}, i.e., $\lambda\in [-\beta_r,\beta_r]$. 
Similarly, for $v<0$, the domain is $\lambda\in [-\beta_l,\beta_l]$ for $-v_{max}<v<0$ while it changes to the domain of the NESS scaled cumulant generating function $\lambda\in [-\beta_l,\beta_r]$ for $v=0$. 
In terms of the large deviation function $I(J_E,v)$ and due to the Legendre duality expressed in Eq.~\eqref{eq:legendreduality}, the presence of these domains translates into asymptotically linear behaviors for large $|J_E|$, with the slopes determined by the boundaries of the domain of the SCGF, given in Eqs.~\eqref{eq:domainbosons} and \eqref{eq:domainbosonsextremes}. 
This is shown in Fig.~\ref{fig:ldevdependentbosons}(a) for $v>0$ and in Fig.~\ref{fig:ldevdependentbosons}(b) for $v<0$. 

The peculiar behavior of the domain of the SCGF in Eq.~\eqref{eq:domainbosons} can be again understood in terms of the quasi-particles picture sketched in Fig.~\ref{fig:quasiparticlesFCS}. Consider, for example, the  case $v>0$: since the transferred energy operator $\Delta e(x,t)$ in Eq.~\eqref{eq:transferredenergyoperator} is a time-integrated quantity
one has to consider the flux of quasi-particles arriving in $x$ within the time interval $(0,t)$, as already noted after Eq.~\eqref{eq:semiclassicalSCGF}. 
Left movers contributing to the expression in Eq.~\eqref{eq:finalresultFCSbosons} are generated initially only within the interval $(x,y_0(k)]$ along the chain, with inverse temperature $\beta_r$. The contribution of right-moving excitations, instead, comes from those initially generated within the intervals $(x_0(k),0]$ and $(0,x]$ along the chain, with different inverse temperatures $\beta_l$ and $\beta_r$, respectively, where $x_0(k)$ and $y_0(k)$ are defined after Eq.~\eqref{eq:fastrandomvariable}. 
In each of these intervals there is a finite probability of generating an arbitrarily large number of bosons 
in the initial state for each mode $k$, according to Eq.~\eqref{eq:semiclassicprob_bosons}. Therefore each interval can behave as an effective reservoir at the corresponding temperature, in the sense that it is able to inject an arbitrarily large number of quasi-particles in the system. 
The domain of the SCGF is then determined by the reservoir of left moving excitations, at inverse temperature $\beta_r$, and by the reservoir of those moving rightwards, with the temperature of the latter being determined by the largest between the temperatures at which the particles in the intervals $(x_0(k),0]$ and those in $(x,y_0(k)]$ are initially generated, in accordance with Eq.~\eqref{eq:domainbosons}. 
This is the physical interpretation of the origin of the behavior displayed in Fig.~\ref{fig:ldevdependentbosons}(a): a similar argument can be repeated for $v<0$ in order to explain the features of Fig.~\ref{fig:ldevdependentbosons}(b). 
In particular, due to the fact $\Delta e(x,t)$ is a time-integrated observable, 
one can conclude that $G(\lambda,v)$ in Eq.~\eqref{eq:SCGF_hydro_euler} as a function of $v$ can be discontinuous in $v=0$ or at the edges $v = \pm v_{max}$ whenever the Hilbert space for each mode $k$ of the excitations is infinite, as it is the case for bosons. If, on the contrary, for every wave vector $k$ the Hilbert space has a finite dimension, as in the fermionic case in Sec.~\ref{sec:FCSTFIC}, these discontinuities are absent. 
%
%
%
%
%

The bosonic large deviation function $I(J_E,v)$ is therefore defined as a function of $J_E$ over the whole real axis for all values of $v$, and fluctuations of the transferred energy $J_E=\Delta e(x,t)/t$ can in principle be arbitrarily large; the physical reason is clear since in this case each mode $k$ is not restricted to be populated by one or zero particles, as in the fermionic case, and as a consequence no bound as in Eq.~\eqref{eq:fermionbound} can be determined.
In particular, the asymptotic linear behavior of the rate function shown in Fig.~\ref{fig:ldevdependentbosons}, according to Eq.~\eqref{eq:large_deviation_principle}, causes the tails of the probability density $p(J_E,v)$ of the transferred energy to be exponentially distributed according to
\begin{equation} 
p(J_E,v) \sim \mbox{exp}[-t \, \beta (v) |J_E|], \label{eq:exponential_distribution_bosons}
\end{equation}
with $\beta(v)$ depending on $v$ consistently with Eqs.~\eqref{eq:domainbosons} and \eqref{eq:domainbosonsextremes}. In the steady state, corresponding to setting $v=0$, one has
\begin{equation}
\beta(v=0) = \beta_l \Theta(J_E) + \beta_r \Theta(-J_E), \label{eq:exponential_tails_bosons}
\end{equation}
in agreement with Ref.~\cite{saito2007fluctuation}, where identical exponential tails have been observed for the probability distribution of the energy current flowing in an harmonic chain connected to two thermal reservoirs at inverse temperatures $\beta_l$ and $\beta_r$.

\section{Conclusions}
\label{finalcomments}

In the present manuscript we considered the energy transport after an inhomogeneous quench of two identical semi-infinite systems initially at thermal equilibrium at different temperatures $\beta_r$ and $\beta_l$ as in Eq.~\eqref{eq:intro-rho0}. 
In particular, 
we focused on the exactly solvable cases of the quantum Ising chain in a transverse field and of the harmonic chain, introduced in Sec.~\ref{firstsection}, which are characterized, respectively, by the fermionic and bosonic excitations in Eqs.~\eqref{eq:post_quench_modes_TFIC} and \eqref{eq:postquenchdiagonal}.
In Sec.~\ref{secondsection} we have discussed the calculation of the energy current $\mathcal{J}^E(v)$ and density $\mathcal{U}(v)$ in the space-time scaling/hydrodynamic limit $x,t \rightarrow \infty$ with fixed $v=x/t$. 
By explicitly evolving the density matrix $\rho(v)$ at the hydrodynamic scale,  in Sec.~\ref{sec:isingtransportsub} 
(see Eqs.~\eqref{eq:hydro_state}, \eqref{eq:hydro_state_proved}, and \eqref{eq:hydro_state_proved_bosons}) we studied the Ising chain, recovering in Eqs.~\eqref{eq:energy_current_hydro_TFIC} and \eqref{eq:energy_density_hydro_TFIC} the results of Refs.~\cite{perfetto2017ballistic,kormos2017inhomogeneous}; in Sec.~\ref{sec:oastransportsub}, instead, we considered the harmonic chain, deriving the analogous statistical properties in Eqs.~\eqref{eq:energycurrentoas_hydro} and \eqref{eq:energydensityoas_hydro}. 
For the latter case, the edge behavior of $\mathcal{J}^E(v)$ as $v \rightarrow \pm v_{max}^{\mp}$ has been analyzed and the Airy kernel has been found to describe the leading sub-diffusive correction to the ballistic profile also for free bosonic systems, see Eqs.~\eqref{eq:scaling_variable_Airy}, \eqref{eq:Airy_kernel_bosons_1}, \eqref{eq_Airy_kernel_bosons_2}, \eqref{eq:Airy_kernel_bosons_3}, and \eqref{eq:Airy_critical}. The expressions of the energy current $\mathcal{J}^E(v)$ an  density $\mathcal{U}(v)$ in the hydrodynamic limit turn out to have a rather universal structure, in the sense that the only appearing model-specific information is the single-particle energy spectrum and the Fermi-Dirac ($f^{+}_{\beta}(k)$) or Bose-Einstein ($f^{-}_{\beta}(k)$) statistics of the quasi particles.

The primary results of this work are presented in Sec.~\ref{thirdsection}. In particular, in Sec.~\ref{sec:SCGF_general_result} the scaled cumulant generating function (SCGF) $G(\lambda,v)$ in the hydrodynamic limit, see Eq.~\eqref{eq:SCGF_hydro_euler}, has been determined, with the result reported in  Eqs.~\eqref{eq:finalresultFCS_general} and \eqref{eq:finalresultFCS_general_v_negative}. The calculation is based on an exponential tilting $\rho(v,\lambda)$ of the density matrix according to the exponential of the time integrated current, as shown by Eqs.~\eqref{eq:tilted_average}, \eqref{eq:non_local_tilting_1}, and \eqref{eq:non_local_tilting_2}. 
Equations~\eqref{eq:finalresultFCS_general} and \eqref{eq:finalresultFCS_general_v_negative} express the SCGF as a function of $v$, therefore extending known results for the Ising chain (see Ref.~\cite{de2013nonequilibrium}) and for the harmonic oscillators (see Ref.~\cite{saito2007fluctuation}) concerning the stationary limit of this function, corresponding to $v=0$.  
In particular, in Eq.~\eqref{eq:generalized_EFR_hydro} we have derived a generalization of the so-called extended fluctuation relation \cite{bernard2013time}, which allows us to calculate the SCGF $G(\lambda,v)$ via an integration over $\lambda$ of the energy current $\mathcal{J}^E(v)$, with suitably modified inverse temperatures $\beta(\lambda)$. In Sec.~\ref{sec:semi_classics} we have provided a simple semi-classical derivation of Eqs.~\eqref{eq:finalresultFCS_general} and \eqref{eq:finalresultFCS_general_v_negative} in terms of  quasi-particles which ballistically propagate along the chain. 
Via the Legendre-Fenchel transform in Eq.~\eqref{eq:legendrefenchel}, the large deviation function $I(J_E,v)$ --- which expresses the asymptotic scaling, in the hydrodynamic limit, of the probability density function $p(J_E,v)$ --- of the transferred energy $\Delta e(x,t)/t$ (see Eq.~\eqref{eq:large_deviation_principle}) has been obtained. 
As far as the large deviation function is concerned, bosons and fermions behave rather differently. 
For fermions, see Sec.~\ref{sec:FCSTFIC} for the Ising chain, $I(J_E,v)$ as a function of $J_E$ is finite only on the closed interval $[-\mathcal{J}^E_{max},\mathcal{J}^E_{max}]$ (see Eq.~\eqref{eq:extremes}), while it is infinite outside it, meaning that $\Delta e(x,t)/t$ cannot exceed the maximum value $\mathcal{J}^E_{max}$. For bosons, see  Sec.~\ref{sec:SCGF_bosons}, instead, $I(J_E,v)$ as a function of $J_E$ is defined on the whole real axis and it shows linear tails (see Eqs.~\eqref{eq:exponential_distribution_bosons} and \eqref{eq:exponential_tails_bosons}), which imply that the probability density of the rare fluctuations is exponentially distributed. In the bosonic case, we have also noted that the large deviation function $I(J_E,v)$ exhibits a discontinuous dependence as a function on $v$ at the values $v=0$ or $v=\pm v_{max}$. This feature is caused by the fact that $\Delta e(x,t)/t$ is a time-integrated observable, 
which accordingly depends on the flux of quasi-particles reaching the point $x$ within the time interval $(0,t)$, and by the fact that each spatial interval within the chain can actually act as a reservoir of particles, since bosons for each mode $k$ can be produced in an arbitrarily large number. 
                 
The technique we used to calculate $G(\lambda,v)$ in Sec.~\ref{sec:SCGF_general_result}, based on biasing the density matrix $\rho(\lambda,v)$ as a function of $\lambda$, can be generalized to the more complex case of interacting integrable models, as done in Refs.~\cite{doyonMyers2020,doyon2020fluctuations} for the calculation of $G(\lambda)$ in homogeneous and stationary states. For inhomogeneous and non-stationary states, such as $\rho_0$ in Eq.~\eqref{eq:intro-rho0}, a general expression for $G(\lambda,v)$ analogous to Eq.~\eqref{eq:SCGF_hydro_euler} is still lacking. However, within the approach based on generalized hydrodynamics \cite{castro2016emergent,bertini2016transport} we think that it should be possible to extend the analysis of Sec.~\ref{sec:SCGF_general_result} to classical and quantum interacting integrable models. Finally, it would be interesting to test our predictions for the transferred energy probability density function $p(J_E,v)$ in ultra-cold atoms experiments, as done, e.g., in Ref.~\cite{Brantut713}, where heat and particle transport could be studied by preparing two identical clouds of atoms at different temperatures in the same spirit as the partitioning protocol analyzed here.

\begin{acknowledgements}
G.P. is indebted to B.~Doyon for useful discussions and collaboration on a closely related project. We are grateful to A.~Dhar, J.~Dubail, M.~Kormos, K.~Saito and J.~Viti for useful discussions. G.P. thanks King's College London for hospitality during completion of this work and the A. Della Riccia Foundation (Florence, Italy) -- INFN for financial support.  
\end{acknowledgements}

\onecolumngrid

\appendix

\section{Non interacting models: details of their solutions}
\label{app:appendix1}
In this Appendix, we provide some additional details on the expressions of the operators entering in the exact solution of the models introduced in Sec.~\ref{firstsection}. 

In particular, for the quantum Ising chain the pre-quench mode operators $\Phi_{r,l}(k)$ can be eventually expressed in terms of the post-quench ones $\Psi_R(k)$, according to Eq.~\eqref{eq:pre_quench_mode_post_quench_mode}: we report here only the final results, derived in Ref.~\cite{perfetto2017ballistic}, which we follow closely. 
The pre-quench modes $\Phi_{r}(k)$ introduced in Eq.~\eqref{eq:HRcont} are defined in terms of the lattice fermionic operators $c_n$ in Eq.~\eqref{eq:Jordan_Wigner} in the thermodynamic limit as
\be
\Phi_{r}(k) = \sum_{n=1}^{\infty} \left[\omega_r^n(k) \, c_n + \xi_r^n(k) \, c_n^{\dagger}\right], \label{eq:pre_quench_right_chain_TFIC}
\ee
for the right chain, and 
\be
\Phi_{l}(k) = \sum_{n=-\infty}^{0} \left[ \omega_l^n(k) \, c_n + \xi_l^n(k) \, c_n^{\dagger}\right], \label{eq:pre_quench_left_chain_TFIC}
\ee
for the left one, where
\be
\omega_{r,l}^n(k) = \frac{A_{r,l}^n(k)+ B_{r,l}^n(k)}{2} , \quad \xi_{r,l}^n(k) = \frac{A_{r,l}^n(k)- B_{r,l}^n(k)}{2}, \label{eq:omega_xi_functions}
\ee
and 
\be
A_r^n(k) = \sqrt{2/\pi} \,\sin(n k - f(k)) , \quad B_r^n(k) = \sqrt{2/\pi} \,\sin(n k), \label{eq:A_B_functions_right} 
\ee 
where $f(k)$ is defined in Eq.~\eqref{eq:Bogoliubov_angle}. 
The functions $A_l^n$ and $B_l^n$ for the left chain are simply related to those of the right chain as 
\be
A_l^n(k) = B_r^{1-n}(k)  \quad \mbox{and} \quad B_l^n(k) = A_r^{1-n}(k). \label{eq_A_B_functions_left}
\ee 
In order to express the operators $\Phi_{r,l}$ in Eqs.~\eqref{eq:pre_quench_right_chain_TFIC} and \eqref{eq:pre_quench_left_chain_TFIC} in terms of $\Psi_R(k)$, we fist need to write the lattice fermionic operators $c_n$ in terms of the post-quench modes as 
\be
c_n = \int_{-\pi}^{\pi} dk \,  \left[ \Psi_R(k)(\omega_R^n(k))^{\ast} + \Psi_R^{\dagger}(k) \xi_R^n(k)\right], \label{eq:lattice_fermions_post_quench}
\ee
where $\omega_R^n$ and $\xi_R^n$ have been defined in Eqs.~\eqref{eq:post_quench_TFIC_functions_1} and \eqref{eq:post_quench_TFIC_functions_2}. 
By inserting Eq.~\eqref{eq:lattice_fermions_post_quench} into Eqs.~\eqref{eq:pre_quench_right_chain_TFIC} and \eqref{eq:pre_quench_left_chain_TFIC}, we get Eq.~\eqref{eq:pre_quench_mode_post_quench_mode}, where the sums over lattice sites can be computed as a geometric series resulting into the coefficients $m_{\pm,\alpha}(k,k')$, with $\alpha \in \{l,r \}$. The latter have been first computed in Ref.~\cite{de2013nonequilibrium} 
and are listed here for completeness 
\begin{eqnarray}
m_{\pm ,l}(k,k') &=& \frac{1}{4 \pi i} \left\{ \frac{e^{-i[f(k)+f(k')]} \pm 1}{1-\mbox{e}^{i(k+k'+i\delta)}} -  \frac{\mbox{e}^{i[f(k')-f(k)]} \pm 1}{1-\mbox{e}^{i(k-k'+ i\delta)}} \right\}, \nonumber \\ 
\label{eq:m_coefficients_Ising_left}
\end{eqnarray}
for the left chain, and 
\begin{eqnarray}
m_{\pm,r}(k,k') &=& \frac{1}{4 \pi i} \left\{ \frac{e^{i[k-f(k')]} \pm e^{i[k-f(k)]}  }{1-\mbox{e}^{i(k'-k+i\delta)}} -  \frac{\mbox{e}^{i[k+f(k')]} \pm \mbox{e}^{i[k-f(k)]} }{1-\mbox{e}^{-i(k+k'-i\delta)}} \right\} , \label{eq:m_coefficients_Ising_right}
\end{eqnarray}
for the right one. 

For the harmonic chain, the exact solution of the left Hamiltonian in Eq.~\eqref{eq:prequenchoscillators} can be obtained by following the same procedure as for the right one with the definition, in the thermodynamic limit, of the operators  $\hat{\phi}_l(k)$, $\hat{p}_l(k)$, where
\begin{eqnarray}
\hat{\phi}_l(k) &=& \; -\sqrt{2/\pi}\, \sum_{x=-\infty}^{0} \sin(k (x-1)) \, \phi_x, \nonumber \\
\hat{p}_l(k) &=& \; -\sqrt{2/\pi}\, \sum_{x=-\infty}^{0} \sin(k (x-1)) \, p_x,  \label{eq:ftransfleft} 
\end{eqnarray}
as a function of which the original lattice operators $\phi_x,p_x$ can be written as 
\begin{eqnarray}
\phi_x &=& -\sqrt{2/\pi}\, \int_{0}^{\pi} dk \, \sin(k (x-1)) \, \hat{\phi}_l(k), \nonumber \\
p_x &=& -\sqrt{2/\pi}\, \int_{0}^{\pi} dk \, \sin(k (x-1)) \, \hat{p}_l(k).  \label{eq:left_functions_oas_almost_modes}
\end{eqnarray}
In particular, we emphasize that $\phi_x$ and $p_x$ in Eq.~\eqref{eq:left_functions_oas_almost_modes} automatically satisfy the boundary conditions for the left chain $\phi_1=p_1 \equiv 0 $ reported right after Eq.~\eqref{eq:fixedwall}. From the operators  $\hat{\phi}_l(k)$ and $\hat{p}_l(k)$ in Eq.~\eqref{eq:ftransfleft}, the bosonic annihilation and creation operators $A_l(k)$ and $A_l^{\dagger}(k)$, respectively, for the left chain can be introduced in the same way as in Eq.~\eqref{eq:bosonicmodes}, i.e.,
\be
A_l(k) = \frac{1}{\sqrt{2 \Omega(k)}}\left[\Omega(k) \hat{\phi}_l(k) + i \hat{p}_l(k)\right], \label{eq:bosonicmodes_left}
\ee
and the Hamiltonian then takes the diagonal form  
\be 
H_l = \int_{0}^{\pi} dk \, \Omega(k) \, A_l^{\dagger}(k) A_l(k). \label{eq:diagonal_left_Hamiltonian} 
\ee

For the harmonic chain the dynamics can be studied according to the same strategy as the quantum Ising chain, i.e., it is useful to write the pre-quench modes $A_{r,l}(k)$ in terms of the post-quench ones $\AA(k)$ in Eq.~\eqref{eq:post_quench_Oas_diagonal}.
This can be done by inserting Eq.~\eqref{eq:ftransf} (or Eq.~\eqref{eq:ftransfleft} for the left chain) into Eq.~\eqref{eq:bosonicmodes} (or Eq.~\eqref{eq:bosonicmodes_left} for the left Hamiltonian) and then by writing the lattice operators $\phi_x$ and $p_x$ as in Eq.~\eqref{eq:Ftransform}. 
The sum over the lattice coordinate can be again computed as a geometric series, with the following result:
\begin{eqnarray}
A_r(k) &=& \int_{-\pi}^{\pi} dk' \, \left[-\AA^{\dagger}(k') \, m_{-,r}(k',k) + \AA(k') \, m_{+,r}^{\ast}(k',k)\right], \nonumber \\
A_l(k) &=& \int_{-\pi}^{\pi} dk' \, \left[\AA^{\dagger}(k') \, m_{-,l}(k',k) + \AA(k')\, m_{+,l}^{\ast}(k',k)\right], \label{eq:pre_post_quench_bosons_right}
\end{eqnarray}
with the following expressions for the coefficients $m_{\pm,l,r}(k,k')$:
\begin{eqnarray}
m_{\pm, l }(k,k') &=& \pm \frac{1}{4 \pi i}\left[ \frac{\mbox{e}^{ik'}}{1-\mbox{e}^{i(k+k'+i\delta)}} -  \frac{\mbox{e}^{-ik'}}{1-\mbox{e}^{i(k-k'+i\delta)}}    \right]\left[\sqrt{\frac{\Omega(k')}{\Omega(k)}} \pm \sqrt{\frac{\Omega(k)}{\Omega(k')}} \right], \nonumber \\
m_{\pm, r }(k,k') &=& \frac{1}{4 \pi i}\left[ \frac{\mbox{e}^{i(k'-k)}}{1-\mbox{e}^{i(k'-k+i\delta)}} -  \frac{\mbox{e}^{-i(k+k')}}{1-\mbox{e}^{-i(k+k'-i\delta)}}    \right]\left[\sqrt{\frac{\Omega(k)}{\Omega(k')}} \pm \sqrt{\frac{\Omega(k')}{\Omega(k)}} \right]. \label{eq:m_coefficients_harmonic_chains}
\end{eqnarray}
It is also possible to invert Eq.~\eqref{eq:pre_post_quench_bosons_right} in order to express $\AA(k)$ as a function of $A_{r,l}(k)$,
\be
\AA(k) = \int_{0}^{\pi} dk' \left[ m_{-,l}(k,k') A^{\dagger}_l(k') + m_{+,l}(k,k') A_{l}(k') + m_{-,r}(k,k') A^{\dagger}_r(k') + m_{+,r}(k,k') A_{r}(k')\right]. \label{eq:post_pre_quench_bosons} 
\ee

\section{Calculation of the energy current in the hydrodynamic limit for the harmonic chain}
\label{app:appendix2}

In this Appendix we report the calculations in the hydrodynamic limit underlying Eq.~\eqref{eq:hydro_state_proved_bosons} for the harmonic chain. We do not provide the analogous derivation of Eq.~\eqref{eq:hydro_state_proved} for the quantum Ising model, as it proceeds as presented here and it is equivalent to the procedure outlined in Refs.~\cite{kormos2017inhomogeneous,perfetto2017ballistic}.

We start by writing the left and right Hamiltonians $H_{\alpha}$, with $\alpha \in \{l,r\}$, in Eqs.~\eqref{eq:diagonalrightbosons} and \eqref{eq:diagonal_left_Hamiltonian}, determining  the initial state $\rho_0$ of Eq.~\eqref{eq:intro-rho0}, as a function of the post-quench modes via Eq.~\eqref{eq:pre_post_quench_bosons_right}. 
In terms of the operators $\AA(k)$, 
the space and time evolution of Eq.~\eqref{eq:hydro_limit} can be easily calculated, because for them it is simply given by Eq.~\eqref{eq:transinvariance2} and $e^{iHt}\AA(k)e^{-iHt}=e^{-i \Omega(k)t}\AA(k)$, 
resulting in
\begin{align}
(P_{tr}^{\dagger})^x e^{-iHt} H_{\alpha} e^{iHt} (P_{tr})^x = \int_{-\pi}^{\pi}\! \!dk'\! \int_{-\pi}^{\pi} \!\! & dk'' \Big[e^{-i \varphi_{x,t}^{+}(k',k'')} I_{--}^{\alpha}(k',k'') \AA(k') \AA^{\dagger}(k'') + e^{i \varphi_{x,t}^{+}(k',k'')} I_{++}^{\alpha}(k',k'')\AA^{\dagger}(k')\AA(k'') \nonumber \\
&-e^{i \varphi_{x,t}^{-}(k',k'')} I_{-+}^{\alpha}(k',k'') \AA(k') \AA(k'') -e^{-i \varphi_{x,t}^{-}(k',k'')} I_{+-}^{\alpha}(k',k'') \AA^{\dagger}(k')\AA^{\dagger}(k'') \Big],
\label{eq:time_evolved_left_right_Hamiltonians} 
\end{align}
with the phases $\varphi_{x,t}^{\pm}(k',k'')$ defined as 
\be
\varphi_{x,t}^{\pm}(k',k'') = [\Omega(k'') \mp \Omega(k')]t \pm x (k' \mp k''), \label{eq:phases_hydro_appendix}
\ee
and the coefficients $I^{\alpha}_{\pm,\pm}(k',k'')$ given by integrals of the overlaps in Eq.~\eqref{eq:m_coefficients_harmonic_chains} as follows
\begin{align}
I_{++}^{\alpha}(k',k'') &= \frac{1}{2}\int_{-\pi}^{\pi} dk \, \Omega(k) \, m_{+, \alpha }(k',k) \, m_{+, \alpha }^{\ast}(k'',k) = \frac{1}{2} \oint_{C_1} \frac{\Omega(-i\mbox{ln}(z)) \, m_{+, \alpha }(k',-i\mbox{ln}(z)) \, m_{+, \alpha }^{\ast}(k'',-i\mbox{ln}(z))}{i z}, \nonumber \\
I_{+-}^{\alpha}(k',k'') &= \frac{1}{2}\int_{-\pi}^{\pi} dk \, \Omega(k) \, m_{+, \alpha }(k',k) \, m_{-, \alpha }(k'',k) = \frac{1}{2} \oint_{C_1} \frac{\Omega(-i\mbox{ln}(z)) \, m_{+, \alpha }(k',-i\mbox{ln}(z)) \, m_{-, \alpha }(k'',-i\mbox{ln}(z))}{i z}, \label{eq:I_integrals_complex_plane}
\end{align}
while $I_{-\pm}^{\alpha}$ can be obtained from $I_{+\pm}^{\alpha}$ by taking the complex conjugate and exchanging $m_{+,\alpha} \leftrightarrows m_{-,\alpha}$. Note that in Eq.~\eqref{eq:I_integrals_complex_plane} we have first extended the integrals from $(0,\pi)$ to $(-\pi,\pi)$ by exploiting the properties $m_{\pm,\alpha}(k',k) = - m_{\pm,\alpha}(k',-k)$, $\Omega(-k)=\Omega(k)$, and then we have introduced the variable $z=e^{ik}$ which transforms the original integral into one along  the circle $C_1$ with unit radius centered at the origin of the complex plane.

In the space-time scaling limit of Eq.~\eqref{eq:hydro_limit} each of the four integrals appearing in Eq.~\eqref{eq:time_evolved_left_right_Hamiltonians} is dominated by the regions in the $(k',k'')$ plane where the phases $\varphi_{x,t}^{\pm}$ are stationary and by the singularities of the integrands $\{I^{\alpha}(k',k'')\}$ occurring in correspondence to these stationary points. 
In particular, the stationary-phase condition for $\varphi_{x,t}^{\pm}(k)$ is  
\begin{equation}
\left \{
         \begin{array}{ll}
          \displaystyle {\frac{\partial \varphi_{x,t}^{\pm}(k',k'')}{\partial k'} = \mp v_g(k')t \pm x = 0},  \\ 
          \\
         \displaystyle {\frac{\partial \varphi_{x,t}^{\pm}(k',k'')}{\partial k''} = v_g(k'')t-x = 0},
          \end{array}
          \right. \label{eq:stationary_phase_system}
\end{equation}
where $v_g(k)$ is the group velocity defined after Eq.~\eqref{eq:hydro_state_proved}. 
Each of this stationary phase conditions has two solutions $k_{\pm}(v)$ if $x/t = v <v_{max}$ (where $v_{max}$ is given by Eq.~\eqref{eq:maxvelocityOAS} for the harmonic chain) such that
\begin{align}
\mbox{cos}(k_{\pm}(v)) &= \frac{v^2}{\omega^2} \pm \sqrt{\frac{v^4}{\omega^4}-\frac{v^2 m^2}{\omega^4} -\frac{2 v^2}{\omega^2} +1}, \nonumber \\ 
&= \frac{v^2}{\omega^2} \pm \frac{1}{\omega^2} \sqrt{(v_{max}^2-v^2)(y^2-v^2)},   \label{eq:systemsolution}
\end{align}
where
\be
y=\frac{m+\sqrt{m^2+4 \omega^2}}{2}; \label{eq:y_constant_v_max}
\ee
accordingly, the system in Eq.~\eqref{eq:stationary_phase_system} admits four pairs of solutions $(k_{+},k_{+})$, $(k_{-},k_{-})$, $(k_{-},k_{+})$, and $(k_{+},k_{-})$. 
The integrands $I^{\alpha}_{\pm\pm}(k',k'')$ are, however, singular only for the stationary points $(k',k'') = (k_{+},k_{+})$ and $(k_{-},k_{-})$, at which $k'=k''$. Accordingly, the integrals in Eq.~\eqref{eq:time_evolved_left_right_Hamiltonians} can be computed, in the hydrodynamic limit, by expanding the integrand around $k' \simeq k''$. The singular part of $I^{\alpha}_{++}(k',k'')$ as $k' \rightarrow k''$ can be extracted from Eq.~\eqref{eq:I_integrals_complex_plane}, with the residue theorem, finding that 
\be
\beta_l \, I^{l}_{++}(k',k'') + \beta_r \, I^{r}_{++}(k',k'') = \frac{\beta_l}{4 \pi i} \frac{\Omega(k)+\Omega(k')}{k''-k'-2i\delta} -  \frac{\beta_r}{4 \pi i} \frac{\Omega(k)+\Omega(k')}{k''-k'+ 2i\delta} + \, \,  \mbox{regular terms as} \, \,  k' \rightarrow k'', \label{eq:I_plus_Integral_appendix}
\ee
where we omitted terms that are regular as $k' \rightarrow k''$ and that are therefore sub-leading in the space-time scaling limit. 
The integrals $I^{\alpha}_{+-}(k',k'')$, $I^{\alpha}_{-+}(k',k'')$, and $I^{\alpha}_{--}(k',k'')$ can be neglected for the same reason as these regular terms, as they are not singular for $k' \rightarrow k''$. 
Notice that the singularities of the matrix elements $m_{\pm,\alpha}(k,k')$, in Eq.~\eqref{eq:m_coefficients_harmonic_chains} for the harmonic chains, are identical to those of the same coefficients in Eqs.~\eqref{eq:m_coefficients_Ising_left} and \eqref{eq:m_coefficients_Ising_right} for the Ising chain. 
The stationary-phase analysis for the fermionic case of the integral in Eq.~\eqref{eq:time_evolved_left_right_Hamiltonians} proceeds therefore in the same way as in the bosonic case outlined here and the expression for $I^{\alpha}_{++}$ is identical to that in Eq.~\eqref{eq:I_plus_Integral_appendix} upon replacing $\Omega(k) \leftrightarrows \varepsilon(k)$, in agreement with the result of Refs.~\cite{perfetto2017ballistic,kormos2017inhomogeneous}.
By inserting Eq.~\eqref{eq:time_evolved_left_right_Hamiltonians} for the right and left Hamiltonians into Eq.~\eqref{eq:intro-rho0} and by taking into account that the only singular contribution as $k' \rightarrow k''$ comes from Eq.~\eqref{eq:I_plus_Integral_appendix}, one obtains Eq.~\eqref{eq:intermediate_result_rho_x,t} (with the replacement $\AA(k) \leftrightarrows \Psi_R(k)$ for the harmonic chain) for the leading space-time dependence $\rho(x,t)$ of the density matrix in the hydrodynamic limit.
The rest of the calculation, as outlined in the main text, follows by changing variables to $Q= k'-k''$ and $K=(k'+k'')/2$ in the double integral in Eq.~\eqref{eq:intermediate_result_rho_x,t}. Expanding $\varphi_{x,t}^+(k',k'')$ around $Q =0$, i.e., 
\be
\varphi_{x,t}^+(k',k'') = \varphi_{x,t}^+(K+Q/2,K-Q/2) = Q(x- v_g(K) \, t) + O(Q^2), \label{eq:phase_expansion_Q}
\ee
and using the integral definition of the Heaviside step function $\Theta = \lim_{\delta
\rightarrow 0^+} \int_{-\infty}^{\infty} \frac{dy}{2 \pi i} \frac{e^{ixy}}{y-i\delta}$, the result in Eqs.~\eqref{eq:hydro_state_proved} and \eqref{eq:hydro_state_proved_bosons} for the time-evolved density matrix $\rho(v)$ in the hydrodynamic limit is eventually found.

Similarly, the calculation of the energy current can be done by writing the operator $j_0^E$ in Eq.~\eqref{eq:oscillatorcurrent} in terms of the post-quench operators $\AA(k)$ by using Eqs.~\eqref{eq:postquenchdiagonal} and \eqref{eq:Ftransform}, i.e.,
\be
j_0^E = \frac{i \omega^2}{4} \int_{-\pi}^{\pi} dk \int_{-\pi}^{\pi} \frac{dk'}{2 \pi} \sqrt{\frac{\Omega(k')}{\Omega(k)}}(e^{-ik}-1)(e^{-ik'}+1) \left[\AA(k) \AA^{\dagger}(-k') -\AA(k)\AA(k') +\AA^{\dagger}(-k) \AA^{\dagger}(-k')-\AA^{\dagger}(-k) \AA(k')\right], \label{eq:current_operator_modes} 
\ee  
and then by exploiting Eqs.~\eqref{eq:mode_1} and \eqref{eq:mode_1_bosons} together with 
\begin{eqnarray}
\mbox{Tr}[\rho(v) \, \AA(k) \, \AA(k')] = \mbox{Tr}[\rho(v) \, \AA^{\dagger}(k) \, \AA^{\dagger}(k')] =0, 
\end{eqnarray}
for expressing the average over $\rho(v)$ of bilinears in the post-quench modes. 
The very same procedure applies to the energy density operator $u_0$ and it is not reported here for brevity. 
The expression of $\mathcal{J}^E (v)$ in Eqs.~\eqref{eq:mode_GHD_bosons} and \eqref{eq:energycurrentoas_hydro} can be written as
\begin{align}
\mathcal{J}^E (v) &= \int_{0}^{\pi} \frac{dk}{2 \pi} \, \Omega(k) \, v_g(k) \, (f^{-}_{\beta_l}(k) - f^{-}_{\beta_r}(k)) \Theta(v_g(k)-|v|) = \int_{\Omega_{min}}^{\Omega_{max}} \frac{d\Omega}{2 \pi} \, \Omega \, (f^{-}_{\beta_l}(\Omega) - f^{-}_{\beta_r}(\Omega)) \Theta(v_g(\Omega)-|v|), \nonumber \\
&= \int_{\Omega_{-}(v)}^{\Omega_{+}(v)} \frac{d\Omega}{2 \pi} \, \Omega \, (f^{-}_{\beta_l}(\Omega) - f^{-}_{\beta_r}(\Omega)) = \int_{\Omega_{-}(v)}^{\Omega_{+}(v)} \frac{d\Omega}{2 \pi} \, \Omega \, \left(\frac{1}{e^{\beta_l \Omega}-1} - \frac{1}{e^{\beta_r \Omega}-1}\right), \label{eq:intermediate_result_current_oas}
\end{align}
where 
\begin{align}
\Omega_{\pm}(v) = \Omega(k_{\mp}(v)) &= \sqrt{m^2+2(\omega^2-v^2)\pm 2\sqrt{(\omega^2-v^2+mv)(\omega^2-v^2-mv)}}, \nonumber \\           				&= \sqrt{m^2+2(\omega^2-v^2) \pm  2\sqrt{(v_{max}^2-v^2)(y^2-v^2)}}, \label{eq:Omega_1_2}
\end{align}
and $k_{\pm}(k)$ is defined in Eq.~\eqref{eq:systemsolution} as the roots of the stationary-phase equations in Eq.~\eqref{eq:stationary_phase_system}. 
Integrating the expression in Eq.~\eqref{eq:intermediate_result_current_oas}, the results reported in Eqs.~\eqref{eq:integrated_oas_current_1} and \eqref{eq:integrated_oas_current_2} are eventually recovered by introducing  the integral representation of the function $Y(x)$ in Eq.~\eqref{eq:integrated_oas_current_3} \cite{NIST:DLMF}
\be
Y(x) = \int_{x}^{\infty} dy \frac{y}{e^{y}-1}. \label{eq:integral_representation_Y_function}
\ee 
In order to determine the edge asymptotic of $\mathcal{J}^E (v)$ in Eqs.~\eqref{eq:integrated_oas_current_1}, \eqref{eq:integrated_oas_current_2} and \eqref{eq:integrated_oas_current_3} as $v \rightarrow \pm v_{max}^{\mp}$ 
we start by noting that $\Omega_+(v)$ and $\Omega_-(v)$ tend to coalesce in this limit to the value
\be
\Omega_0 = \Omega_{\pm}(v_{max}) = \sqrt{m \sqrt{m^2+4 \omega^2}}, \label{eq:Omega_asymptotic}
\ee
and therefore 
\be
\Omega_{-}(v) - \Omega_{+}(v) = -2 \sqrt{v_{max}^2-v^2} + \mathcal{O}\left((v_{max}-|v|)^{3/2} \right) \label{eq:Omega_1_minus_2_asymptotic};
\ee 
correspondingly, for the function $\mathcal{Y}(\beta,v)$ in Eq.~\eqref{eq:integrated_oas_current_2}, one can write
\begin{align} 
\mathcal{Y}(\beta,v) &= \frac{Y(\beta \Omega_{-})-Y(\beta \Omega_{+})}{2 \pi \beta^2} = - \frac{1}{2 \pi \beta^2} \beta[\Omega_{+}(v)-\Omega_{-}(v)]Y'(\beta \Omega_0) + \mathcal{O}\left((v_{max}-|v|)^{3/2} \right) \nonumber \\ 
&= \frac{1}{2 \pi}[\Omega_{+}(v)-\Omega_{-}(v)] \, \Omega_0 \, f^{-}_{\beta}(\Omega_0) + \mathcal{O}\left((v_{max}-|v|)^{3/2} \right), \label{eq:Omega_asymptotic_2}
\end{align}
where the last step follows from the integral representation of $Y(x)$ in Eq.~\eqref{eq:integral_representation_Y_function}. Inserting Eq.~\eqref{eq:Omega_1_minus_2_asymptotic} into Eq.~\eqref{eq:Omega_asymptotic_2} and eventually into Eq.~\eqref{eq:integrated_oas_current_1}, the result in Eq.~\eqref{eq:edge_ballistic_bosons} is obtained, i.e.,
\be
\mathcal{J}^E = C_1 \sqrt{v_{max}^2-v^2}+ \mathcal{O}\left((v_{max}-|v|)^{3/2} \right),
\ee 
with $C_1$ given by
\be
C_1 = \frac{\Omega_0}{\pi} [f^{-}_{\beta_l}(\Omega_0)-f^{-}_{\beta_r}(\Omega_0)]= \frac{\Omega_0}{\pi} \left(\frac{1}{e^{\beta_l \Omega_0}-1}-\frac{1}{e^{\beta_r \Omega_0}-1} \right). \label{eq:C_1_constant}
\ee
As emphasized in the main text, when the mass $m$ is set to zero, the edge behavior is still expressed by Eq.~\eqref{eq:edge_ballistic_bosons}: indeed, from Eq.~\eqref{eq:C_1_constant}, we see that $f^{-}_{\beta}(\Omega_0) \rightarrow 1/(\beta \Omega_0)$
as $m \rightarrow 0$ and therefore Eq.~\eqref{eq:edge_ballistic_bosons} remains valid with $C_1= 1/ \pi(1/\beta_l - 1/\beta_r)$.

\section{Fine structure of the edge of the propagating front for the harmonic chain: the Airy kernel}
\label{app:appendix2.5}

In order to study the sub-diffusive 
corrections to the edge behavior of the energy current $\mathcal{J}^E$, expressed by  Eqs.~\eqref{eq:Airy_kernel_bosons_1}, \eqref{eq_Airy_kernel_bosons_2}, \eqref{eq:Airy_kernel_bosons_3}, and \eqref{eq:Airy_critical} of the main text, it is simpler to take the hydrodynamic limit by evolving directly in space and time the operator in Eq.~\eqref{eq:current_operator_modes} and eventually taking the trace over the initial state $\rho_0$ in Eq.~\eqref{eq:intro-rho0}; this can be done by writing the post-quench operators $\AA(k)$ in terms of the pre-quench ones $\AA_{r,l}(k)$ via Eq.~\eqref{eq:post_pre_quench_bosons}. The procedure is completely analogous to the one followed in the main the text, where the space-time scaling limit is first taken on the density matrix $\rho(v)$, as done in Refs.~\cite{perfetto2017ballistic,kormos2017inhomogeneous} for the quantum Ising model. Accordingly, we report here only the main steps for the harmonic chain.

Following the procedure outlined above, it turns out that the dominant term in the hydrodynamic limit is
\begin{equation}
\mathcal{J}^E(x,t) = \frac{i \omega^2}{4} \int_{-\pi}^{\pi} \!\! dk \int_{-\pi}^{\pi}\!\! \frac{dk'}{2 \pi} \, e^{\varphi_{x,t}^{+}(k,k')} \, \left[ I^{r}_{++}(k,k') +  I^{l}_{++}(k,k')\right] g(k,k'),
\label{eq:current_first_step_Airy}
\end{equation}
where we have defined for brevity
\begin{equation}
g(k,k') = (e^{-ik}-1)(e^{ik'}+1)\sqrt{\frac{\Omega(k')}{\Omega(k)}} - (e^{ik'}-1)(e^{-ik}+1)\sqrt{\frac{\Omega(k)}{\Omega(k')}}, \label{eq:g_function_Airy}
\end{equation}
and $I^{r,l}_{++}$ are given by Eq.~\eqref{eq:I_plus_Integral_appendix} after replacing $\Omega(k)$ and $\Omega(k')$ in the numerator with $f^{-}_{\beta_{r,l}}(k)$ and $f^{-}_{\beta_{r,l}}(k')$, respectively:
\begin{align}
I^{l}_{++}(k,k') = \frac{1}{4 \pi i} \frac{f^{-}_{\beta_{l}}(k)+f^{-}_{\beta_{l}}(k')}{k'-k-2i\delta} 
\quad \mbox{and}\quad
I^{r}_{++}(k,k')= -\frac{1}{4 \pi i} \frac{f^{-}_{\beta_{r}}(k)+f^{-}_{\beta_r}(k')}{k'-k + 2i\delta}. \label{eq:I_integral_bose_functions} 
\end{align}
From the previous expression, introducing the variables $Q=k-k'$, $K=(k+k')/2$ and expanding the phase $\varphi_{x,t}^{+}$ to first order in $Q$ as in Eq.~\eqref{eq:phase_expansion_Q}, one readily obtains Eqs.~\eqref{eq:energycurrentoas_hydro} and \eqref{eq:intermediate_result_current_oas} as detailed in Appendix \ref{app:appendix2}.
Here, however, we are interested in the behavior of $\mathcal{J}^E(x,t)$ for $x \simeq v_{max}t$: in this case, higher-order terms in the expansion of the phase $\varphi_{x,t}^{+}$ become important and the profile at the edge of the propagating front qualitatively changes with respect to the one predicted at the ballistic scale
in Eq.~\eqref{eq:edge_ballistic_bosons}. In particular, the two stationary points $k_{\pm}(v)$ in Eq.~\eqref{eq:systemsolution} merge into a unique solution $k_s$ as $x \simeq v_{max}t$, obtained by setting $v=v_{max}$ into Eq.~\eqref{eq:systemsolution}, where the group velocity is maximum $v_g(k_s)=v_{max}$. The second derivative of  $\varphi_{x,t}^{+}$ therefore vanishes and the leading correction to the ballistic profile is obtained by expanding to the third order in $k-k_s$:
\be
\varphi_{x,t}^{+}(k,k') = (k-k_s)(x-v_{max}t) +\frac{(k-k_s)^3}{3!}v_{max}t + (v_{max}t-x)(k'-k_s) -\frac{(k'-k_s)^3}{3!}v_{max}t + \mathcal{O}((k-k_s)^4),  \label{eq:Airy_expansion_phase},
\ee  
where we used the fact that $\Omega''(k_s)=0$, $\Omega^{(3)}(k_s)=-v_{max}$, with $\Omega(k)$ is given in Eq.~\eqref{eq:oasspectrum}. 
In order to evaluate the integral in Eq.~\eqref{eq:current_first_step_Airy} from a saddle-point approximation around $k_s$ it is then useful to make the change of variables $\tilde{k}=k-k_s$, $\tilde{k}'=k'-k_s$ and write
\begin{equation}
\mathcal{J}^E(x,t) = \frac{i \omega^2}{4} \int_{-\infty}^{\infty} \frac{d\tilde{k}}{2 \pi} \int_{-\infty}^{\infty} \frac{d\tilde{k}'}{2 \pi}  \, e^{\varphi_{x,t}^{+}(\tilde{k},\tilde{k}')} g(\tilde{k},\tilde{k}') \left(  \frac{f_{\beta_l}^{-}(\tilde{k})+f_{\beta_l}^{-}(\tilde{k}')}{2 i \, (\tilde{k}'-\tilde{k}-2i\delta)} -\frac{f_{\beta_r}^{-}(\tilde{k})+f_{\beta_r}^{-}(\tilde{k}')}{2 i \, (\tilde{k}'-\tilde{k}+2i\delta)}\right) ,  \label{eq:current_second_step_Airy}
\end{equation} 
where we have extended the integrals to the whole real line as the regions with large $\tilde{k}$ and $\tilde{k}'$ do not contribute. In the previous expression $\varphi_{x,t}^{+}(\tilde{k},\tilde{k}')$ is given in Eq.~\eqref{eq:Airy_expansion_phase}.
Since the cubic term in Eq.~\eqref{eq:Airy_expansion_phase} is expected to be the dominant one, it is convenient to introduce the variables 
\be
K=\left(\frac{v_{max}t}{2}\right)^{1/3} \tilde{k}, \quad Q = \left(\frac{v_{max}t}{2}\right)^{1/3} \tilde{k}', \label{eq:rescaled_integration_variables_Airy}
\ee
and the scaling variable $X$ in Eq.~\eqref{eq:scaling_variable_Airy}. 
For the first contribution on the r.h.s.~of Eq.~\eqref{eq:current_second_step_Airy} we find, after expanding the integrand around the saddle-point $k_s=0$,
\begin{multline}
\frac{i \omega^2}{4} \left(\frac{2}{v_{max}t}\right)^{1/3} \int_{-\infty}^{\infty} \frac{dK}{2 \pi} \int_{-\infty}^{\infty} \frac{dQ}{2 \pi} \, \frac{e^{iKX+iK^3/3-iQX-iQ^3/3} g(K,Q) [f_{\beta_l}^{-}(K)+f_{\beta_l}^{-}(Q)]}{2 i \, (Q-K-2i\delta)}=  \\
= \frac{i \omega^2}{4} \left(\frac{2}{v_{max}t}\right)^{1/3} g(k_s,k_s) f_{\beta_l}^{-}(k_s) K^A(X,X) = \left(\frac{2}{v_{max}t}\right)^{1/3} \Omega(k_s) v_{max} \, f_{\beta_l}^{-}(k_s) K^A(X,X),  \label{eq:Airy_kernel_left}
\end{multline}
where we have used the integral representation of the Airy kernel \cite{tracy1994level}
\be
K^A(X,Y) = \int_{-\infty}^{\infty} \frac{dK}{2 \pi} \int_{-\infty}^{\infty} \frac{dQ}{2 \pi} \frac{e^{iKY+iK^3/3-iQX-iQ^3/3}}{i(Q-K-i\delta)}. \label{eq:Airy_kernel_integral_definition} 
\ee
For the second contribution, instead, a bit more care is needed: in fact, one can notice that the sign of the infinitesimal displacement $\delta$ in the integrand is opposite to that present in the definition of the Airy kernel in Eq.~\eqref{eq:Airy_kernel_integral_definition}. The $\tilde{k}'$ integral in Eq.~\eqref{eq:current_second_step_Airy} therefore avoids the pole at $\tilde{k}'=\tilde{k}$ from above and, by using the residue theorem, one can pull the integration contour below the pole, thereby changing the sign of $\delta$ in Eq.~\eqref{eq:current_second_step_Airy}, at the price of subtracting the residue at $\tilde{k}'=\tilde{k}$. The latter is easily computed to be
\be
-\int_{-\pi}^{\pi} \frac{dk}{2 \pi}\omega^2 \mbox{sin}(k) f_{\beta_r}^{-}(k) = -\int_{-\pi}^{\pi} \frac{dk}{2 \pi} \Omega(k) v_g(k) f_{\beta_r}^{-}(k) =0, \label{eq:equilibrium_current_value} 
\ee  
and  corresponds to the equilibrium value of the energy current for $v>v_{max}$, which is zero because, in the initial equilibrium state in Eq.~\eqref{eq:intro-rho0}, the current vanishes. 
This fact applies also to other physical quantities, for example the energy density, and the residue of the integral at $\tilde{k}=\tilde{k}'$ gives the equilibrium value of the observable outside the light-cone; this constant has to be added to the Airy kernel to give the correct edge-profile\footnote{We thank M.~Kormos for suggesting this procedure to determine this additive constant to the Airy kernel.}.
Accordingly, the second contribution in Eq.~\eqref{eq:current_second_step_Airy} can be analyzed in the same way as we did for the first one in Eq.~\eqref{eq:Airy_kernel_left}:
\begin{multline}
\frac{i \omega^2}{4} \left(\frac{2}{v_{max}t}\right)^{1/3} \int_{-\infty}^{\infty} \frac{dK}{2 \pi} \int_{-\infty}^{\infty} \frac{dQ}{2 \pi} \, \frac{e^{iKX+iK^3/3-iQX-iQ^3/3} g(K,Q) [f_{\beta_r}^{-}(K)+f_{\beta_r}^{-}(Q)]}{2 i \, (Q-K-2i\delta)}=  \\
= \frac{i \omega^2}{4} \left(\frac{2}{v_{max}t}\right)^{1/3} g(k_s,k_s) f_{\beta_r}^{-}(k_s) K^A(X,X) = \left(\frac{2}{v_{max}t}\right)^{1/3} \Omega(k_s) v_{max} \, f_{\beta_r}^{-}(k_s) K^A(X,X).  \label{eq:Airy_kernel_right}
\end{multline}
Inserting Eqs.~\eqref{eq:Airy_kernel_left} and \eqref{eq:Airy_kernel_right} into Eq.~\eqref{eq:current_second_step_Airy},  the results in Eqs.~\eqref{eq:Airy_kernel_bosons_1}, \eqref{eq_Airy_kernel_bosons_2}, and \eqref{eq:Airy_kernel_bosons_3} of the main text immediately follow. When the mass $m$ is set to zero, as mentioned at the end of Appendix \ref{app:appendix2},
$f^{-}_{\beta}(\Omega(k_s)) \rightarrow 1/(\beta \Omega(k_s))$ 
and Eq.~\eqref{eq:Airy_critical} is obtained, with the scaling variable $X$ in Eq.~\eqref{eq:scaling_variable_critical} as a consequence of the fact that for $m=0$
\be
\lim_{k \rightarrow 0^{\pm}} \Omega^{(3)}(k) = \mp v_{max}/4. \label{eq:discontinuity_third_derivative_criticality}
\ee

%
%

\section{Scaled cumulant generating function in the hydrodynamic limit: semi-classical derivation}

In this Appendix we report the main steps of the derivation of Eqs.~\eqref{eq:finalresultFCS_general}  and \eqref{eq:reservoir_FCS_general} within the semi-classical description of Sec.~\ref{sec:semi_classics}. 

Considering the case $v_g(k)>v$ and $v>0$, one inserts Eq.~\eqref{eq:fastrandomvariable} for $\Delta e(x,t;k)$ into Eq.~\eqref{eq:semiclassicalFCS}, with $x_0(k)=x-v_g(k)t<0$, $y_0(k)=x+v_g(k)t>0$, with the result
\begin{align}
g(\lambda,x,t;k) &=  \Bigg[ \prod_{x_0(k)<y<0} \langle e^{-\lambda \varepsilon(k) n_{\beta(y)}(k)} \rangle_{sc} \Bigg] \Bigg[ \prod_{0<y<x} \langle e^{-\lambda \varepsilon(k) n_{\beta(y)}(k)} \rangle_{sc} \Bigg] \Bigg[ \prod_{x<y<y_0(k)} \langle e^{\lambda \varepsilon(k) n_{\beta(y)}(k)} \rangle_{sc} \Bigg] \nonumber \\
               &= \left[\langle e^{-\lambda \varepsilon(k) n_{\beta_l}(k) } \rangle_{sc} \right]^{-x_0(k)} \left[\langle e^{-\lambda \varepsilon(k) n_{\beta_r}(k) } \rangle_{sc} \right]^x  \left[\langle e^{\lambda \varepsilon(k) n_{\beta_r}(k)} \rangle_{sc} \right]^{v_g(k)t}, \label{eq:first_step_SCGF_semiclassic}
\end{align}
where the subscript \textit{``sc''} denotes the semi-classical average as explained for Eq.~\eqref{eq:semiclassicalFCS} in the main text. From Eq.~\eqref{eq:Bernoulliaverage} for the fermionic case and Eq.~\eqref{eq:extremesdomain} for the bosonic one, $g(\lambda,x,t;k)$ is readily computed  
\begin{multline}
g(\lambda,x,t;k) =  \exp\Big(-x_0(k)[F(\beta_l \varepsilon(k))-F((\beta_l+\lambda) \varepsilon(k))]\Big) \, \exp\Big(x[F(\beta_r \varepsilon(k))-F((\beta_r + \lambda) \varepsilon(k))]\Big) \\
 \times \exp\Big(v_g(k)t[F(\beta_r \varepsilon(k))-F((\beta_r - \lambda) \varepsilon(k))]\Big). \label{eq:second_step_SCGF_semiclassic}
\end{multline}
By inserting the logarithm of the previous expression into Eq.~\eqref{eq:semiclassicalSCGF} and by taking the hydrodynamic limit we eventually find
\begin{align}
G(\lambda,v) = \int_{0}^{\pi} \frac{dk}{2 \pi} \Theta(v_g(k)-v) \Big\{ (v_g & (k)-v)[F(\beta_l \varepsilon(k))-F((\beta_l+\lambda) \varepsilon(k))] + v  [F(\beta_r \varepsilon(k))-F((\beta_r+\lambda) \varepsilon(k))]  \nonumber \\
&+  v_g(k) [F(\beta_r \varepsilon(k))-F((\beta_r-\lambda) \varepsilon(k))]\Big\}, \label{eq:vpositivepartialresult}
\end{align}
where we inserted the Heaviside step function to explicitly enforce the constraint $v_g(k)>v$. In the case $v_g(k)<v$ one can proceed similarly, obtaining  
\begin{align}
g(\lambda,x,t;k) &=  \Bigg[ \prod_{x_0(k)<y<x} \langle e^{-\lambda \varepsilon(k) n_{\beta(y)}(k)} \rangle_{sc} \Bigg] \Bigg[\prod_{x<y<y_0(k)} \langle e^{\lambda \varepsilon(k) n_{\beta(y)}(k)} \rangle_{sc} \Bigg] = \Big[\langle e^{-\lambda \varepsilon(k) n_{\beta_r}(k) } \rangle_{sc} \, \, \langle e^{\lambda \varepsilon(k) n_{\beta_r}(k) } \rangle_{sc} \Big]^{v_g(k)t}     \nonumber \\
&=   \exp \Big( v_g(k)t[F(\beta_r \varepsilon(k))-F((\beta_r+\lambda) \varepsilon(k))]\Big)\, \exp\Big(v_g(k)t [F(\beta_r \varepsilon(k))-F((\beta_r - \lambda) \varepsilon(k))] \Big) ,    			\label{eq:v_negative_semiclassics_first_step}
\end{align}
and for the SCGF from Eq.~\eqref{eq:semiclassicalSCGF} 
\begin{align}
G(\lambda,v) &= \int_{0}^{\pi}\frac{dk}{2 \pi} v_g(k) \Theta(v-v_g(k)) \Big\{[F(\beta_r \, \varepsilon(k)) - F((\beta_r+\lambda) \, \varepsilon(k))] 
+ [F(\beta_r \, \varepsilon(k)) - F((\beta_r-\lambda) \, \varepsilon(k))] \Big\}, \label{eq:Levitovreservoirexplicit}
\end{align}
where, as above, the Heaviside step function has been introduced in order to enforce the constraint $v_g(k)<v$. 
Adding the expressions in Eqs.~\eqref{eq:vpositivepartialresult} and \eqref{eq:Levitovreservoirexplicit}, using that $\Theta(v-v_g(k))=1-\Theta(v_g(k)-v)$, and performing the change of variable $k \rightarrow \varepsilon(k)$ ($\Omega(k)$ in the bosonic case), the results in Eqs.~\eqref{eq:finalresultFCS_general} and \eqref{eq:reservoir_FCS_general} are eventually obtained. For $v<0$, the calculations are totally analogous to those described above, leading to Eq.~\eqref{eq:finalresultFCS_general_v_negative}. 


\label{app:appendix3}
\twocolumngrid
  \bibliography{bibliography}

\begin{thebibliography}{104}%
\makeatletter
\providecommand \@ifxundefined [1]{%
 \@ifx{#1\undefined}
}%
\providecommand \@ifnum [1]{%
 \ifnum #1\expandafter \@firstoftwo
 \else \expandafter \@secondoftwo
 \fi
}%
\providecommand \@ifx [1]{%
 \ifx #1\expandafter \@firstoftwo
 \else \expandafter \@secondoftwo
 \fi
}%
\providecommand \natexlab [1]{#1}%
\providecommand \enquote  [1]{``#1''}%
\providecommand \bibnamefont  [1]{#1}%
\providecommand \bibfnamefont [1]{#1}%
\providecommand \citenamefont [1]{#1}%
\providecommand \href@noop [0]{\@secondoftwo}%
\providecommand \href [0]{\begingroup \@sanitize@url \@href}%
\providecommand \@href[1]{\@@startlink{#1}\@@href}%
\providecommand \@@href[1]{\endgroup#1\@@endlink}%
\providecommand \@sanitize@url [0]{\catcode `\\12\catcode `\$12\catcode
  `\&12\catcode `\#12\catcode `\^12\catcode `\_12\catcode `\%12\relax}%
\providecommand \@@startlink[1]{}%
\providecommand \@@endlink[0]{}%
\providecommand \url  [0]{\begingroup\@sanitize@url \@url }%
\providecommand \@url [1]{\endgroup\@href {#1}{\urlprefix }}%
\providecommand \urlprefix  [0]{URL }%
\providecommand \Eprint [0]{\href }%
\providecommand \doibase [0]{http://dx.doi.org/}%
\providecommand \selectlanguage [0]{\@gobble}%
\providecommand \bibinfo  [0]{\@secondoftwo}%
\providecommand \bibfield  [0]{\@secondoftwo}%
\providecommand \translation [1]{[#1]}%
\providecommand \BibitemOpen [0]{}%
\providecommand \bibitemStop [0]{}%
\providecommand \bibitemNoStop [0]{.\EOS\space}%
\providecommand \EOS [0]{\spacefactor3000\relax}%
\providecommand \BibitemShut  [1]{\csname bibitem#1\endcsname}%
\let\auto@bib@innerbib\@empty
\bibitem [{\citenamefont {Greiner}\ \emph {et~al.}(2002)\citenamefont
  {Greiner}, \citenamefont {Mandel}, \citenamefont {H{\"a}nsch},\ and\
  \citenamefont {Bloch}}]{greiner2002collapse}%
  \BibitemOpen
  \bibfield  {author} {\bibinfo {author} {\bibfnamefont {M.}~\bibnamefont
  {Greiner}}, \bibinfo {author} {\bibfnamefont {O.}~\bibnamefont {Mandel}},
  \bibinfo {author} {\bibfnamefont {T.~W.}\ \bibnamefont {H{\"a}nsch}}, \ and\
  \bibinfo {author} {\bibfnamefont {I.}~\bibnamefont {Bloch}},\ }\href
  {\doibase https://doi.org/10.1038/nature00968} {\bibfield  {journal}
  {\bibinfo  {journal} {Nature}\ }\textbf {\bibinfo {volume} {419}},\ \bibinfo
  {pages} {51} (\bibinfo {year} {2002})}\BibitemShut {NoStop}%
\bibitem [{\citenamefont {Kinoshita}\ \emph {et~al.}(2006)\citenamefont
  {Kinoshita}, \citenamefont {Wenger},\ and\ \citenamefont
  {Weiss}}]{kinoshita2006quantum}%
  \BibitemOpen
  \bibfield  {author} {\bibinfo {author} {\bibfnamefont {T.}~\bibnamefont
  {Kinoshita}}, \bibinfo {author} {\bibfnamefont {T.}~\bibnamefont {Wenger}}, \
  and\ \bibinfo {author} {\bibfnamefont {D.~S.}\ \bibnamefont {Weiss}},\ }\href
  {\doibase https://doi.org/10.1038/nature04693} {\bibfield  {journal}
  {\bibinfo  {journal} {Nature}\ }\textbf {\bibinfo {volume} {440}},\ \bibinfo
  {pages} {900} (\bibinfo {year} {2006})}\BibitemShut {NoStop}%
\bibitem [{\citenamefont {Hofferberth}\ \emph {et~al.}(2007)\citenamefont
  {Hofferberth}, \citenamefont {Lesanovsky}, \citenamefont {Fischer},
  \citenamefont {Schumm},\ and\ \citenamefont
  {Schmiedmayer}}]{hofferberth2007non}%
  \BibitemOpen
  \bibfield  {author} {\bibinfo {author} {\bibfnamefont {S.}~\bibnamefont
  {Hofferberth}}, \bibinfo {author} {\bibfnamefont {I.}~\bibnamefont
  {Lesanovsky}}, \bibinfo {author} {\bibfnamefont {B.}~\bibnamefont {Fischer}},
  \bibinfo {author} {\bibfnamefont {T.}~\bibnamefont {Schumm}}, \ and\ \bibinfo
  {author} {\bibfnamefont {J.}~\bibnamefont {Schmiedmayer}},\ }\href {\doibase
  doi:10.1038/nature06149} {\bibfield  {journal} {\bibinfo  {journal} {Nature}\
  }\textbf {\bibinfo {volume} {449}},\ \bibinfo {pages} {324} (\bibinfo {year}
  {2007})}\BibitemShut {NoStop}%
\bibitem [{\citenamefont {Calabrese}\ \emph {et~al.}(2016)\citenamefont
  {Calabrese}, \citenamefont {Essler},\ and\ \citenamefont
  {Mussardo}}]{1742-5468-2016-6-064001}%
  \BibitemOpen
  \bibfield  {author} {\bibinfo {author} {\bibfnamefont {P.}~\bibnamefont
  {Calabrese}}, \bibinfo {author} {\bibfnamefont {F.~H.~L.}\ \bibnamefont
  {Essler}}, \ and\ \bibinfo {author} {\bibfnamefont {G.}~\bibnamefont
  {Mussardo}},\ }\href {\doibase 10.1088/1742-5468/2016/06/064001} {\bibfield
  {journal} {\bibinfo  {journal} {J. Stat. Mech.: Theory Exp.}\ }\textbf
  {\bibinfo {volume} {2016}},\ \bibinfo {pages} {064001} (\bibinfo {year}
  {2016})}\BibitemShut {NoStop}%
\bibitem [{\citenamefont {Eisert}\ \emph {et~al.}(2015)\citenamefont {Eisert},
  \citenamefont {Friesdorf},\ and\ \citenamefont
  {Gogolin}}]{eisert2015quantum}%
  \BibitemOpen
  \bibfield  {author} {\bibinfo {author} {\bibfnamefont {J.}~\bibnamefont
  {Eisert}}, \bibinfo {author} {\bibfnamefont {M.}~\bibnamefont {Friesdorf}}, \
  and\ \bibinfo {author} {\bibfnamefont {C.}~\bibnamefont {Gogolin}},\ }\href
  {\doibase https://doi.org/10.1038/nphys3215} {\bibfield  {journal} {\bibinfo
  {journal} {Nat. Phys.}\ }\textbf {\bibinfo {volume} {11}},\ \bibinfo {pages}
  {124} (\bibinfo {year} {2015})}\BibitemShut {NoStop}%
\bibitem [{\citenamefont {Polkovnikov}\ \emph {et~al.}(2011)\citenamefont
  {Polkovnikov}, \citenamefont {Sengupta}, \citenamefont {Silva},\ and\
  \citenamefont {Vengalattore}}]{polkovnikov2011colloquium}%
  \BibitemOpen
  \bibfield  {author} {\bibinfo {author} {\bibfnamefont {A.}~\bibnamefont
  {Polkovnikov}}, \bibinfo {author} {\bibfnamefont {K.}~\bibnamefont
  {Sengupta}}, \bibinfo {author} {\bibfnamefont {A.}~\bibnamefont {Silva}}, \
  and\ \bibinfo {author} {\bibfnamefont {M.}~\bibnamefont {Vengalattore}},\
  }\href {\doibase 10.1103/RevModPhys.83.863} {\bibfield  {journal} {\bibinfo
  {journal} {Rev. Mod. Phys.}\ }\textbf {\bibinfo {volume} {83}},\ \bibinfo
  {pages} {863} (\bibinfo {year} {2011})}\BibitemShut {NoStop}%
\bibitem [{\citenamefont {Altman}(2016)}]{altman2015nonBIS}%
  \BibitemOpen
  \bibfield  {author} {\bibinfo {author} {\bibfnamefont {E.}~\bibnamefont
  {Altman}},\ }in\ \href {\doibase 10.1093/acprof:oso/9780198768166.003.0001}
  {\emph {\bibinfo {booktitle} {Strongly Interacting Quantum Systems out of
  Equilibrium: Lecture Notes of the Les Houches Summer School}}},\
  Vol.~\bibinfo {volume} {99},\ \bibinfo {editor} {edited by\ \bibinfo {editor}
  {\bibfnamefont {T.}~\bibnamefont {Giamarchi}}, \bibinfo {editor}
  {\bibfnamefont {A.~J.}\ \bibnamefont {Millis}}, \bibinfo {editor}
  {\bibfnamefont {O.}~\bibnamefont {Parcollet}}, \bibinfo {editor}
  {\bibfnamefont {H.}~\bibnamefont {Saleur}}, \ and\ \bibinfo {editor}
  {\bibfnamefont {L.~F.}\ \bibnamefont {Cugliandolo}}}\ (\bibinfo  {publisher}
  {Oxford University Press},\ \bibinfo {address} {Oxford},\ \bibinfo {year}
  {2016})\BibitemShut {NoStop}%
\bibitem [{\citenamefont {Calabrese}(2015)}]{calabrese2015non}%
  \BibitemOpen
  \bibfield  {author} {\bibinfo {author} {\bibfnamefont {P.}~\bibnamefont
  {Calabrese}},\ }in\ \href {\doibase
  https://doi.org/10.1051/epjconf/20159008001} {\emph {\bibinfo {booktitle}
  {EPJ Web of Conferences}}},\ Vol.~\bibinfo {volume} {90}\ (\bibinfo
  {organization} {EDP Sciences},\ \bibinfo {year} {2015})\ p.\ \bibinfo {pages}
  {08001}\BibitemShut {NoStop}%
\bibitem [{\citenamefont {Calabrese}\ and\ \citenamefont
  {Cardy}(2006)}]{calabrese2006time}%
  \BibitemOpen
  \bibfield  {author} {\bibinfo {author} {\bibfnamefont {P.}~\bibnamefont
  {Calabrese}}\ and\ \bibinfo {author} {\bibfnamefont {J.}~\bibnamefont
  {Cardy}},\ }\href {\doibase 10.1103/PhysRevLett.96.136801} {\bibfield
  {journal} {\bibinfo  {journal} {Phys. Rev. Lett.}\ }\textbf {\bibinfo
  {volume} {96}},\ \bibinfo {pages} {136801} (\bibinfo {year}
  {2006})}\BibitemShut {NoStop}%
\bibitem [{\citenamefont {Calabrese}\ and\ \citenamefont
  {Cardy}(2007)}]{calabrese2007quantum}%
  \BibitemOpen
  \bibfield  {author} {\bibinfo {author} {\bibfnamefont {P.}~\bibnamefont
  {Calabrese}}\ and\ \bibinfo {author} {\bibfnamefont {J.}~\bibnamefont
  {Cardy}},\ }\href {\doibase https://doi.org/10.1088/1742-5468/2007/06/P06008}
  {\bibfield  {journal} {\bibinfo  {journal} {J. Stat. Mech.: Theory Exp.}\
  }\textbf {\bibinfo {volume} {2007}},\ \bibinfo {pages} {P06008} (\bibinfo
  {year} {2007})}\BibitemShut {NoStop}%
\bibitem [{\citenamefont {Calabrese}\ \emph {et~al.}(2011)\citenamefont
  {Calabrese}, \citenamefont {Essler},\ and\ \citenamefont
  {Fagotti}}]{calabrese2011quantum0}%
  \BibitemOpen
  \bibfield  {author} {\bibinfo {author} {\bibfnamefont {P.}~\bibnamefont
  {Calabrese}}, \bibinfo {author} {\bibfnamefont {F.~H.}\ \bibnamefont
  {Essler}}, \ and\ \bibinfo {author} {\bibfnamefont {M.}~\bibnamefont
  {Fagotti}},\ }\href {\doibase 10.1103/PhysRevLett.106.227203} {\bibfield
  {journal} {\bibinfo  {journal} {Phys. Rev. Lett.}\ }\textbf {\bibinfo
  {volume} {106}},\ \bibinfo {pages} {227203} (\bibinfo {year}
  {2011})}\BibitemShut {NoStop}%
\bibitem [{\citenamefont {Calabrese}\ \emph
  {et~al.}(2012{\natexlab{a}})\citenamefont {Calabrese}, \citenamefont
  {Essler},\ and\ \citenamefont {Fagotti}}]{calabrese2012quantum1}%
  \BibitemOpen
  \bibfield  {author} {\bibinfo {author} {\bibfnamefont {P.}~\bibnamefont
  {Calabrese}}, \bibinfo {author} {\bibfnamefont {F.~H.}\ \bibnamefont
  {Essler}}, \ and\ \bibinfo {author} {\bibfnamefont {M.}~\bibnamefont
  {Fagotti}},\ }\href {\doibase
  https://doi.org/10.1088/1742-5468/2012/07/P07016} {\bibfield  {journal}
  {\bibinfo  {journal} {J. Stat. Mech.: Theory Exp.}\ }\textbf {\bibinfo
  {volume} {2012}},\ \bibinfo {pages} {P07016} (\bibinfo {year}
  {2012}{\natexlab{a}})}\BibitemShut {NoStop}%
\bibitem [{\citenamefont {Calabrese}\ \emph
  {et~al.}(2012{\natexlab{b}})\citenamefont {Calabrese}, \citenamefont
  {Essler},\ and\ \citenamefont {Fagotti}}]{calabrese2012quantum2}%
  \BibitemOpen
  \bibfield  {author} {\bibinfo {author} {\bibfnamefont {P.}~\bibnamefont
  {Calabrese}}, \bibinfo {author} {\bibfnamefont {F.~H.}\ \bibnamefont
  {Essler}}, \ and\ \bibinfo {author} {\bibfnamefont {M.}~\bibnamefont
  {Fagotti}},\ }\href {\doibase
  https://doi.org/10.1088/1742-5468/2012/07/P07022} {\bibfield  {journal}
  {\bibinfo  {journal} {J. Stat. Mech.: Theory Exp.}\ }\textbf {\bibinfo
  {volume} {2012}},\ \bibinfo {pages} {P07022} (\bibinfo {year}
  {2012}{\natexlab{b}})}\BibitemShut {NoStop}%
\bibitem [{\citenamefont {Fagotti}\ and\ \citenamefont
  {Essler}(2013)}]{fagotti2013reduced}%
  \BibitemOpen
  \bibfield  {author} {\bibinfo {author} {\bibfnamefont {M.}~\bibnamefont
  {Fagotti}}\ and\ \bibinfo {author} {\bibfnamefont {F.~H.}\ \bibnamefont
  {Essler}},\ }\href {\doibase 10.1103/PhysRevB.87.245107} {\bibfield
  {journal} {\bibinfo  {journal} {Phys. Rev. B}\ }\textbf {\bibinfo {volume}
  {87}},\ \bibinfo {pages} {245107} (\bibinfo {year} {2013})}\BibitemShut
  {NoStop}%
\bibitem [{\citenamefont {Vidmar}\ and\ \citenamefont
  {Rigol}(2016)}]{vidmar2016generalized}%
  \BibitemOpen
  \bibfield  {author} {\bibinfo {author} {\bibfnamefont {L.}~\bibnamefont
  {Vidmar}}\ and\ \bibinfo {author} {\bibfnamefont {M.}~\bibnamefont {Rigol}},\
  }\href {\doibase https://doi.org/10.1088/1742-5468/2016/06/064007} {\bibfield
   {journal} {\bibinfo  {journal} {J. Stat. Mech.: Theory Exp.}\ }\textbf
  {\bibinfo {volume} {2016}},\ \bibinfo {pages} {064007} (\bibinfo {year}
  {2016})}\BibitemShut {NoStop}%
\bibitem [{\citenamefont {{\v{Z}}nidari{\v{c}}}(2010)}]{znidaric2010matrix}%
  \BibitemOpen
  \bibfield  {author} {\bibinfo {author} {\bibfnamefont {M.}~\bibnamefont
  {{\v{Z}}nidari{\v{c}}}},\ }\href {\doibase 10.1088/1751-8113/43/41/415004}
  {\bibfield  {journal} {\bibinfo  {journal} {J. Phys. A: Math. Theor.}\
  }\textbf {\bibinfo {volume} {43}},\ \bibinfo {pages} {415004} (\bibinfo
  {year} {2010})}\BibitemShut {NoStop}%
\bibitem [{\citenamefont {Prosen}(2011)}]{Prosen2011openXXZ}%
  \BibitemOpen
  \bibfield  {author} {\bibinfo {author} {\bibfnamefont {T.}~\bibnamefont
  {Prosen}},\ }\href {\doibase 10.1103/PhysRevLett.107.137201} {\bibfield
  {journal} {\bibinfo  {journal} {Phys. Rev. Lett.}\ }\textbf {\bibinfo
  {volume} {107}},\ \bibinfo {pages} {137201} (\bibinfo {year}
  {2011})}\BibitemShut {NoStop}%
\bibitem [{\citenamefont {{\v{Z}}nidari{\v{c}}}\ \emph
  {et~al.}(2011)\citenamefont {{\v{Z}}nidari{\v{c}}}, \citenamefont
  {{\v{Z}}unkovi{\v{c}}},\ and\ \citenamefont
  {Prosen}}]{vznidarivc2011transport}%
  \BibitemOpen
  \bibfield  {author} {\bibinfo {author} {\bibfnamefont {M.}~\bibnamefont
  {{\v{Z}}nidari{\v{c}}}}, \bibinfo {author} {\bibfnamefont {B.}~\bibnamefont
  {{\v{Z}}unkovi{\v{c}}}}, \ and\ \bibinfo {author} {\bibfnamefont
  {T.}~\bibnamefont {Prosen}},\ }\href {\doibase 10.1103/PhysRevE.84.051115}
  {\bibfield  {journal} {\bibinfo  {journal} {Phys. Rev. E}\ }\textbf {\bibinfo
  {volume} {84}},\ \bibinfo {pages} {051115} (\bibinfo {year}
  {2011})}\BibitemShut {NoStop}%
\bibitem [{\citenamefont {Ilievski}\ and\ \citenamefont
  {Prosen}(2014)}]{ILIEVSKI2014}%
  \BibitemOpen
  \bibfield  {author} {\bibinfo {author} {\bibfnamefont {E.}~\bibnamefont
  {Ilievski}}\ and\ \bibinfo {author} {\bibfnamefont {T.}~\bibnamefont
  {Prosen}},\ }\href {\doibase https://doi.org/10.1016/j.nuclphysb.2014.03.016}
  {\bibfield  {journal} {\bibinfo  {journal} {Nucl. Phys. B}\ }\textbf
  {\bibinfo {volume} {882}},\ \bibinfo {pages} {485 } (\bibinfo {year}
  {2014})}\BibitemShut {NoStop}%
\bibitem [{\citenamefont {Carollo}\ \emph {et~al.}(2017)\citenamefont
  {Carollo}, \citenamefont {Garrahan}, \citenamefont {Lesanovsky},\ and\
  \citenamefont {P\'erez-Espigares}}]{Carollo2017fluctuating}%
  \BibitemOpen
  \bibfield  {author} {\bibinfo {author} {\bibfnamefont {F.}~\bibnamefont
  {Carollo}}, \bibinfo {author} {\bibfnamefont {J.~P.}\ \bibnamefont
  {Garrahan}}, \bibinfo {author} {\bibfnamefont {I.}~\bibnamefont
  {Lesanovsky}}, \ and\ \bibinfo {author} {\bibfnamefont {C.}~\bibnamefont
  {P\'erez-Espigares}},\ }\href {\doibase 10.1103/PhysRevE.96.052118}
  {\bibfield  {journal} {\bibinfo  {journal} {Phys. Rev. E}\ }\textbf {\bibinfo
  {volume} {96}},\ \bibinfo {pages} {052118} (\bibinfo {year}
  {2017})}\BibitemShut {NoStop}%
\bibitem [{\citenamefont {Carollo}\ \emph {et~al.}(2018)\citenamefont
  {Carollo}, \citenamefont {Garrahan},\ and\ \citenamefont
  {Lesanovsky}}]{Carollo2018fluctuations}%
  \BibitemOpen
  \bibfield  {author} {\bibinfo {author} {\bibfnamefont {F.}~\bibnamefont
  {Carollo}}, \bibinfo {author} {\bibfnamefont {J.~P.}\ \bibnamefont
  {Garrahan}}, \ and\ \bibinfo {author} {\bibfnamefont {I.}~\bibnamefont
  {Lesanovsky}},\ }\href {\doibase 10.1103/PhysRevB.98.094301} {\bibfield
  {journal} {\bibinfo  {journal} {Phys. Rev. B}\ }\textbf {\bibinfo {volume}
  {98}},\ \bibinfo {pages} {094301} (\bibinfo {year} {2018})}\BibitemShut
  {NoStop}%
\bibitem [{\citenamefont {Schwab}\ \emph {et~al.}(2000)\citenamefont {Schwab},
  \citenamefont {Henriksen}, \citenamefont {Worlock},\ and\ \citenamefont
  {Roukes}}]{schwab2000measurement}%
  \BibitemOpen
  \bibfield  {author} {\bibinfo {author} {\bibfnamefont {K.}~\bibnamefont
  {Schwab}}, \bibinfo {author} {\bibfnamefont {E.}~\bibnamefont {Henriksen}},
  \bibinfo {author} {\bibfnamefont {J.}~\bibnamefont {Worlock}}, \ and\
  \bibinfo {author} {\bibfnamefont {M.~L.}\ \bibnamefont {Roukes}},\ }\href
  {\doibase https://doi.org/10.1038/35010065} {\bibfield  {journal} {\bibinfo
  {journal} {Nature}\ }\textbf {\bibinfo {volume} {404}},\ \bibinfo {pages}
  {974} (\bibinfo {year} {2000})}\BibitemShut {NoStop}%
\bibitem [{\citenamefont {Jezouin}\ \emph {et~al.}(2013)\citenamefont
  {Jezouin}, \citenamefont {Parmentier}, \citenamefont {Anthore}, \citenamefont
  {Gennser}, \citenamefont {Cavanna}, \citenamefont {Jin},\ and\ \citenamefont
  {Pierre}}]{jezouin2013quantum}%
  \BibitemOpen
  \bibfield  {author} {\bibinfo {author} {\bibfnamefont {S.}~\bibnamefont
  {Jezouin}}, \bibinfo {author} {\bibfnamefont {F.}~\bibnamefont {Parmentier}},
  \bibinfo {author} {\bibfnamefont {A.}~\bibnamefont {Anthore}}, \bibinfo
  {author} {\bibfnamefont {U.}~\bibnamefont {Gennser}}, \bibinfo {author}
  {\bibfnamefont {A.}~\bibnamefont {Cavanna}}, \bibinfo {author} {\bibfnamefont
  {Y.}~\bibnamefont {Jin}}, \ and\ \bibinfo {author} {\bibfnamefont
  {F.}~\bibnamefont {Pierre}},\ }\href {\doibase
  https://doi.org/10.1126/science.1241912} {\bibfield  {journal} {\bibinfo
  {journal} {Science}\ }\textbf {\bibinfo {volume} {342}},\ \bibinfo {pages}
  {601} (\bibinfo {year} {2013})}\BibitemShut {NoStop}%
\bibitem [{\citenamefont {Spohn}\ and\ \citenamefont
  {Lebowitz}(1977)}]{spohn1977stationary}%
  \BibitemOpen
  \bibfield  {author} {\bibinfo {author} {\bibfnamefont {H.}~\bibnamefont
  {Spohn}}\ and\ \bibinfo {author} {\bibfnamefont {J.~L.}\ \bibnamefont
  {Lebowitz}},\ }\href {\doibase https://doi.org/10.1007/BF01614132} {\bibfield
   {journal} {\bibinfo  {journal} {Commun. Math. Phys.}\ }\textbf {\bibinfo
  {volume} {54}},\ \bibinfo {pages} {97} (\bibinfo {year} {1977})}\BibitemShut
  {NoStop}%
\bibitem [{\citenamefont {Vasseur}\ and\ \citenamefont
  {Moore}(2016)}]{vasseur2016nonequilibrium}%
  \BibitemOpen
  \bibfield  {author} {\bibinfo {author} {\bibfnamefont {R.}~\bibnamefont
  {Vasseur}}\ and\ \bibinfo {author} {\bibfnamefont {J.~E.}\ \bibnamefont
  {Moore}},\ }\href {\doibase https://doi.org/10.1088/1742-5468/2016/06/064010}
  {\bibfield  {journal} {\bibinfo  {journal} {J. Stat. Mech.: Theory Exp.}\
  }\textbf {\bibinfo {volume} {2016}},\ \bibinfo {pages} {064010} (\bibinfo
  {year} {2016})}\BibitemShut {NoStop}%
\bibitem [{\citenamefont {Bernard}\ and\ \citenamefont
  {Doyon}(2016)}]{bernard2016conformal}%
  \BibitemOpen
  \bibfield  {author} {\bibinfo {author} {\bibfnamefont {D.}~\bibnamefont
  {Bernard}}\ and\ \bibinfo {author} {\bibfnamefont {B.}~\bibnamefont
  {Doyon}},\ }\href {\doibase https://doi.org/10.1088/1742-5468/2016/06/064005}
  {\bibfield  {journal} {\bibinfo  {journal} {J. Stat. Mech.: Theory Exp.}\
  }\textbf {\bibinfo {volume} {2016}},\ \bibinfo {pages} {064005} (\bibinfo
  {year} {2016})}\BibitemShut {NoStop}%
\bibitem [{\citenamefont {Bernard}\ and\ \citenamefont
  {Doyon}(2012{\natexlab{a}})}]{bernard2012energy}%
  \BibitemOpen
  \bibfield  {author} {\bibinfo {author} {\bibfnamefont {D.}~\bibnamefont
  {Bernard}}\ and\ \bibinfo {author} {\bibfnamefont {B.}~\bibnamefont
  {Doyon}},\ }\href {\doibase https://doi.org/10.1088/1751-8113/45/36/362001}
  {\bibfield  {journal} {\bibinfo  {journal} {J. Phys. A: Math. Theor.}\
  }\textbf {\bibinfo {volume} {45}},\ \bibinfo {pages} {362001} (\bibinfo
  {year} {2012}{\natexlab{a}})}\BibitemShut {NoStop}%
\bibitem [{\citenamefont {Bernard}\ and\ \citenamefont
  {Doyon}(2015)}]{bernard2015non}%
  \BibitemOpen
  \bibfield  {author} {\bibinfo {author} {\bibfnamefont {D.}~\bibnamefont
  {Bernard}}\ and\ \bibinfo {author} {\bibfnamefont {B.}~\bibnamefont
  {Doyon}},\ }\href {\doibase https://doi.org/10.1007/s00023-014-0314-8}
  {\bibfield  {journal} {\bibinfo  {journal} {Ann. Henri Poincar{\'e}}\
  }\textbf {\bibinfo {volume} {16}},\ \bibinfo {pages} {113} (\bibinfo {year}
  {2015})}\BibitemShut {NoStop}%
\bibitem [{\citenamefont {Bernard}\ and\ \citenamefont
  {Doyon}(2013)}]{bernard2013time}%
  \BibitemOpen
  \bibfield  {author} {\bibinfo {author} {\bibfnamefont {D.}~\bibnamefont
  {Bernard}}\ and\ \bibinfo {author} {\bibfnamefont {B.}~\bibnamefont
  {Doyon}},\ }\href {\doibase https://doi.org/10.1088/1751-8113/46/37/372001}
  {\bibfield  {journal} {\bibinfo  {journal} {J. Phys. A: Math. Theor.}\
  }\textbf {\bibinfo {volume} {46}},\ \bibinfo {pages} {372001} (\bibinfo
  {year} {2013})}\BibitemShut {NoStop}%
\bibitem [{\citenamefont {De~Luca}\ \emph {et~al.}(2013)\citenamefont
  {De~Luca}, \citenamefont {Viti}, \citenamefont {Bernard},\ and\ \citenamefont
  {Doyon}}]{de2013nonequilibrium}%
  \BibitemOpen
  \bibfield  {author} {\bibinfo {author} {\bibfnamefont {A.}~\bibnamefont
  {De~Luca}}, \bibinfo {author} {\bibfnamefont {J.}~\bibnamefont {Viti}},
  \bibinfo {author} {\bibfnamefont {D.}~\bibnamefont {Bernard}}, \ and\
  \bibinfo {author} {\bibfnamefont {B.}~\bibnamefont {Doyon}},\ }\href
  {\doibase 10.1103/PhysRevB.88.134301} {\bibfield  {journal} {\bibinfo
  {journal} {Phys. Rev. B}\ }\textbf {\bibinfo {volume} {88}},\ \bibinfo
  {pages} {134301} (\bibinfo {year} {2013})}\BibitemShut {NoStop}%
\bibitem [{\citenamefont {De~Luca}\ \emph {et~al.}(2015)\citenamefont
  {De~Luca}, \citenamefont {Martelloni},\ and\ \citenamefont
  {Viti}}]{de2015stationary}%
  \BibitemOpen
  \bibfield  {author} {\bibinfo {author} {\bibfnamefont {A.}~\bibnamefont
  {De~Luca}}, \bibinfo {author} {\bibfnamefont {G.}~\bibnamefont {Martelloni}},
  \ and\ \bibinfo {author} {\bibfnamefont {J.}~\bibnamefont {Viti}},\ }\href
  {\doibase 10.1103/PhysRevA.91.021603} {\bibfield  {journal} {\bibinfo
  {journal} {Phys. Rev. A}\ }\textbf {\bibinfo {volume} {91}},\ \bibinfo
  {pages} {021603} (\bibinfo {year} {2015})}\BibitemShut {NoStop}%
\bibitem [{\citenamefont {Doyon}\ \emph {et~al.}(2015)\citenamefont {Doyon},
  \citenamefont {Lucas}, \citenamefont {Schalm},\ and\ \citenamefont
  {Bhaseen}}]{doyon2015non}%
  \BibitemOpen
  \bibfield  {author} {\bibinfo {author} {\bibfnamefont {B.}~\bibnamefont
  {Doyon}}, \bibinfo {author} {\bibfnamefont {A.}~\bibnamefont {Lucas}},
  \bibinfo {author} {\bibfnamefont {K.}~\bibnamefont {Schalm}}, \ and\ \bibinfo
  {author} {\bibfnamefont {M.}~\bibnamefont {Bhaseen}},\ }\href {\doibase
  https://doi.org/10.1088/1751-8113/48/9/095002} {\bibfield  {journal}
  {\bibinfo  {journal} {J. Phys. A: Math. Theor.}\ }\textbf {\bibinfo {volume}
  {48}},\ \bibinfo {pages} {095002} (\bibinfo {year} {2015})}\BibitemShut
  {NoStop}%
\bibitem [{\citenamefont {Antal}\ \emph {et~al.}(1999)\citenamefont {Antal},
  \citenamefont {R{\'a}cz}, \citenamefont {R{\'a}kos},\ and\ \citenamefont
  {Sch{\"u}tz}}]{antal1999transport}%
  \BibitemOpen
  \bibfield  {author} {\bibinfo {author} {\bibfnamefont {T.}~\bibnamefont
  {Antal}}, \bibinfo {author} {\bibfnamefont {Z.}~\bibnamefont {R{\'a}cz}},
  \bibinfo {author} {\bibfnamefont {A.}~\bibnamefont {R{\'a}kos}}, \ and\
  \bibinfo {author} {\bibfnamefont {G.}~\bibnamefont {Sch{\"u}tz}},\ }\href
  {\doibase 10.1103/PhysRevE.59.4912} {\bibfield  {journal} {\bibinfo
  {journal} {Phys. Rev. E}\ }\textbf {\bibinfo {volume} {59}},\ \bibinfo
  {pages} {4912} (\bibinfo {year} {1999})}\BibitemShut {NoStop}%
\bibitem [{\citenamefont {Karevski}(2002)}]{karevski2002scaling}%
  \BibitemOpen
  \bibfield  {author} {\bibinfo {author} {\bibfnamefont {D.}~\bibnamefont
  {Karevski}},\ }\href {\doibase https://doi.org/10.1140/epjb/e20020139}
  {\bibfield  {journal} {\bibinfo  {journal} {Eur. Phys. J. B}\ }\textbf
  {\bibinfo {volume} {27}},\ \bibinfo {pages} {147} (\bibinfo {year}
  {2002})}\BibitemShut {NoStop}%
\bibitem [{\citenamefont {Platini}\ and\ \citenamefont
  {Karevski}(2005)}]{platini2005scaling}%
  \BibitemOpen
  \bibfield  {author} {\bibinfo {author} {\bibfnamefont {T.}~\bibnamefont
  {Platini}}\ and\ \bibinfo {author} {\bibfnamefont {D.}~\bibnamefont
  {Karevski}},\ }\href {\doibase https://doi.org/10.1140/epjb/e2005-00402-2}
  {\bibfield  {journal} {\bibinfo  {journal} {Eur. Phys. J. B}\ }\textbf
  {\bibinfo {volume} {48}},\ \bibinfo {pages} {225} (\bibinfo {year}
  {2005})}\BibitemShut {NoStop}%
\bibitem [{\citenamefont {Collura}\ and\ \citenamefont
  {Martelloni}(2014)}]{collura2014non}%
  \BibitemOpen
  \bibfield  {author} {\bibinfo {author} {\bibfnamefont {M.}~\bibnamefont
  {Collura}}\ and\ \bibinfo {author} {\bibfnamefont {G.}~\bibnamefont
  {Martelloni}},\ }\href {\doibase
  https://doi.org/10.1088/1742-5468/2014/08/P08006} {\bibfield  {journal}
  {\bibinfo  {journal} {J. Stat. Mech.: Theory Exp.}\ }\textbf {\bibinfo
  {volume} {2014}},\ \bibinfo {pages} {P08006} (\bibinfo {year}
  {2014})}\BibitemShut {NoStop}%
\bibitem [{\citenamefont {Collura}\ and\ \citenamefont
  {Karevski}(2014)}]{collura2014quantum}%
  \BibitemOpen
  \bibfield  {author} {\bibinfo {author} {\bibfnamefont {M.}~\bibnamefont
  {Collura}}\ and\ \bibinfo {author} {\bibfnamefont {D.}~\bibnamefont
  {Karevski}},\ }\href {\doibase 10.1103/PhysRevB.89.214308} {\bibfield
  {journal} {\bibinfo  {journal} {Phys. Rev. B}\ }\textbf {\bibinfo {volume}
  {89}},\ \bibinfo {pages} {214308} (\bibinfo {year} {2014})}\BibitemShut
  {NoStop}%
\bibitem [{\citenamefont {Allegra}\ \emph {et~al.}(2016)\citenamefont
  {Allegra}, \citenamefont {Dubail}, \citenamefont {St{\'e}phan},\ and\
  \citenamefont {Viti}}]{allegra2016inhomogeneous}%
  \BibitemOpen
  \bibfield  {author} {\bibinfo {author} {\bibfnamefont {N.}~\bibnamefont
  {Allegra}}, \bibinfo {author} {\bibfnamefont {J.}~\bibnamefont {Dubail}},
  \bibinfo {author} {\bibfnamefont {J.-M.}\ \bibnamefont {St{\'e}phan}}, \ and\
  \bibinfo {author} {\bibfnamefont {J.}~\bibnamefont {Viti}},\ }\href {\doibase
  10.1088/1742-5468/2016/05/053108} {\bibfield  {journal} {\bibinfo  {journal}
  {J. Stat. Mech.: Theory Exp.}\ }\textbf {\bibinfo {volume} {2016}},\ \bibinfo
  {pages} {053108} (\bibinfo {year} {2016})}\BibitemShut {NoStop}%
\bibitem [{\citenamefont {Viti}\ \emph {et~al.}(2016)\citenamefont {Viti},
  \citenamefont {St{\'e}phan}, \citenamefont {Dubail},\ and\ \citenamefont
  {Haque}}]{viti2016inhomogeneous}%
  \BibitemOpen
  \bibfield  {author} {\bibinfo {author} {\bibfnamefont {J.}~\bibnamefont
  {Viti}}, \bibinfo {author} {\bibfnamefont {J.-M.}\ \bibnamefont
  {St{\'e}phan}}, \bibinfo {author} {\bibfnamefont {J.}~\bibnamefont {Dubail}},
  \ and\ \bibinfo {author} {\bibfnamefont {M.}~\bibnamefont {Haque}},\ }\href
  {\doibase https://doi.org/10.1209/0295-5075/115/40011} {\bibfield  {journal}
  {\bibinfo  {journal} {EPL}\ }\textbf {\bibinfo {volume} {115}},\ \bibinfo
  {pages} {40011} (\bibinfo {year} {2016})}\BibitemShut {NoStop}%
\bibitem [{\citenamefont {Bertini}\ and\ \citenamefont
  {Fagotti}(2016)}]{bertini2016determination}%
  \BibitemOpen
  \bibfield  {author} {\bibinfo {author} {\bibfnamefont {B.}~\bibnamefont
  {Bertini}}\ and\ \bibinfo {author} {\bibfnamefont {M.}~\bibnamefont
  {Fagotti}},\ }\href {\doibase 10.1103/PhysRevLett.117.130402} {\bibfield
  {journal} {\bibinfo  {journal} {Phys. Rev. Lett.}\ }\textbf {\bibinfo
  {volume} {117}},\ \bibinfo {pages} {130402} (\bibinfo {year}
  {2016})}\BibitemShut {NoStop}%
\bibitem [{\citenamefont {Eisler}\ \emph {et~al.}(2016)\citenamefont {Eisler},
  \citenamefont {Maislinger},\ and\ \citenamefont
  {Evertz}}]{eisler2016universal}%
  \BibitemOpen
  \bibfield  {author} {\bibinfo {author} {\bibfnamefont {V.}~\bibnamefont
  {Eisler}}, \bibinfo {author} {\bibfnamefont {F.}~\bibnamefont {Maislinger}},
  \ and\ \bibinfo {author} {\bibfnamefont {H.~G.}\ \bibnamefont {Evertz}},\
  }\href {\doibase 10.21468/SciPostPhys.1.2.014} {\bibfield  {journal}
  {\bibinfo  {journal} {SciPost Phys.}\ }\textbf {\bibinfo {volume} {1}},\
  \bibinfo {pages} {014} (\bibinfo {year} {2016})}\BibitemShut {NoStop}%
\bibitem [{\citenamefont {Perfetto}\ and\ \citenamefont
  {Gambassi}(2017)}]{perfetto2017ballistic}%
  \BibitemOpen
  \bibfield  {author} {\bibinfo {author} {\bibfnamefont {G.}~\bibnamefont
  {Perfetto}}\ and\ \bibinfo {author} {\bibfnamefont {A.}~\bibnamefont
  {Gambassi}},\ }\href {\doibase 10.1103/PhysRevE.96.012138} {\bibfield
  {journal} {\bibinfo  {journal} {Phys. Rev. E}\ }\textbf {\bibinfo {volume}
  {96}},\ \bibinfo {pages} {012138} (\bibinfo {year} {2017})}\BibitemShut
  {NoStop}%
\bibitem [{\citenamefont {Kormos}(2017)}]{kormos2017inhomogeneous}%
  \BibitemOpen
  \bibfield  {author} {\bibinfo {author} {\bibfnamefont {M.}~\bibnamefont
  {Kormos}},\ }\href {\doibase 10.21468/SciPostPhys.3.3.020} {\bibfield
  {journal} {\bibinfo  {journal} {SciPost Phys.}\ }\textbf {\bibinfo {volume}
  {3}},\ \bibinfo {pages} {020} (\bibinfo {year} {2017})}\BibitemShut {NoStop}%
\bibitem [{\citenamefont {Ljubotina}\ \emph {et~al.}(2019)\citenamefont
  {Ljubotina}, \citenamefont {Sotiriadis},\ and\ \citenamefont
  {Prosen}}]{10.21468/SciPostPhys.6.1.004}%
  \BibitemOpen
  \bibfield  {author} {\bibinfo {author} {\bibfnamefont {M.}~\bibnamefont
  {Ljubotina}}, \bibinfo {author} {\bibfnamefont {S.}~\bibnamefont
  {Sotiriadis}}, \ and\ \bibinfo {author} {\bibfnamefont {T.}~\bibnamefont
  {Prosen}},\ }\href {\doibase 10.21468/SciPostPhys.6.1.004} {\bibfield
  {journal} {\bibinfo  {journal} {SciPost Phys.}\ }\textbf {\bibinfo {volume}
  {6}},\ \bibinfo {pages} {4} (\bibinfo {year} {2019})}\BibitemShut {NoStop}%
\bibitem [{\citenamefont {Mitra}(2018)}]{mitra2018quantum}%
  \BibitemOpen
  \bibfield  {author} {\bibinfo {author} {\bibfnamefont {A.}~\bibnamefont
  {Mitra}},\ }\href {\doibase
  https://doi.org/10.1146/annurev-conmatphys-031016-025451} {\bibfield
  {journal} {\bibinfo  {journal} {Annu. Rev. Condens. Matter Phys.}\ }\textbf
  {\bibinfo {volume} {9}},\ \bibinfo {pages} {245} (\bibinfo {year}
  {2018})}\BibitemShut {NoStop}%
\bibitem [{\citenamefont {Castro-Alvaredo}\ \emph {et~al.}(2016)\citenamefont
  {Castro-Alvaredo}, \citenamefont {Doyon},\ and\ \citenamefont
  {Yoshimura}}]{castro2016emergent}%
  \BibitemOpen
  \bibfield  {author} {\bibinfo {author} {\bibfnamefont {O.~A.}\ \bibnamefont
  {Castro-Alvaredo}}, \bibinfo {author} {\bibfnamefont {B.}~\bibnamefont
  {Doyon}}, \ and\ \bibinfo {author} {\bibfnamefont {T.}~\bibnamefont
  {Yoshimura}},\ }\href {\doibase 10.1103/PhysRevX.6.041065} {\bibfield
  {journal} {\bibinfo  {journal} {Phys. Rev. X}\ }\textbf {\bibinfo {volume}
  {6}},\ \bibinfo {pages} {041065} (\bibinfo {year} {2016})}\BibitemShut
  {NoStop}%
\bibitem [{\citenamefont {Bertini}\ \emph {et~al.}(2016)\citenamefont
  {Bertini}, \citenamefont {Collura}, \citenamefont {De~Nardis},\ and\
  \citenamefont {Fagotti}}]{bertini2016transport}%
  \BibitemOpen
  \bibfield  {author} {\bibinfo {author} {\bibfnamefont {B.}~\bibnamefont
  {Bertini}}, \bibinfo {author} {\bibfnamefont {M.}~\bibnamefont {Collura}},
  \bibinfo {author} {\bibfnamefont {J.}~\bibnamefont {De~Nardis}}, \ and\
  \bibinfo {author} {\bibfnamefont {M.}~\bibnamefont {Fagotti}},\ }\href
  {\doibase 10.1103/PhysRevLett.117.207201} {\bibfield  {journal} {\bibinfo
  {journal} {Phys. Rev. Lett.}\ }\textbf {\bibinfo {volume} {117}},\ \bibinfo
  {pages} {207201} (\bibinfo {year} {2016})}\BibitemShut {NoStop}%
\bibitem [{\citenamefont {Doyon}(2020)}]{doyon2019lecture}%
  \BibitemOpen
  \bibfield  {author} {\bibinfo {author} {\bibfnamefont {B.}~\bibnamefont
  {Doyon}},\ }\href {\doibase 10.21468/SciPostPhysLectNotes.18} {\bibfield
  {journal} {\bibinfo  {journal} {SciPost Phys. Lect. Notes}\ ,\ \bibinfo
  {pages} {18}} (\bibinfo {year} {2020})}\BibitemShut {NoStop}%
\bibitem [{\citenamefont {Fagotti}(2016)}]{fagotti2016charges}%
  \BibitemOpen
  \bibfield  {author} {\bibinfo {author} {\bibfnamefont {M.}~\bibnamefont
  {Fagotti}},\ }\href {\doibase 10.1088/1751-8121/50/3/034005} {\bibfield
  {journal} {\bibinfo  {journal} {J. Phys. A: Math. Theor.}\ }\textbf {\bibinfo
  {volume} {50}},\ \bibinfo {pages} {034005} (\bibinfo {year}
  {2016})}\BibitemShut {NoStop}%
\bibitem [{\citenamefont {Doyon}\ and\ \citenamefont
  {Spohn}(2017)}]{doyon2017dynamics}%
  \BibitemOpen
  \bibfield  {author} {\bibinfo {author} {\bibfnamefont {B.}~\bibnamefont
  {Doyon}}\ and\ \bibinfo {author} {\bibfnamefont {H.}~\bibnamefont {Spohn}},\
  }\href {\doibase 10.1088/1742-5468/aa7abf} {\bibfield  {journal} {\bibinfo
  {journal} {J. Stat. Mech.: Theory Exp.}\ }\textbf {\bibinfo {volume}
  {2017}},\ \bibinfo {pages} {073210} (\bibinfo {year} {2017})}\BibitemShut
  {NoStop}%
\bibitem [{\citenamefont {Piroli}\ \emph {et~al.}(2017)\citenamefont {Piroli},
  \citenamefont {De~Nardis}, \citenamefont {Collura}, \citenamefont {Bertini},\
  and\ \citenamefont {Fagotti}}]{piroli2017transport}%
  \BibitemOpen
  \bibfield  {author} {\bibinfo {author} {\bibfnamefont {L.}~\bibnamefont
  {Piroli}}, \bibinfo {author} {\bibfnamefont {J.}~\bibnamefont {De~Nardis}},
  \bibinfo {author} {\bibfnamefont {M.}~\bibnamefont {Collura}}, \bibinfo
  {author} {\bibfnamefont {B.}~\bibnamefont {Bertini}}, \ and\ \bibinfo
  {author} {\bibfnamefont {M.}~\bibnamefont {Fagotti}},\ }\href {\doibase
  10.1103/PhysRevB.96.115124} {\bibfield  {journal} {\bibinfo  {journal} {Phys.
  Rev. B}\ }\textbf {\bibinfo {volume} {96}},\ \bibinfo {pages} {115124}
  (\bibinfo {year} {2017})}\BibitemShut {NoStop}%
\bibitem [{\citenamefont {Bulchandani}\ \emph {et~al.}(2017)\citenamefont
  {Bulchandani}, \citenamefont {Vasseur}, \citenamefont {Karrasch},\ and\
  \citenamefont {Moore}}]{bulchandani2017solvable}%
  \BibitemOpen
  \bibfield  {author} {\bibinfo {author} {\bibfnamefont {V.~B.}\ \bibnamefont
  {Bulchandani}}, \bibinfo {author} {\bibfnamefont {R.}~\bibnamefont
  {Vasseur}}, \bibinfo {author} {\bibfnamefont {C.}~\bibnamefont {Karrasch}}, \
  and\ \bibinfo {author} {\bibfnamefont {J.~E.}\ \bibnamefont {Moore}},\ }\href
  {\doibase 10.1103/PhysRevLett.119.220604} {\bibfield  {journal} {\bibinfo
  {journal} {Phys. Rev. Lett.}\ }\textbf {\bibinfo {volume} {119}},\ \bibinfo
  {pages} {220604} (\bibinfo {year} {2017})}\BibitemShut {NoStop}%
\bibitem [{\citenamefont {Collura}\ \emph {et~al.}(2018)\citenamefont
  {Collura}, \citenamefont {De~Luca},\ and\ \citenamefont
  {Viti}}]{collura2018analytic}%
  \BibitemOpen
  \bibfield  {author} {\bibinfo {author} {\bibfnamefont {M.}~\bibnamefont
  {Collura}}, \bibinfo {author} {\bibfnamefont {A.}~\bibnamefont {De~Luca}}, \
  and\ \bibinfo {author} {\bibfnamefont {J.}~\bibnamefont {Viti}},\ }\href
  {\doibase 10.1103/PhysRevB.97.081111} {\bibfield  {journal} {\bibinfo
  {journal} {Phys. Rev. B}\ }\textbf {\bibinfo {volume} {97}},\ \bibinfo
  {pages} {081111} (\bibinfo {year} {2018})}\BibitemShut {NoStop}%
\bibitem [{\citenamefont {Bastianello}\ \emph
  {et~al.}(2018{\natexlab{a}})\citenamefont {Bastianello}, \citenamefont
  {Doyon}, \citenamefont {Watts},\ and\ \citenamefont
  {Yoshimura}}]{AlviseGHD2018}%
  \BibitemOpen
  \bibfield  {author} {\bibinfo {author} {\bibfnamefont {A.}~\bibnamefont
  {Bastianello}}, \bibinfo {author} {\bibfnamefont {B.}~\bibnamefont {Doyon}},
  \bibinfo {author} {\bibfnamefont {G.}~\bibnamefont {Watts}}, \ and\ \bibinfo
  {author} {\bibfnamefont {T.}~\bibnamefont {Yoshimura}},\ }\href {\doibase
  10.21468/SciPostPhys.4.6.045} {\bibfield  {journal} {\bibinfo  {journal}
  {SciPost Phys.}\ }\textbf {\bibinfo {volume} {4}},\ \bibinfo {pages} {45}
  (\bibinfo {year} {2018}{\natexlab{a}})}\BibitemShut {NoStop}%
\bibitem [{\citenamefont {De~Nardis}\ \emph {et~al.}(2018)\citenamefont
  {De~Nardis}, \citenamefont {Bernard},\ and\ \citenamefont
  {Doyon}}]{DeNardis2018}%
  \BibitemOpen
  \bibfield  {author} {\bibinfo {author} {\bibfnamefont {J.}~\bibnamefont
  {De~Nardis}}, \bibinfo {author} {\bibfnamefont {D.}~\bibnamefont {Bernard}},
  \ and\ \bibinfo {author} {\bibfnamefont {B.}~\bibnamefont {Doyon}},\ }\href
  {\doibase 10.1103/PhysRevLett.121.160603} {\bibfield  {journal} {\bibinfo
  {journal} {Phys. Rev. Lett.}\ }\textbf {\bibinfo {volume} {121}},\ \bibinfo
  {pages} {160603} (\bibinfo {year} {2018})}\BibitemShut {NoStop}%
\bibitem [{\citenamefont {Bertini}\ \emph {et~al.}(2019)\citenamefont
  {Bertini}, \citenamefont {Piroli},\ and\ \citenamefont
  {Kormos}}]{kormossemiclassic2}%
  \BibitemOpen
  \bibfield  {author} {\bibinfo {author} {\bibfnamefont {B.}~\bibnamefont
  {Bertini}}, \bibinfo {author} {\bibfnamefont {L.}~\bibnamefont {Piroli}}, \
  and\ \bibinfo {author} {\bibfnamefont {M.}~\bibnamefont {Kormos}},\ }\href
  {\doibase 10.1103/PhysRevB.100.035108} {\bibfield  {journal} {\bibinfo
  {journal} {Phys. Rev. B}\ }\textbf {\bibinfo {volume} {100}},\ \bibinfo
  {pages} {035108} (\bibinfo {year} {2019})}\BibitemShut {NoStop}%
\bibitem [{\citenamefont {Doyon}(2018)}]{DoyonLargeScaleCorrelations2018}%
  \BibitemOpen
  \bibfield  {author} {\bibinfo {author} {\bibfnamefont {B.}~\bibnamefont
  {Doyon}},\ }\href {\doibase 10.21468/SciPostPhys.5.5.054} {\bibfield
  {journal} {\bibinfo  {journal} {SciPost Phys.}\ }\textbf {\bibinfo {volume}
  {5}},\ \bibinfo {pages} {54} (\bibinfo {year} {2018})}\BibitemShut {NoStop}%
\bibitem [{\citenamefont {Bertini}\ \emph {et~al.}(2018)\citenamefont
  {Bertini}, \citenamefont {Fagotti}, \citenamefont {Piroli},\ and\
  \citenamefont {Calabrese}}]{bertini2018entanglement}%
  \BibitemOpen
  \bibfield  {author} {\bibinfo {author} {\bibfnamefont {B.}~\bibnamefont
  {Bertini}}, \bibinfo {author} {\bibfnamefont {M.}~\bibnamefont {Fagotti}},
  \bibinfo {author} {\bibfnamefont {L.}~\bibnamefont {Piroli}}, \ and\ \bibinfo
  {author} {\bibfnamefont {P.}~\bibnamefont {Calabrese}},\ }\href {\doibase
  10.1088/1751-8121/aad82e} {\bibfield  {journal} {\bibinfo  {journal} {J.
  Phys. A: Math. Theor.}\ }\textbf {\bibinfo {volume} {51}},\ \bibinfo {pages}
  {39LT01} (\bibinfo {year} {2018})}\BibitemShut {NoStop}%
\bibitem [{\citenamefont {Alba}\ \emph {et~al.}(2019)\citenamefont {Alba},
  \citenamefont {Bertini},\ and\ \citenamefont {Fagotti}}]{EeEvolutionGHD2019}%
  \BibitemOpen
  \bibfield  {author} {\bibinfo {author} {\bibfnamefont {V.}~\bibnamefont
  {Alba}}, \bibinfo {author} {\bibfnamefont {B.}~\bibnamefont {Bertini}}, \
  and\ \bibinfo {author} {\bibfnamefont {M.}~\bibnamefont {Fagotti}},\ }\href
  {\doibase 10.21468/SciPostPhys.7.1.005} {\bibfield  {journal} {\bibinfo
  {journal} {SciPost Phys.}\ }\textbf {\bibinfo {volume} {7}},\ \bibinfo
  {pages} {5} (\bibinfo {year} {2019})}\BibitemShut {NoStop}%
\bibitem [{\citenamefont {M{\o}ller}\ \emph {et~al.}(2020)\citenamefont
  {M{\o}ller}, \citenamefont {Perfetto}, \citenamefont {Doyon},\ and\
  \citenamefont {Schmiedmayer}}]{moller2020correlations}%
  \BibitemOpen
  \bibfield  {author} {\bibinfo {author} {\bibfnamefont {F.~S.}\ \bibnamefont
  {M{\o}ller}}, \bibinfo {author} {\bibfnamefont {G.}~\bibnamefont {Perfetto}},
  \bibinfo {author} {\bibfnamefont {B.}~\bibnamefont {Doyon}}, \ and\ \bibinfo
  {author} {\bibfnamefont {J.}~\bibnamefont {Schmiedmayer}},\ }\href
  {https://arxiv.org/abs/2007.00527} {\bibfield  {journal} {\bibinfo  {journal}
  {arXiv:2007.00527}\ } (\bibinfo {year} {2020})}\BibitemShut {NoStop}%
\bibitem [{\citenamefont {Touchette}(2009)}]{touchette2009large}%
  \BibitemOpen
  \bibfield  {author} {\bibinfo {author} {\bibfnamefont {H.}~\bibnamefont
  {Touchette}},\ }\href {\doibase
  https://doi.org/10.1016/j.physrep.2009.05.002} {\bibfield  {journal}
  {\bibinfo  {journal} {Phys. Rep.}\ }\textbf {\bibinfo {volume} {478}},\
  \bibinfo {pages} {1} (\bibinfo {year} {2009})}\BibitemShut {NoStop}%
\bibitem [{\citenamefont {Lamacraft}\ and\ \citenamefont
  {Fendley}(2008)}]{lamacraft2008order}%
  \BibitemOpen
  \bibfield  {author} {\bibinfo {author} {\bibfnamefont {A.}~\bibnamefont
  {Lamacraft}}\ and\ \bibinfo {author} {\bibfnamefont {P.}~\bibnamefont
  {Fendley}},\ }\href {\doibase 10.1103/PhysRevLett.100.165706} {\bibfield
  {journal} {\bibinfo  {journal} {Phys. Rev. Lett.}\ }\textbf {\bibinfo
  {volume} {100}},\ \bibinfo {pages} {165706} (\bibinfo {year}
  {2008})}\BibitemShut {NoStop}%
\bibitem [{\citenamefont {Eisler}\ and\ \citenamefont
  {R{\'a}cz}(2013)}]{eisler2013full}%
  \BibitemOpen
  \bibfield  {author} {\bibinfo {author} {\bibfnamefont {V.}~\bibnamefont
  {Eisler}}\ and\ \bibinfo {author} {\bibfnamefont {Z.}~\bibnamefont
  {R{\'a}cz}},\ }\href {\doibase 10.1103/PhysRevLett.110.060602} {\bibfield
  {journal} {\bibinfo  {journal} {Phys. Rev. Lett.}\ }\textbf {\bibinfo
  {volume} {110}},\ \bibinfo {pages} {060602} (\bibinfo {year}
  {2013})}\BibitemShut {NoStop}%
\bibitem [{\citenamefont {Groha}\ \emph {et~al.}(2018)\citenamefont {Groha},
  \citenamefont {Essler},\ and\ \citenamefont {Calabrese}}]{groha2018full}%
  \BibitemOpen
  \bibfield  {author} {\bibinfo {author} {\bibfnamefont {S.}~\bibnamefont
  {Groha}}, \bibinfo {author} {\bibfnamefont {F.~H.~L.}\ \bibnamefont
  {Essler}}, \ and\ \bibinfo {author} {\bibfnamefont {P.}~\bibnamefont
  {Calabrese}},\ }\href {\doibase 10.21468/SciPostPhys.4.6.043} {\bibfield
  {journal} {\bibinfo  {journal} {SciPost Phys.}\ }\textbf {\bibinfo {volume}
  {4}},\ \bibinfo {pages} {43} (\bibinfo {year} {2018})}\BibitemShut {NoStop}%
\bibitem [{\citenamefont {Collura}\ and\ \citenamefont
  {Essler}(2020)}]{essler2020order}%
  \BibitemOpen
  \bibfield  {author} {\bibinfo {author} {\bibfnamefont {M.}~\bibnamefont
  {Collura}}\ and\ \bibinfo {author} {\bibfnamefont {F.~H.~L.}\ \bibnamefont
  {Essler}},\ }\href {\doibase 10.1103/PhysRevB.101.041110} {\bibfield
  {journal} {\bibinfo  {journal} {Phys. Rev. B}\ }\textbf {\bibinfo {volume}
  {101}},\ \bibinfo {pages} {041110} (\bibinfo {year} {2020})}\BibitemShut
  {NoStop}%
\bibitem [{\citenamefont {Collura}(2019)}]{Collura2020}%
  \BibitemOpen
  \bibfield  {author} {\bibinfo {author} {\bibfnamefont {M.}~\bibnamefont
  {Collura}},\ }\href {\doibase 10.21468/SciPostPhys.7.6.072} {\bibfield
  {journal} {\bibinfo  {journal} {SciPost Phys.}\ }\textbf {\bibinfo {volume}
  {7}},\ \bibinfo {pages} {72} (\bibinfo {year} {2019})}\BibitemShut {NoStop}%
\bibitem [{\citenamefont {Calabrese}\ \emph {et~al.}(2020)\citenamefont
  {Calabrese}, \citenamefont {Collura}, \citenamefont {Di~Giulio},\ and\
  \citenamefont {Murciano}}]{calabrese2020full}%
  \BibitemOpen
  \bibfield  {author} {\bibinfo {author} {\bibfnamefont {P.}~\bibnamefont
  {Calabrese}}, \bibinfo {author} {\bibfnamefont {M.}~\bibnamefont {Collura}},
  \bibinfo {author} {\bibfnamefont {G.}~\bibnamefont {Di~Giulio}}, \ and\
  \bibinfo {author} {\bibfnamefont {S.}~\bibnamefont {Murciano}},\ }\href
  {https://arxiv.org/abs/2002.04367} {\bibfield  {journal} {\bibinfo  {journal}
  {arXiv:2002.04367}\ } (\bibinfo {year} {2020})}\BibitemShut {NoStop}%
\bibitem [{\citenamefont {Tortora}\ \emph {et~al.}(2020)\citenamefont
  {Tortora}, \citenamefont {Calabrese},\ and\ \citenamefont
  {Collura}}]{tortora2020relaxation}%
  \BibitemOpen
  \bibfield  {author} {\bibinfo {author} {\bibfnamefont {R.~J.~V.}\
  \bibnamefont {Tortora}}, \bibinfo {author} {\bibfnamefont {P.}~\bibnamefont
  {Calabrese}}, \ and\ \bibinfo {author} {\bibfnamefont {M.}~\bibnamefont
  {Collura}},\ }\href {https://arxiv.org/abs/2005.01679} {\bibfield  {journal}
  {\bibinfo  {journal} {arXiv:2005.01679}\ } (\bibinfo {year}
  {2020})}\BibitemShut {NoStop}%
\bibitem [{\citenamefont {Armijo}\ \emph {et~al.}(2010)\citenamefont {Armijo},
  \citenamefont {Jacqmin}, \citenamefont {Kheruntsyan},\ and\ \citenamefont
  {Bouchoule}}]{armijo2010probing}%
  \BibitemOpen
  \bibfield  {author} {\bibinfo {author} {\bibfnamefont {J.}~\bibnamefont
  {Armijo}}, \bibinfo {author} {\bibfnamefont {T.}~\bibnamefont {Jacqmin}},
  \bibinfo {author} {\bibfnamefont {K.}~\bibnamefont {Kheruntsyan}}, \ and\
  \bibinfo {author} {\bibfnamefont {I.}~\bibnamefont {Bouchoule}},\ }\href
  {\doibase 10.1103/PhysRevLett.105.230402} {\bibfield  {journal} {\bibinfo
  {journal} {Phys. Rev. Lett.}\ }\textbf {\bibinfo {volume} {105}},\ \bibinfo
  {pages} {230402} (\bibinfo {year} {2010})}\BibitemShut {NoStop}%
\bibitem [{\citenamefont {Pietraszewicz}\ and\ \citenamefont
  {Deuar}(2017)}]{pietraszewicz2017mesoscopic}%
  \BibitemOpen
  \bibfield  {author} {\bibinfo {author} {\bibfnamefont {J.}~\bibnamefont
  {Pietraszewicz}}\ and\ \bibinfo {author} {\bibfnamefont {P.}~\bibnamefont
  {Deuar}},\ }\href {\doibase 10.1088/1367-2630/aa91c5} {\bibfield  {journal}
  {\bibinfo  {journal} {New J. Phys.}\ }\textbf {\bibinfo {volume} {19}},\
  \bibinfo {pages} {123010} (\bibinfo {year} {2017})}\BibitemShut {NoStop}%
\bibitem [{\citenamefont {Lovas}\ \emph {et~al.}(2017)\citenamefont {Lovas},
  \citenamefont {D{\'o}ra}, \citenamefont {Demler},\ and\ \citenamefont
  {Zar{\'a}nd}}]{lovas2017full}%
  \BibitemOpen
  \bibfield  {author} {\bibinfo {author} {\bibfnamefont {I.}~\bibnamefont
  {Lovas}}, \bibinfo {author} {\bibfnamefont {B.}~\bibnamefont {D{\'o}ra}},
  \bibinfo {author} {\bibfnamefont {E.}~\bibnamefont {Demler}}, \ and\ \bibinfo
  {author} {\bibfnamefont {G.}~\bibnamefont {Zar{\'a}nd}},\ }\href {\doibase
  10.1103/PhysRevA.95.053621} {\bibfield  {journal} {\bibinfo  {journal} {Phys.
  Rev. A}\ }\textbf {\bibinfo {volume} {95}},\ \bibinfo {pages} {053621}
  (\bibinfo {year} {2017})}\BibitemShut {NoStop}%
\bibitem [{\citenamefont {Bastianello}\ \emph
  {et~al.}(2018{\natexlab{b}})\citenamefont {Bastianello}, \citenamefont
  {Piroli},\ and\ \citenamefont {Calabrese}}]{bastianello2018exact}%
  \BibitemOpen
  \bibfield  {author} {\bibinfo {author} {\bibfnamefont {A.}~\bibnamefont
  {Bastianello}}, \bibinfo {author} {\bibfnamefont {L.}~\bibnamefont {Piroli}},
  \ and\ \bibinfo {author} {\bibfnamefont {P.}~\bibnamefont {Calabrese}},\
  }\href {\doibase 10.1103/PhysRevLett.120.190601} {\bibfield  {journal}
  {\bibinfo  {journal} {Phys. Rev. Lett.}\ }\textbf {\bibinfo {volume} {120}},\
  \bibinfo {pages} {190601} (\bibinfo {year} {2018}{\natexlab{b}})}\BibitemShut
  {NoStop}%
\bibitem [{\citenamefont {Bastianello}\ and\ \citenamefont
  {Piroli}(2018)}]{bastianello2018sinh}%
  \BibitemOpen
  \bibfield  {author} {\bibinfo {author} {\bibfnamefont {A.}~\bibnamefont
  {Bastianello}}\ and\ \bibinfo {author} {\bibfnamefont {L.}~\bibnamefont
  {Piroli}},\ }\href {\doibase 10.1088/1742-5468/aaeb48} {\bibfield  {journal}
  {\bibinfo  {journal} {J. Stat. Mech.: Theory Exp.}\ }\textbf {\bibinfo
  {volume} {2018}},\ \bibinfo {pages} {113104} (\bibinfo {year}
  {2018})}\BibitemShut {NoStop}%
\bibitem [{\citenamefont {Arzamasovs}\ and\ \citenamefont
  {Gangardt}(2019)}]{arzamasovs2019full}%
  \BibitemOpen
  \bibfield  {author} {\bibinfo {author} {\bibfnamefont {M.}~\bibnamefont
  {Arzamasovs}}\ and\ \bibinfo {author} {\bibfnamefont {D.~M.}\ \bibnamefont
  {Gangardt}},\ }\href {\doibase 10.1103/PhysRevLett.122.120401} {\bibfield
  {journal} {\bibinfo  {journal} {Phys. Rev. Lett.}\ }\textbf {\bibinfo
  {volume} {122}},\ \bibinfo {pages} {120401} (\bibinfo {year}
  {2019})}\BibitemShut {NoStop}%
\bibitem [{\citenamefont {Silva}(2008)}]{silva2008statistics}%
  \BibitemOpen
  \bibfield  {author} {\bibinfo {author} {\bibfnamefont {A.}~\bibnamefont
  {Silva}},\ }\href {\doibase 10.1103/PhysRevLett.101.120603} {\bibfield
  {journal} {\bibinfo  {journal} {Phys. Rev. Lett.}\ }\textbf {\bibinfo
  {volume} {101}},\ \bibinfo {pages} {120603} (\bibinfo {year}
  {2008})}\BibitemShut {NoStop}%
\bibitem [{\citenamefont {Gambassi}\ and\ \citenamefont
  {Silva}(2012)}]{gambassi2012large}%
  \BibitemOpen
  \bibfield  {author} {\bibinfo {author} {\bibfnamefont {A.}~\bibnamefont
  {Gambassi}}\ and\ \bibinfo {author} {\bibfnamefont {A.}~\bibnamefont
  {Silva}},\ }\href {\doibase 10.1103/PhysRevLett.109.250602} {\bibfield
  {journal} {\bibinfo  {journal} {Phys. Rev. Lett.}\ }\textbf {\bibinfo
  {volume} {109}},\ \bibinfo {pages} {250602} (\bibinfo {year}
  {2012})}\BibitemShut {NoStop}%
\bibitem [{\citenamefont {Sotiriadis}\ \emph {et~al.}(2013)\citenamefont
  {Sotiriadis}, \citenamefont {Gambassi},\ and\ \citenamefont
  {Silva}}]{sotiriadis2013statistics}%
  \BibitemOpen
  \bibfield  {author} {\bibinfo {author} {\bibfnamefont {S.}~\bibnamefont
  {Sotiriadis}}, \bibinfo {author} {\bibfnamefont {A.}~\bibnamefont
  {Gambassi}}, \ and\ \bibinfo {author} {\bibfnamefont {A.}~\bibnamefont
  {Silva}},\ }\href {\doibase 10.1103/PhysRevE.87.052129} {\bibfield  {journal}
  {\bibinfo  {journal} {Phys. Rev. E}\ }\textbf {\bibinfo {volume} {87}},\
  \bibinfo {pages} {052129} (\bibinfo {year} {2013})}\BibitemShut {NoStop}%
\bibitem [{\citenamefont {Smacchia}\ and\ \citenamefont
  {Silva}(2013)}]{smacchia2013work}%
  \BibitemOpen
  \bibfield  {author} {\bibinfo {author} {\bibfnamefont {P.}~\bibnamefont
  {Smacchia}}\ and\ \bibinfo {author} {\bibfnamefont {A.}~\bibnamefont
  {Silva}},\ }\href {\doibase 10.1103/PhysRevE.88.042109} {\bibfield  {journal}
  {\bibinfo  {journal} {Phys. Rev. E}\ }\textbf {\bibinfo {volume} {88}},\
  \bibinfo {pages} {042109} (\bibinfo {year} {2013})}\BibitemShut {NoStop}%
\bibitem [{\citenamefont {Rotondo}\ \emph {et~al.}(2018)\citenamefont
  {Rotondo}, \citenamefont {Min{\'a}{\v{r}}}, \citenamefont {Garrahan},
  \citenamefont {Lesanovsky},\ and\ \citenamefont
  {Marcuzzi}}]{rotondo2018singularities}%
  \BibitemOpen
  \bibfield  {author} {\bibinfo {author} {\bibfnamefont {P.}~\bibnamefont
  {Rotondo}}, \bibinfo {author} {\bibfnamefont {J.}~\bibnamefont
  {Min{\'a}{\v{r}}}}, \bibinfo {author} {\bibfnamefont {J.~P.}\ \bibnamefont
  {Garrahan}}, \bibinfo {author} {\bibfnamefont {I.}~\bibnamefont
  {Lesanovsky}}, \ and\ \bibinfo {author} {\bibfnamefont {M.}~\bibnamefont
  {Marcuzzi}},\ }\href {\doibase 10.1103/PhysRevB.98.184303} {\bibfield
  {journal} {\bibinfo  {journal} {Phys. Rev. B}\ }\textbf {\bibinfo {volume}
  {98}},\ \bibinfo {pages} {184303} (\bibinfo {year} {2018})}\BibitemShut
  {NoStop}%
\bibitem [{\citenamefont {Palmai}\ and\ \citenamefont
  {Sotiriadis}(2014)}]{palmai2014quench}%
  \BibitemOpen
  \bibfield  {author} {\bibinfo {author} {\bibfnamefont {T.}~\bibnamefont
  {Palmai}}\ and\ \bibinfo {author} {\bibfnamefont {S.}~\bibnamefont
  {Sotiriadis}},\ }\href {\doibase 10.1103/PhysRevE.90.052102} {\bibfield
  {journal} {\bibinfo  {journal} {Phys. Rev. E}\ }\textbf {\bibinfo {volume}
  {90}},\ \bibinfo {pages} {052102} (\bibinfo {year} {2014})}\BibitemShut
  {NoStop}%
\bibitem [{\citenamefont {Rylands}\ and\ \citenamefont
  {Andrei}(2019{\natexlab{a}})}]{rylands2019loschmidt}%
  \BibitemOpen
  \bibfield  {author} {\bibinfo {author} {\bibfnamefont {C.}~\bibnamefont
  {Rylands}}\ and\ \bibinfo {author} {\bibfnamefont {N.}~\bibnamefont
  {Andrei}},\ }\href {\doibase 10.1103/PhysRevB.99.085133} {\bibfield
  {journal} {\bibinfo  {journal} {Phys. Rev. B}\ }\textbf {\bibinfo {volume}
  {99}},\ \bibinfo {pages} {085133} (\bibinfo {year}
  {2019}{\natexlab{a}})}\BibitemShut {NoStop}%
\bibitem [{\citenamefont {Perfetto}\ \emph {et~al.}(2019)\citenamefont
  {Perfetto}, \citenamefont {Piroli},\ and\ \citenamefont
  {Gambassi}}]{perfetto2019quench}%
  \BibitemOpen
  \bibfield  {author} {\bibinfo {author} {\bibfnamefont {G.}~\bibnamefont
  {Perfetto}}, \bibinfo {author} {\bibfnamefont {L.}~\bibnamefont {Piroli}}, \
  and\ \bibinfo {author} {\bibfnamefont {A.}~\bibnamefont {Gambassi}},\ }\href
  {\doibase 10.1103/PhysRevE.100.032114} {\bibfield  {journal} {\bibinfo
  {journal} {Phys. Rev. E}\ }\textbf {\bibinfo {volume} {100}},\ \bibinfo
  {pages} {032114} (\bibinfo {year} {2019})}\BibitemShut {NoStop}%
\bibitem [{\citenamefont {Rylands}\ and\ \citenamefont
  {Andrei}(2019{\natexlab{b}})}]{rylands2019quantum}%
  \BibitemOpen
  \bibfield  {author} {\bibinfo {author} {\bibfnamefont {C.}~\bibnamefont
  {Rylands}}\ and\ \bibinfo {author} {\bibfnamefont {N.}~\bibnamefont
  {Andrei}},\ }\href {\doibase 10.1103/PhysRevB.100.064308} {\bibfield
  {journal} {\bibinfo  {journal} {Phys. Rev. B}\ }\textbf {\bibinfo {volume}
  {100}},\ \bibinfo {pages} {064308} (\bibinfo {year}
  {2019}{\natexlab{b}})}\BibitemShut {NoStop}%
\bibitem [{\citenamefont {Levitov}\ and\ \citenamefont
  {Lesovik}(1993)}]{levitov1993charge}%
  \BibitemOpen
  \bibfield  {author} {\bibinfo {author} {\bibfnamefont {L.~S.}\ \bibnamefont
  {Levitov}}\ and\ \bibinfo {author} {\bibfnamefont {G.~B.}\ \bibnamefont
  {Lesovik}},\ }\href
  {http://www.jetpletters.ac.ru/ps/1186/article_17907.shtml} {\bibfield
  {journal} {\bibinfo  {journal} {JETP Lett.}\ }\textbf {\bibinfo {volume}
  {58}},\ \bibinfo {pages} {230} (\bibinfo {year} {1993})}\BibitemShut
  {NoStop}%
\bibitem [{\citenamefont {Levitov}\ and\ \citenamefont
  {Lesovik}(1994)}]{levitov1994quantum}%
  \BibitemOpen
  \bibfield  {author} {\bibinfo {author} {\bibfnamefont {L.~S.}\ \bibnamefont
  {Levitov}}\ and\ \bibinfo {author} {\bibfnamefont {G.~B.}\ \bibnamefont
  {Lesovik}},\ }\href {https://arxiv.org/abs/cond-mat/9401004} {\bibfield
  {journal} {\bibinfo  {journal} {arXiv:cond-mat/9401004}\ } (\bibinfo {year}
  {1994})}\BibitemShut {NoStop}%
\bibitem [{\citenamefont {Levitov}\ \emph {et~al.}(1996)\citenamefont
  {Levitov}, \citenamefont {Lee},\ and\ \citenamefont
  {Lesovik}}]{levitov1996electron}%
  \BibitemOpen
  \bibfield  {author} {\bibinfo {author} {\bibfnamefont {L.~S.}\ \bibnamefont
  {Levitov}}, \bibinfo {author} {\bibfnamefont {H.}~\bibnamefont {Lee}}, \ and\
  \bibinfo {author} {\bibfnamefont {G.~B.}\ \bibnamefont {Lesovik}},\ }\href
  {\doibase https://doi.org/10.1063/1.531672} {\bibfield  {journal} {\bibinfo
  {journal} {J. Math. Phys.}\ }\textbf {\bibinfo {volume} {37}},\ \bibinfo
  {pages} {4845} (\bibinfo {year} {1996})}\BibitemShut {NoStop}%
\bibitem [{\citenamefont {Klich}(2003)}]{klich2003elementary}%
  \BibitemOpen
  \bibfield  {author} {\bibinfo {author} {\bibfnamefont {I.}~\bibnamefont
  {Klich}},\ }in\ \href {\doibase https://doi.org/10.1007/978-94-010-0089-5_19}
  {\emph {\bibinfo {booktitle} {Quantum Noise in Mesoscopic Physics}}}\
  (\bibinfo  {publisher} {Springer},\ \bibinfo {year} {2003})\ pp.\ \bibinfo
  {pages} {397--402}\BibitemShut {NoStop}%
\bibitem [{\citenamefont {Sch{\"o}nhammer}(2007)}]{schonhammer2007full}%
  \BibitemOpen
  \bibfield  {author} {\bibinfo {author} {\bibfnamefont {K.}~\bibnamefont
  {Sch{\"o}nhammer}},\ }\href {\doibase 10.1103/PhysRevB.75.205329} {\bibfield
  {journal} {\bibinfo  {journal} {Phys. Rev. B}\ }\textbf {\bibinfo {volume}
  {75}},\ \bibinfo {pages} {205329} (\bibinfo {year} {2007})}\BibitemShut
  {NoStop}%
\bibitem [{\citenamefont {Bernard}\ and\ \citenamefont
  {Doyon}(2012{\natexlab{b}})}]{bernard2012full}%
  \BibitemOpen
  \bibfield  {author} {\bibinfo {author} {\bibfnamefont {D.}~\bibnamefont
  {Bernard}}\ and\ \bibinfo {author} {\bibfnamefont {B.}~\bibnamefont
  {Doyon}},\ }\href {\doibase https://doi.org/10.1063/1.4763471} {\bibfield
  {journal} {\bibinfo  {journal} {J. Math. Phys.}\ }\textbf {\bibinfo {volume}
  {53}},\ \bibinfo {pages} {122302} (\bibinfo {year}
  {2012}{\natexlab{b}})}\BibitemShut {NoStop}%
\bibitem [{\citenamefont {Klich}(2014)}]{klich2014note}%
  \BibitemOpen
  \bibfield  {author} {\bibinfo {author} {\bibfnamefont {I.}~\bibnamefont
  {Klich}},\ }\href {\doibase 10.1088/1742-5468/2014/11/P11006} {\bibfield
  {journal} {\bibinfo  {journal} {J. Stat. Mech.: Theory Exp.}\ }\textbf
  {\bibinfo {volume} {2014}},\ \bibinfo {pages} {P11006} (\bibinfo {year}
  {2014})}\BibitemShut {NoStop}%
\bibitem [{\citenamefont {Yoshimura}(2018)}]{yoshimura2018full}%
  \BibitemOpen
  \bibfield  {author} {\bibinfo {author} {\bibfnamefont {T.}~\bibnamefont
  {Yoshimura}},\ }\href {\doibase https://doi.org/10.1088/1751-8121/aae769}
  {\bibfield  {journal} {\bibinfo  {journal} {J. Phys. A: Math. Theor.}\
  }\textbf {\bibinfo {volume} {51}},\ \bibinfo {pages} {475002} (\bibinfo
  {year} {2018})}\BibitemShut {NoStop}%
\bibitem [{\citenamefont {Gamayun}\ \emph
  {et~al.}(2020{\natexlab{a}})\citenamefont {Gamayun}, \citenamefont
  {Lychkovskiy},\ and\ \citenamefont {Caux}}]{Gamayun1}%
  \BibitemOpen
  \bibfield  {author} {\bibinfo {author} {\bibfnamefont {O.}~\bibnamefont
  {Gamayun}}, \bibinfo {author} {\bibfnamefont {O.}~\bibnamefont
  {Lychkovskiy}}, \ and\ \bibinfo {author} {\bibfnamefont {J.-S.}\ \bibnamefont
  {Caux}},\ }\href {\doibase 10.21468/SciPostPhys.8.3.036} {\bibfield
  {journal} {\bibinfo  {journal} {SciPost Phys.}\ }\textbf {\bibinfo {volume}
  {8}},\ \bibinfo {pages} {36} (\bibinfo {year}
  {2020}{\natexlab{a}})}\BibitemShut {NoStop}%
\bibitem [{\citenamefont {Gamayun}\ \emph
  {et~al.}(2020{\natexlab{b}})\citenamefont {Gamayun}, \citenamefont
  {Slobodeniuk}, \citenamefont {Caux},\ and\ \citenamefont
  {Lychkovskiy}}]{Gamayun2}%
  \BibitemOpen
  \bibfield  {author} {\bibinfo {author} {\bibfnamefont {O.}~\bibnamefont
  {Gamayun}}, \bibinfo {author} {\bibfnamefont {A.}~\bibnamefont
  {Slobodeniuk}}, \bibinfo {author} {\bibfnamefont {J.-S.}\ \bibnamefont
  {Caux}}, \ and\ \bibinfo {author} {\bibfnamefont {O.}~\bibnamefont
  {Lychkovskiy}},\ }\href {https://arxiv.org/abs/2006.02400} {\bibfield
  {journal} {\bibinfo  {journal} {arXiv:2006.02400}\ } (\bibinfo {year}
  {2020}{\natexlab{b}})}\BibitemShut {NoStop}%
\bibitem [{\citenamefont {Saito}\ and\ \citenamefont
  {Dhar}(2007)}]{saito2007fluctuation}%
  \BibitemOpen
  \bibfield  {author} {\bibinfo {author} {\bibfnamefont {K.}~\bibnamefont
  {Saito}}\ and\ \bibinfo {author} {\bibfnamefont {A.}~\bibnamefont {Dhar}},\
  }\href {\doibase 10.1103/PhysRevLett.99.180601} {\bibfield  {journal}
  {\bibinfo  {journal} {Phys. Rev. Lett.}\ }\textbf {\bibinfo {volume} {99}},\
  \bibinfo {pages} {180601} (\bibinfo {year} {2007})}\BibitemShut {NoStop}%
\bibitem [{\citenamefont {Myers}\ \emph {et~al.}(2020)\citenamefont {Myers},
  \citenamefont {Bhaseen}, \citenamefont {Harris},\ and\ \citenamefont
  {Doyon}}]{doyonMyers2020}%
  \BibitemOpen
  \bibfield  {author} {\bibinfo {author} {\bibfnamefont {J.}~\bibnamefont
  {Myers}}, \bibinfo {author} {\bibfnamefont {M.~J.}\ \bibnamefont {Bhaseen}},
  \bibinfo {author} {\bibfnamefont {R.~J.}\ \bibnamefont {Harris}}, \ and\
  \bibinfo {author} {\bibfnamefont {B.}~\bibnamefont {Doyon}},\ }\href
  {\doibase 10.21468/SciPostPhys.8.1.007} {\bibfield  {journal} {\bibinfo
  {journal} {SciPost Phys.}\ }\textbf {\bibinfo {volume} {8}},\ \bibinfo
  {pages} {7} (\bibinfo {year} {2020})}\BibitemShut {NoStop}%
\bibitem [{\citenamefont {Doyon}\ and\ \citenamefont
  {Myers}(2020)}]{doyon2020fluctuations}%
  \BibitemOpen
  \bibfield  {author} {\bibinfo {author} {\bibfnamefont {B.}~\bibnamefont
  {Doyon}}\ and\ \bibinfo {author} {\bibfnamefont {J.}~\bibnamefont {Myers}},\
  }\href {\doibase 10.1007/s00023-019-00860-w} {\bibfield  {journal} {\bibinfo
  {journal} {Ann. Henri Poincar{\'e}}\ }\textbf {\bibinfo {volume} {21}},\
  \bibinfo {pages} {255} (\bibinfo {year} {2020})}\BibitemShut {NoStop}%
\bibitem [{\citenamefont {Sachdev}(2007)}]{sachdev2007quantum}%
  \BibitemOpen
  \bibfield  {author} {\bibinfo {author} {\bibfnamefont {S.}~\bibnamefont
  {Sachdev}},\ }\href {\doibase https://doi.org/10.1002/9780470022184.hmm108}
  {\emph {\bibinfo {title} {Quantum phase transitions}}}\ (\bibinfo
  {publisher} {Wiley Online Library},\ \bibinfo {year} {2007})\BibitemShut
  {NoStop}%
\bibitem [{\citenamefont {Lievens}\ \emph {et~al.}(2008)\citenamefont
  {Lievens}, \citenamefont {Stoilova},\ and\ \citenamefont {Van~der
  Jeugt}}]{lievens2008linear}%
  \BibitemOpen
  \bibfield  {author} {\bibinfo {author} {\bibfnamefont {S.}~\bibnamefont
  {Lievens}}, \bibinfo {author} {\bibfnamefont {N.}~\bibnamefont {Stoilova}}, \
  and\ \bibinfo {author} {\bibfnamefont {J.}~\bibnamefont {Van~der Jeugt}},\
  }\href {\doibase 10.1063/1.2948894} {\bibfield  {journal} {\bibinfo
  {journal} {J. Math. Phys.}\ }\textbf {\bibinfo {volume} {49}},\ \bibinfo
  {pages} {073502} (\bibinfo {year} {2008})}\BibitemShut {NoStop}%
\bibitem [{\citenamefont {Fagotti}(2017)}]{Fagotti2017Higher}%
  \BibitemOpen
  \bibfield  {author} {\bibinfo {author} {\bibfnamefont {M.}~\bibnamefont
  {Fagotti}},\ }\href {\doibase 10.1103/PhysRevB.96.220302} {\bibfield
  {journal} {\bibinfo  {journal} {Phys. Rev. B}\ }\textbf {\bibinfo {volume}
  {96}},\ \bibinfo {pages} {220302} (\bibinfo {year} {2017})}\BibitemShut
  {NoStop}%
\bibitem [{\citenamefont {Tracy}\ and\ \citenamefont
  {Widom}(1994)}]{tracy1994level}%
  \BibitemOpen
  \bibfield  {author} {\bibinfo {author} {\bibfnamefont {C.~A.}\ \bibnamefont
  {Tracy}}\ and\ \bibinfo {author} {\bibfnamefont {H.}~\bibnamefont {Widom}},\
  }\href {\doibase 10.1007/BF02100489} {\bibfield  {journal} {\bibinfo
  {journal} {Commun. Math. Phys.}\ }\textbf {\bibinfo {volume} {159}},\
  \bibinfo {pages} {151} (\bibinfo {year} {1994})}\BibitemShut {NoStop}%
\bibitem [{{\relax DLMF}()}]{NIST:DLMF}%
  \BibitemOpen
  {\relax DLMF},\ \href {http://dlmf.nist.gov/} {\enquote {\bibinfo {title}
  {{\it NIST Digital Library of Mathematical Functions}},}\ }\bibinfo
  {howpublished} {Release 1.0.27 of 2020-06-15} (\bibinfo {year} {2020}),\
  \bibinfo {note} {{F}.~W.~J. Olver, A.~B. {Olde Daalhuis}, D.~W. Lozier, B.~I.
  Schneider, R.~F. Boisvert, C.~W. Clark, B.~R. Miller, B.~V. Saunders, H.~S.
  Cohl, and M.~A. McClain, eds.}\BibitemShut {Stop}%
\bibitem [{\citenamefont {Huang}(2009)}]{huang2009introduction}%
  \BibitemOpen
  \bibfield  {author} {\bibinfo {author} {\bibfnamefont {K.}~\bibnamefont
  {Huang}},\ }\href@noop {} {\emph {\bibinfo {title} {Introduction to
  statistical physics}}}\ (\bibinfo  {publisher} {CRC press},\ \bibinfo {year}
  {2009})\BibitemShut {NoStop}%
\bibitem [{\citenamefont {Kormos}\ \emph {et~al.}(2018)\citenamefont {Kormos},
  \citenamefont {Moca},\ and\ \citenamefont {Zar\'and}}]{kormossemiclassic1}%
  \BibitemOpen
  \bibfield  {author} {\bibinfo {author} {\bibfnamefont {M.}~\bibnamefont
  {Kormos}}, \bibinfo {author} {\bibfnamefont {C.~P.}\ \bibnamefont {Moca}}, \
  and\ \bibinfo {author} {\bibfnamefont {G.}~\bibnamefont {Zar\'and}},\ }\href
  {\doibase 10.1103/PhysRevE.98.032105} {\bibfield  {journal} {\bibinfo
  {journal} {Phys. Rev. E}\ }\textbf {\bibinfo {volume} {98}},\ \bibinfo
  {pages} {032105} (\bibinfo {year} {2018})}\BibitemShut {NoStop}%
\bibitem [{\citenamefont {Brantut}\ \emph {et~al.}(2013)\citenamefont
  {Brantut}, \citenamefont {Grenier}, \citenamefont {Meineke}, \citenamefont
  {Stadler}, \citenamefont {Krinner}, \citenamefont {Kollath}, \citenamefont
  {Esslinger},\ and\ \citenamefont {Georges}}]{Brantut713}%
  \BibitemOpen
  \bibfield  {author} {\bibinfo {author} {\bibfnamefont {J.-P.}\ \bibnamefont
  {Brantut}}, \bibinfo {author} {\bibfnamefont {C.}~\bibnamefont {Grenier}},
  \bibinfo {author} {\bibfnamefont {J.}~\bibnamefont {Meineke}}, \bibinfo
  {author} {\bibfnamefont {D.}~\bibnamefont {Stadler}}, \bibinfo {author}
  {\bibfnamefont {S.}~\bibnamefont {Krinner}}, \bibinfo {author} {\bibfnamefont
  {C.}~\bibnamefont {Kollath}}, \bibinfo {author} {\bibfnamefont
  {T.}~\bibnamefont {Esslinger}}, \ and\ \bibinfo {author} {\bibfnamefont
  {A.}~\bibnamefont {Georges}},\ }\href {\doibase 10.1126/science.1242308}
  {\bibfield  {journal} {\bibinfo  {journal} {Science}\ }\textbf {\bibinfo
  {volume} {342}},\ \bibinfo {pages} {713} (\bibinfo {year}
  {2013})}\BibitemShut {NoStop}%
\end{thebibliography}%
\end{document}